\documentclass[aps,pra,superscriptaddress,twocolumn,showpacs]{revtex4-1}

\usepackage[utf8]{inputenc}
\usepackage[T1]{fontenc}
\usepackage[german,english]{babel}

\usepackage{amsmath}
\usepackage{mathtools}
\usepackage{amssymb}
\usepackage{amsthm}
\usepackage{microtype} 
\usepackage{quoting} 
\quotingsetup{font=small}

\usepackage[pdftex]{graphicx}
\usepackage{fancyhdr}
\usepackage{bm}
\usepackage{comment}
\usepackage{booktabs}
\usepackage[normalem]{ulem} 
\usepackage{eurosym} 
\usepackage{braket}
\usepackage{bbold} 
\usepackage{emptypage}
\usepackage{epstopdf}
\usepackage{verbatim}
\usepackage{enumitem}
\usepackage{float} 

\graphicspath{{Figures/PNG/}{Figures/PDF/}{Figures/EPS/}{Figures/TEX/}{Figures/}} 

\usepackage[usenames,dvipsnames]{xcolor} 
\definecolor{islamicgreen}{rgb}{0.0, 0.56, 0.0}

\newcommand{\silcom}[1]{\textcolor{islamicgreen}{#1}}

\newcommand{\beq}{\begin{equation}}
\newcommand{\eeq}{\end{equation}}
\newcommand{\heff}{\hbar_{\text{eff}}}
\newcommand{\cl}{\text{cl}}

\newcommand {\nn} \nonumber

\renewcommand{\ket}[1]{\left|#1\right\rangle}
\renewcommand{\bra}[1]{\left\langle #1\right|}

\DeclarePairedDelimiter{\abs}{\lvert}{\rvert} 

\DeclareMathOperator{\Tr}{Tr}
\DeclareMathOperator{\Min}{Min}
\DeclareMathOperator{\Max}{Max}
\DeclareMathOperator{\arccoth}{arccoth}

\renewcommand{\sim}{\thicksim}

{\left\lbrace\begin{array}{@{}l@{}}}%
{\end{array}\right.}

\usepackage[bookmarks=true,colorlinks,linkcolor=blue,urlcolor=blue,citecolor=blue]{hyperref} 

\def\ro{\hat{\rho}}

\begin{document}

\title{
{Bridging} entanglement dynamics and chaos in {semiclassical} 
systems 
} 

\author{Alessio Lerose} 
\affiliation{SISSA --- International School for Advanced Studies, via Bonomea 265, I-34136 Trieste, Italy}
\affiliation{INFN --- Istituto Nazionale di Fisica Nucleare, Sezione di Trieste, I-34136 Trieste, Italy}
\affiliation{Department of Theoretical Physics,
University of Geneva,
Quai Ernest-Ansermet 30,
1205 Geneva, Switzerland}

\author{Silvia Pappalardi} 
\affiliation{SISSA --- International School for Advanced Studies, via Bonomea 265, I-34136 Trieste, Italy}
\affiliation{Abdus Salam ICTP --- International Center for Theoretical Physics, Strada Costiera 11, I-34151 Trieste, Italy}

\date{\today} 

\begin{abstract}
It is widely recognized  that entanglement generation and dynamical chaos are intimately related in semiclassical models via the process of decoherence. In this work, we propose a unifying framework which directly connects the bipartite and multipartite entanglement growth to the quantifiers of classical and quantum chaos. 
In the semiclassical regime, the dynamics of the von Neumann entanglement entropy, the spin squeezing, the quantum Fisher information and the out-of-time-order square commutator are governed by the divergence of nearby phase-space trajectories via the local Lyapunov spectrum, as suggested by previous conjectures in the literature.
{General} analytical predictions are confirmed by detailed numerical calculations for two paradigmatic models, relevant in atomic and optical experiments, which exhibit a regular-to-chaotic transition: the quantum kicked top and the Dicke model. 
\end{abstract}


\maketitle

\section{Introduction}
%
Entanglement as ``\emph{the} characteristic trait of quantum mechanics'' is arguably one of the most puzzling properties of composite quantum systems, ``the one that enforces its entire departure from classical lines of thought'' \cite{Schrdinger1935}.
However, 
the dynamics of quantum systems can often exhibit a semiclassical behavior:
When a system is initialized in a localized wavepacket, quantum observables obey the classical equation of motion at short times, in the spirit of the Ehrenfest theorem \cite{wheeler1998remarks,BerryLecture87,Gutzwillerbook,Haake2010}.
{Accordingly,} understanding how to reconcile quantum entanglement with such semiclassical dynamics has been under debate since the beginning of quantum mechanics. 

An early insight was proposed by the seminal work of Zurek and Paz on decoherence in open systems \cite{Zurek1994, zurek1995quantum}. These authors conjectured that {in a system coupled to an environment}, the rate of entropy growth 
is equal to the sum of the positive Lyapunov exponents, the classical Kolmogorov-Sinai entropy rate \cite{KS1,KS2,Pesin}.
A large body of numerical and analytical studies \cite{zarum_quantum-classical_1998,
Furuya1998, miller_entropy_1999, miller_signatures_1999,Pattanayak1999, 
Monteoliva_2000,Gong2003pra, Gong2003prl, Alicki_2004, Angelo_2005, petitjean_lyapunov_2006, Ribeiro_2010, Romero_2008, Casati_2012, Bonanca11, Souza_2014}
proved consistent with the Zurek-Paz surmise, 
establishing that the transient entanglement generation is associated with decoherence and suggesting further relationships between semiclassical entanglement dynamics and the chaoticity of the underlying trajectories. 
Related work focused on understanding the emergence of quantum irreversibility and decoherence through 
the dynamics of the purity and the Loschmidt echo \cite{ znidaric_fidelity_2003, jacquod_semiclassical_2004}.
%
More recently, the interest in entanglement properties of many-particle systems spread to several theoretical research communities, ranging from statistical physics \cite{Calabrese2005} and condensed matter theory \cite{Latorre2005} to quantum information \cite{amico2008entanglement,horodeckireview} and high-energy physics \cite{wittenrmp18,ryutakayanagi}. 
In this context, the Zurek-Paz conjecture  
has  recently been laid on firm mathematical grounds by Bianchi \emph{et al.} in Ref.\cite{Bianchi2018}, see also Refs.\cite{Asplund2016, BianchiModakRigolEntanglementBosons}.

The technological advances of the last decades in the field of
ultracold-atom physics have allowed for probing the coherent quantum
dynamics of large ensembles of particles on unprecedented time scales \cite{Greiner2002a,Greiner2002b,Bloch2008,Trotzky2008,Cheneau2012,Kauf16,Jorg15,rauer17}.
Interestingly, many atomic, molecular and optical quantum systems,
such as Bose-Einstein condensates \cite{Albiez2005m,strobel2014fisher}, cavity-QED setups \cite{Leroux2010implementa,Spin1Monika} and
trapped ions \cite{Bohnet2016, Grttner2017}, can be described by collective uniform interactions between their $N$ elementary degrees of freedom, 
which give rise to 
a controlled emergence of semiclassical  dynamical behavior in the limit of large $N$ \cite{SciollaBiroliMF}.
Such systems thereby offer a natural playground for experimental efforts toward a deeper understanding of the entanglement growth in the semiclassical regime and beyond.

%

Recent theoretical and experimental studies on quantum information spreading 
have also focused on the concepts of \emph{multipartite entanglement} and \emph{scrambling}. 
The former, as witnessed by the quantum Fisher information (QFI) ~\cite{helstrom1969quantum, Tth2014, pezze2014quantum}, quantifies the number of entangled elements of a composite quantum system. It plays a central role in quantum information theory together with spin squeezing \cite{Braunstein1994, PETZ2011, Pezz2018, Hyllus2012,Tth2012}, and it is currently attracting 
interest because of its relation to thermal susceptibilities, in and out of equilibrium \cite{Hauke2016,Gabbrielli2018, pappalardi2017multipartite, Brenes2020}.
On the other hand, scrambling characterizes quantum chaotic properties in terms of
the growth in time of the square commutator of non-equal time observables, or the closely related out-of-time-order correlators (OTOC) ~\cite{kitaevTalk}. Introduced because of their connection  with the divergence of nearby trajectories in the classical limit ~\cite{kitaevTalk, larkin1969quasiclassical, cotler2018out}, OTOCs are now the focus of a great attention  over  various communities \cite{sekino2008fast, hosur2016chaos, Nahum2018}.
%
%
Despite numerical and analytical investigations suggested the possibility of a
connection among all entanglement and chaos quantifiers \cite{Grttner2018OTO,  lewis2019unifying}, the formulation of a universal semiclassical framework is presently still incomplete.  


In this paper, we present a systematic and unifying approach connecting the bipartite and multipartite entanglement growth to the quantifiers of classical and quantum chaos, which applies whenever a quantum system is characterized by an emergent semiclassical limit.
We target many-particle systems with collective interactions initialized in quasiclassical 
states and let to evolve in isolation.
The quantum fluctuations around the limiting classical trajectory remain under control until the so-called Ehrenfest time scale, which diverges in the thermodynamic limit. 
By expanding the Hamiltonian in terms of the instantaneous quantum fluctuations,
we show that their dynamics determine
all the quantifiers of entanglement and chaos introduced above.
This allows to write down explicit analytical expression for the von Neumann entanglement entropy, the quantum Fisher information, the spin squeezing and the square commutator in the semiclassical regime. 

Following standard semiclassical arguments,
the time-evolving correlation matrix of the quantum fluctuations coincides with the classical Oseledets multiplicative matrix, which encodes the local divergence of nearby semiclassical trajectories via the finite-time Lyapunov spectrum.
Accordingly, the transient growth of the quantum entanglement and chaos quantifiers before saturation is dictated by the nature of the underlying classical phase-space.
In the absence of semiclassical chaos, the entanglement entropy grows logarithmically in time, while the multipartite and the square commutator grow quadratically. Contrarily, whenever chaos is present, the entanglement entropy grows linearly with a slope equal to the sum of the {largest} local Lyapunov exponents (in agreement with the Zurek-Paz conjecture), whereas the quantum Fisher information and the square commutator grow exponentially fast in time with a rate given by twice the local largest Lyapunov exponent.
The same occurs for unstable trajectories in integrable systems, cf. Ref.\cite{SacredLog}.

Our analysis is corroborated by detailed numerical computations in paradigmatic many-body collective quantum systems of current experimental relevance, which undergo an order/chaos transition, namely the quantum kicked top \cite{Haake1987, Haake2010} and the Dicke model \cite{dicke_coherence_1954, deAguiar1992}.
We find excellent agreement with the analytical predictions in all dynamical regimes.
In particular, we observe and rationalize strong deviations from the asymptotic Lyapunov exponents, particularly apparent in regimes with mixed regular-chaotic phase space or with dynamical instabilities.


\medskip

The rest of the paper is organized as follows. Sec.\ref{sec_res} contains a brief summary of the main results of the paper.
In Sec.\ref{sec:models}, we review the semiclassical behavior of quantum systems with collective interactions; we discuss the relevant class of initial states under analysis; we introduce the quantum kicked top and the Dicke model. 
In Sec.\ref{sec_quantities}, we define the indicators of entanglement and chaos on which our analysis is focused: the von Neumann entanglement entropy, the quantum Fisher information (QFI), the spin squeezing, the unequal-time square commutator (OTOC) and the classical Lyapunov spectrum.
In Sec.\ref{sec_theory}, we present our analysis: After rederiving the general dynamics of quantum fluctuations around a semiclassical trajectory, we show how the entanglement measures can be explicitly related to that. 
In Secs.\ref{sec_KT} and \ref{sec_Dicke}, we numerically study the quantum kicked top and the Dicke model. 
Finally, in Sec.\ref{sec_conclusions} we 
present our conclusions and perspectives.

\begin{table}
\begin{tabular}{lccc}
\toprule
Classical trajectory& Stable & Regular  & Chaotic\\
&&&(Unstable)  \\
\midrule
entanglement entropy \cite{Zurek1994, Bianchi2018} 
& oscillations & $\ln t$ & $\Lambda_{K}\, t$ \\
quantum Fisher information
& oscillations 
&$t^2$ & $e^{2 \lambda t}$ \\
square commutator 
& oscillations 
&$t^2$ & $e^{2 \lambda t}$ \\
Ehrenfest time scale 
& $\mathcal{O}(\sqrt{N})$ &$\mathcal{O}(\sqrt{N})$ & $\mathcal{O}(\ln N)$\\
\bottomrule
\end{tabular}
\label{tab:esempio}
\caption{
Summary of  the dynamical behavior of entanglement and chaos quantifiers of $N$-particle collective systems in the semiclassical regime.
The growth of the entanglement quantifiers and the square commutator depends on the nature of the limiting classical trajectory in the $2n$-dimensional phase space (stable configuration, regular or chaotic), up to the Ehrenfest time. 
Here, $\lambda \equiv \lambda_1$ is the maximum Lyapunov exponent, and $\Lambda_{K}= \sum_{k =1}^{2K} \lambda_k$ is the sum of the $2K$ largest  Lyapunov exponents, where $K$ is the number of degrees of freedom associated with the considered subsystem.
For $K$=$n/2$, one has the classical Kolmogorov-Sinai entropy rate $\Lambda_{\text{KS}}=\sum_{k \, : \, \lambda_k>0} \lambda_k$. 
}
\end{table}

\section{Summary}
\label{sec_res}
In this work, we analyze the relation between entanglement growth and chaos in $N$-particle quantum systems characterized by a 
classical limit 
in terms of $n$ effective degrees of freedom, 
such as spin models with collective uniform interactions. 
We study the quantum unitary evolution of an initially coherent state in the semiclassical regime, namely before the 
Ehrenfest time $T_{\text{Eh}}(N)$, which slowly diverges as $N\to\infty$.

The starting point of our analysis is the established semiclassical argument that the instantaneous quantum fluctuations 
 around the classical trajectory, denoted $\delta \hat {\boldsymbol
 \xi}
 $ are quantified via the time-dependent correlation matrix
\beq
\label{eq_Gsumma}
\left[G(t)\right]_{ij} = \frac 1 2
\Big\langle
\delta \hat \xi_i (t)\delta \hat \xi_j (t) + \delta \hat \xi_j (t) \delta \hat \xi_i(t)
\Big\rangle 
\eeq 
with $i,j=1,\dots,2n$,
which, in turn, is equivalent to the classical Oseledets multipicative matrix, whose eigenvalues define the local Lyapunov spectrum.

The content of this work can be summarized as follows:
\begin{enumerate}
%
\item 
All the relevant information on the out-of-equilibrium bipartite and multipartite entanglement growth is encoded in the dynamics of the quantum fluctuations:
The entanglement entropy, the quantum Fisher information density and the out-of-time square commutator can be written explicitly in terms of $G(t)$ in Eq.\eqref{eq_Gsumma}. 
It follows that these quantities grow as dictated by the nature of the underlying classical trajectories, see Table \ref{tab:esempio} for a summary. 
\item The correct semiclassical identification holds between the growth rate of the quantum entanglement and the \emph{finite-time} Lyapunov spectrum, rather than the proper asymptotic one.
 Such discrepancy may be particularly severe in the case of underlying mixed phase-space, intermediate between integrability and fully developed chaos. This is shown explicitly for the kicked top and the Dicke model, where we find perfect agreement between the semiclassical theory and exact finite-size numerical computations.
 \end{enumerate}

\begin{figure}[t]
\centering
\includegraphics[width=0.45\textwidth]{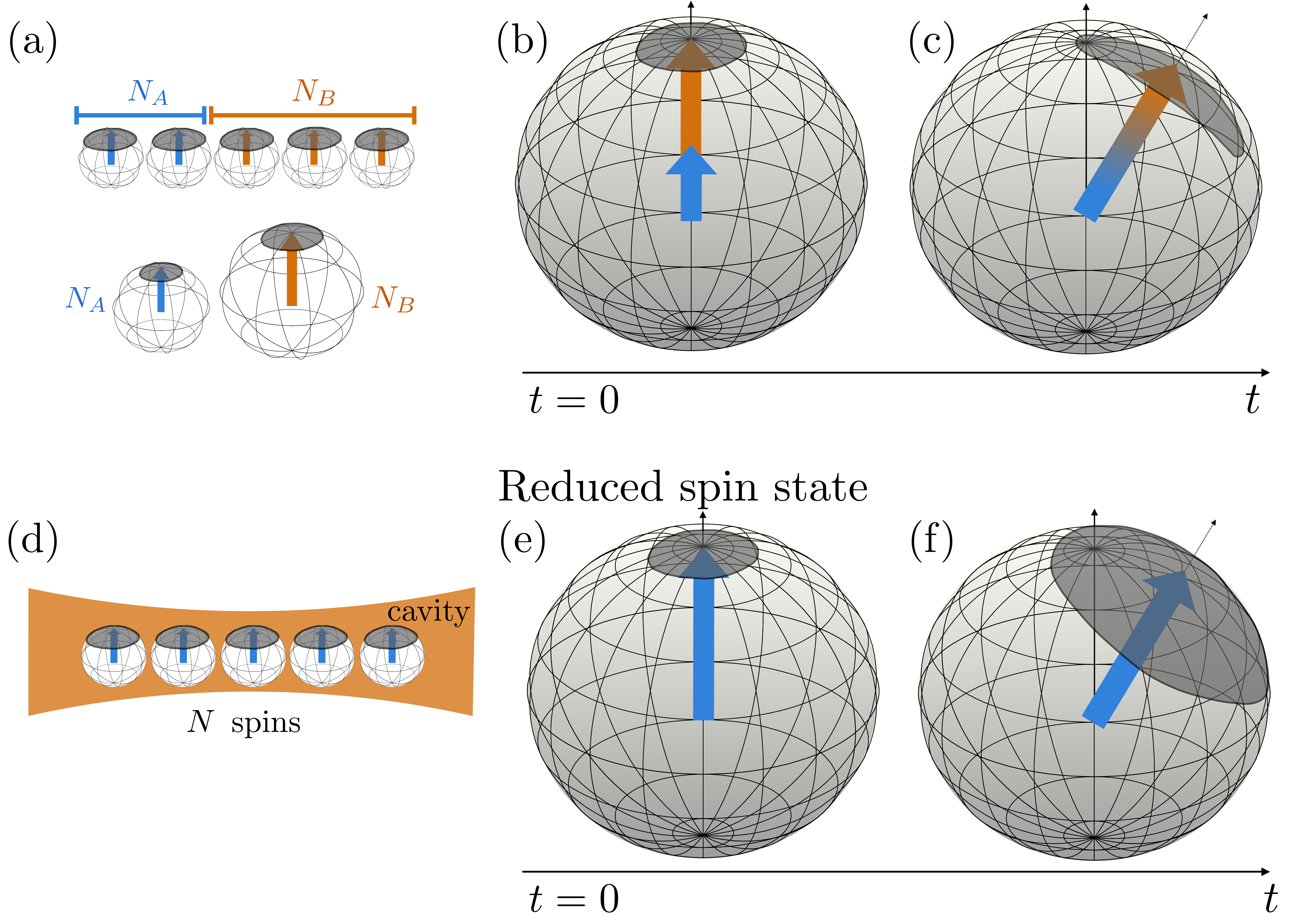}
\caption{
Illustration of the temporal growth of quantum fluctuations and the associated entanglement generation in spin systems with direct interactions (a-c) or with interactions mediated by bosonic ``cavity'' modes (d-f). The collective spin can be represented through an extended grey region on the Bloch sphere surrounding the point identified by its average polarization $\braket{\hat{\mathbf S}(t)}$.
The grey region represents the uncertainty of the collective spin polarization due to quantum fluctuations. 
Top panels (a-c): spin-spin interactions lead to a progressive stretching of the spin fluctuations, or spin squeezing, which determines the growth of bipartite (between subsets of spins $A$ and $B$) and of multipartite entanglement.
Bottom panels (d-f): the entanglement between the spins and the bosonic cavity mode can be read out from the area covered by spin fluctuations on the Bloch sphere.
 }
\label{fig_entDyn}
\end{figure}

One 
appealing and experimentally natural 
consequence is depicted in Fig.\ref{fig_entDyn}. 
In fact, the central result for quadratic bosonic Hamiltonians of Refs. \cite{Bianchi2018,BianchiModakRigolEntanglementBosons}  (see Sec.\ref{sec_theory}) states that the \emph{entanglement entropy of a subsystem $A$, $S_A(t)$, asymptotically coincides with the logarithm of the phase space volume} spanned by the quantum fluctuations of the subsystem degrees of freedom. Hence, entanglement increases because of the growth in time of this 
{``reduced''} volume, while the global phase-space volume is always conserved.
(Notice the interesting correspondence with the quantum Liouville theorem of Ref.\cite{Zhuang2019} in the operator-spreading perspective.) 
%
 This picture corresponds to the well known identification of entanglement generation with the decoherence of the subsystem, illustrated in Fig.\ref{fig_entDyn}.
In an isolated spin system, such as the quantum kicked top, the uncertainty growth in the collective spin of a subset turns out to be dictated by the stretching of the global quantum fluctuations on the Bloch sphere, referred to as spin squeezing \cite{SacredLog}. 
Consequently, all the entanglement and chaos indicators in this semiclassical regime can be reduced to the rate of spin squeezing, which is accessible via standard experimental tools \cite{SacredLog}. 
 Concerning spin-boson systems, such as the Dicke model, the bipartite entanglement entropy between the spins and the boson can be read out from the growth of the volume spanned by the collective spin fluctuations.
 In fact, as illustrated in Fig.\ref{fig_entDyn}, the area covered by spin fluctuations progressively expands during the nonequilibrium evolution,  due to the growth of the entanglement with the cavity mode.
 (This is in contrast to an isolated spin system, where the area spanned by collective spin fluctuations gets stretched in time but is conserved --- compare to Fig.\ref{fig_entDyn}(b-c).) 
 This principle has already been exploited to access bipartite entanglement between the nuclear and the electronic spin in experiments with single atoms \cite{ghose_chaos_2008}. Similar ideas have also been applied to access  entanglement dynamics and chaos in experiments with trapped-ion systems described by the Dicke model  \cite{safavi-naini_verification_2018,lewis2019unifying}.

\section{Models}
\label{sec:models}

In this introductory section, we briefly recall the well-known properties of quantum many-particle systems with collective interactions, with emphasis on their limiting semiclassical description.
We then describe the initial states considered throughout this work. 
We conclude by introducing two paradigmatic models belonging to this class, which will be used to illustrate our analysis: 
the quantum kicked top and the Dicke model. %
This section mostly reviews standard material in the literature. 

\subsection{Collective quantum many-body systems}
\label{sec_models}
We recall how the permutational symmetries 
allow for exactly mapping collective quantum models to systems of few degrees of freedom  characterized by a vanishingly small effective Planck's constant in the thermodynamic limit \cite{SciollaBiroliMF}.

We consider an Hamiltonian $\hat H$ characterizing a uniform all-to-all interaction of $N$ elementary constituents, such as spins or particles. The symmetry under permutations of the degrees of freedom makes the mean-field treatment of the quantum dynamics exact for large $N$.
To show how the semiclassical description emerges, 
we consider an ensemble of $N$ identical $q$-level quantum systems.
A basis of the many-body Hilbert space can be constructed as the tensor product of identical single-unit bases $\{ \Ket{\alpha} \}$ with $\alpha=1,\dots,q$.
%
%
Typical states prepared in experiments may be simple products  of identical single-body states, or ground states. 
Such states generally
belong to the totally-symmetric subspace (TSS) of the many-body Hilbert space simultaneously invariant under all permutations \footnote{Unless permutational symmetry is spontaneously broken or fragmentation phenomena take place \cite{VidalLMGAntiferro}. }.
Due to the symmetry of $\hat H$, the time-evolved state never leaves the TSS.
A basis of the TSS 
can be labeled by the numbers $N_1,\dots,N_q$  of units occupying each level $\alpha=1,\dots,q$ (with $\sum_{\alpha=1}^q N_\alpha = N$).
The dimension of the TSS,
\beq
\text{dim TSS } = \binom{N+q-1}{q-1} \quad  \underset{N\to\infty}{\thicksim} \quad \frac{N^{q-1}}{(q-1)!}\ ,
\eeq
is only polynomially large in $N$, which allows for the exact numerical analysis of large systems. It was shown by Sciolla and Biroli in Ref.\cite{SciollaBiroliMF} that the dynamics of symmetric observables within the TSS is semiclassical in the thermodynamic limit. 
This result is based on the smoothness of the matrix elements of $\hat H$ between two TSS states with respect to small changes in the occupation numbers $N_\alpha \to N_\alpha \pm 1, \pm 2, \dots$ (see Appendix \ref{App:MF_cl} for details).
The Schr{\"o}dinger equation for the TSS wavefunction 
is governed by the effective Hamiltonian $\hat H = N \,  \mathcal{H}_{\cl}(\hat{ 
\mathbf{q}}, \hat{ 
\mathbf{p}})$ 
expressed in terms of the conjugated canonical operators
\beq
\frac{N_\alpha}{N} \mapsto \hat q_\alpha, \qquad -i 
\frac{\partial}{\partial  N_\alpha} \mapsto \hat p_\alpha \ ,
\eeq
with an effective Planck's constant 
\beq
\hbar_{\text{eff}} \equiv \frac{1}{N} \qquad \text{($\hbar=1$ in our units)} \ ,
\eeq
that approaches zero in the thermodynamic limit.
Thus, the quantum dynamics of the original system of all-to-all interacting $q$-level units starting from a quasiclassical state, is equivalent to the semiclassical dynamics of $n=q-1$ collective degrees of freedom, governed by the Hamilton equations generated by $\mathcal{H}_{\cl}$.

A more detailed discussion of the dynamics of quantum fluctuations around the classical limit, and of the range of validity of the semiclassical description, will be reviewed in Sec.\ref{sec_theory} below.

\subsection{Infinite-range spin systems}
\label{sec:infiniteSpinSpyst}

In the specific case a system of $N$ interacting spins-$1/2$ or qubits, the limiting semiclassical description can be formulated in a more direct and intuitive way, by considering that the TSS coincides with the Dicke manifold of maximal collective spin $S=N/2$, whereby the behavior of collective spin operators approaches the classical limit.
In fact, consider general spin models with arbitrary all-to-all multi-body interactions, described by a Hamiltonian of the form
\beq
\label{eq:Hgeneral}
\hat H = -  \sum_{p=1,2,\dots} \Bigg\{ \sum_{\mu_1, \dots, \mu_p = x,y,z} \frac{J_{\mu_1 \dots \mu_p}}{N^{p-1}} \sum_{i_1\ne \dots \ne i_p}^N \hat s^{\mu_1}_{i_1} \dots \hat s^{\mu_p}_{i_p} \Bigg\}
\eeq
where $\mathbf{\hat{s}}_i$, $i=1,2,\dots,N$ are quantum spins-$s$. The rescaling factor $1/N^{p-1}$ is such that the energy contribution of all $p$-body interactions is extensive.
These Hamiltonians can be written in terms of the collective spin of the system 
\begin{equation}
\label{Stot}
\mathbf{\hat{S}}= \sum_{i=1}^N \mathbf{\hat{s}}_i \ ,
\end{equation}
as \footnote{In going from Eq. \eqref{eq:Hgeneral} to Eq. \eqref{eq:Hgeneralcollective}, one needs to add terms with equal indices in the sums. 
Such terms are immaterial for $s=1/2$, while they provide corrections to the coefficients of order $1/N$ in higher-spin systems. 
This small modification does not alter the subsequent analysis and, accordingly, we will simply ignore it.}
\beq
\label{eq:Hgeneralcollective}
\frac{\hat H}{N} =   -  \sum_{p\ge1} \Bigg\{ \frac{1}{N^p}\sum_{\mu_1, \dots, \mu_p = x,y,z}J_{\mu_1 \dots \mu_p}  \; \hat S^{\mu_1} \dots \, \hat S^{\mu_p} \Bigg\}  \ .
\eeq
  The collective spin's magnitude $\abs{\mathbf {\hat S}}=\sqrt{S(S+1)}$ with $S=Ns, Ns-1,Ns-2,\dots$ is extensive and conserved,
$
\Big[\abs{\mathbf {\hat S}}^2,\hat H\Big] = 0
$.
 The  ground state typically  belongs to   the maximal total spin sector, characterized by the maximal spin projection $S = Ns$ (see, e.g., Ref.\cite{VidalLMGAntiferro}).  

For such states with maximal spin,  the thermodynamic limit $N\to\infty$ is equivalent to the semiclassical limit, or, in loose terms, to a classical continuous spin $ \langle \mathbf{\hat S} \rangle / N$ of (conserved) length $ s$.
In fact, these reduced spin variables satisfy a commutation relation of the form
$
\big[ \hat S^{\mu}/N , \hat S^{\nu}/N \big] = (i/N) \; \epsilon_{\mu\nu\rho} \; \hat S^{\rho}/N 
$,
whence one sees that Eq.\eqref{eq:Hgeneralcollective} defines a semiclassical system with an effective Planck's constant $\hbar_{\text{eff}} \equiv 1 /N$ which vanishes in the thermodynamic limit $N\to\infty$. 
The limiting classical Hamiltonian $\hat H/N \to \mathcal{H}_{\text{cl}}$ thus reads 
\beq
\label{eq:Hgeneralcl}
\mathcal{H}_{\text{cl}}(\vec{\mathcal{S}}) =    -  \sum_{\mu_1} J_{\mu_1} \; \mathcal{S}^{\mu_1} 
                        -  \sum_{\mu_1,\mu_2} J_{\mu_1\mu_2} \;  \mathcal{S}^{\mu_1} \mathcal{S}^{\mu_2} 
                        - \dots \ ,
\eeq
where now $\mathbf{\hat{S}}/N \to \vec{\mathcal{S}}$ represents a classical spin on the sphere of radius $s$ which can be parametrized by spherical coordinates: choosing the $z$ direction as the polar axis, $\vec {\mathcal S} = s\hat { \mathbf Z}$ with
\beq
\label{eq:rotating}
\hat {\mathbf Z }= \, 
\begin{pmatrix}
\sin \theta \cos \phi \\
\sin \theta \sin \phi  \\
\cos \theta
\end{pmatrix} \ .
\eeq
The rigorous meaning of the classical limit is that, as ${N\to\infty}$, the ground state expectation values $\braket{\mathbf{\hat{S}}}_{\text{GS}}/N$ of the spin components converge to the minimum point $\vec{\mathcal{S}}^*$ of the classical Hamiltonian $\mathcal{H}_{\text{cl}}$ on the sphere, with vanishingly small quantum fluctuations, and their nonequilibrium evolution $\braket{\mathbf {\hat S}(t)}/N$ upon varying in time some parameter $J=J(t)$ in the Hamiltonian is described by the classical trajectory $\vec{\mathcal{S}}(t)$ on the sphere governed by $\mathcal{H}_{\text{cl}}$, i.e.,
$
\dot{\vec{\mathcal{S}}} = \big\{ \vec{\mathcal{S}}, \mathcal{H}_{\text{cl}} \big\} 
$,
with the Poisson brackets $\{ \mathcal{S}^\mu,\mathcal{S}^\nu \} = \epsilon_{\mu\nu\rho} \mathcal{S}^\rho$. 
{This time-evolution can be recast in terms of the spherical angles $\theta(t),\,\phi(t)$ defined in Eq.\eqref{eq:rotating}.}

If $s>1/2$, a permutationally invariant Hamiltonian may feature additional ``self-interaction'' terms with ${j_1=j_2}$ in Eq.\eqref{eq:Hgeneral}, e.g., energy contributions proportional to 
$
 \sum_{j=1}^N (\hat s_j^z)^2 
$.
Such terms break the conservation of the collective spin magnitude. 
In this case the dynamics take place in the full TSS, which is strictly larger than the Dicke manifold, in agreement with the general mapping of Ref.\cite{SciollaBiroliMF} reviewed above.

\subsection{Beyond global permutational symmetry}
\label{sec_generality}

The semiclassical approach reviewed in the previous Secs.\ref{sec_models} and \ref{sec:infiniteSpinSpyst}  applies to a much wider class of states and models than discussed therein.

One natural extension 
consists of a {composite system of $M$ collective subsystems}, possibly composed of different kinds of degrees of freedom.
Provided the interactions couple the various subsystems uniformly in their elementary units, i.e., via collective operators only, the global system has a semiclassical description.
In fact, when each subsystem is large, the global system will be described by $\sum_{m=1}^M  (q_m-1)$ semiclassical collective degrees of freedom. 
The Dicke model, where $N$ spins interact collectively with a cavity mode (see also Sec.\ref{sec_dicke} below),
can be viewed as an example, as well as the two-species kicked top \cite{miller_signatures_1999}.

A second, closely related generalization,
is represented by {non-symmetric states which partially break the full permutational symmetry}.
Such states may be 
obtained by bringing together a number $M\ll N$ of  initially separated subsystems.
In this case, the full permutational symmetry breaks down into the product of smaller permutational symmetries  acting separately on each subsystem. 
While the full system evolves outside of its TSS, the restricted symmetry allows a description of the dynamics within the product of the TSSs of the $M$ individual subsystems.
The semiclassical theory can thereby be applied in the thermodynamic limit, and one ends up with a few-body semiclassical system described by $M \times (q-1)$ collective degrees of freedom.
In this case, the  Hamiltonian depends on these variables only via the $q-1$ global collective combinations, leaving all the $(M-1)\times (q-1)$  remaining coordinates frozen in their initial values.
A simple example 
is given by a permutationally invariant system of $N$ spins-$1/2$  initially in a random product state $\ket{\dots \nearrow\nearrow\nearrow \swarrow \nearrow\swarrow\swarrow\nearrow \dots }$ of spins pointing up or down along a given axis. Such a state is far away from the Dicke manifold of maximal total spin length $N/2$. 
Grouping together the spins pointing in the same direction into two subsystems $A$ and $B$, with $N_A$ and $N_B$ spins respectively, the global system may be viewed as two interacting collective spins $\hat{\mathbf S}_A$, $\hat{\mathbf S}_B$, of length $N_A/2$ and $N_B/2$ respectively, initially pointing in opposite directions.
In agreement with the above observation, the motion of  the the two spins is not independent: the Hamiltonian generates a nonlinear collective precession,
and the angle between 
$\hat{\mathbf S}_A$ and $\hat{\mathbf S}_B$  is a constant of motion.

\subsection{Quasiclassical initial states}
\label{sec_is}

We consider systems initialized in 
pure nonentangled  states, such as uncorrelated product states. 
These states are routinely prepared in cold-atom experiments via standard techniques.  
In systems described by interacting spins,
a natural class of  nonentangled states is given by 
fully polarized states, in which all spins point along a common direction.
For  composite systems, we will consider uncorrelated products of coherent states.
Weakly entangled initial states may be treated on equal footing.

Such initial states have a semiclassical nature, as their classical phase-space representations via the Wigner function \cite{littlejohn, TWA} correspond to narrow Gaussian distributions centered around a point with a \emph{small variance} of quantum fluctuations of order $\mathcal O(\heff)$.
For example, a system of $N$ spins fully polarized in the $z$ direction has 
\label{eq:ISGaussian}
\begin{align}
\left \langle \frac {\hat{\mathbf S}}N \right \rangle  = s \,
\begin{pmatrix}
0 \\
0 \\
1 \\
\end{pmatrix} \ , \quad 
\left \langle \frac {\delta \hat{\mathbf S}^2}{N^2} \right \rangle  = {\frac 1 2 }s\,
\begin{pmatrix}
\heff \\
\heff \\
0 \\
\end{pmatrix}
 \ ,
\end{align}
(with $\delta \hat {\mathbf S} = \hat{\mathbf S} - \braket{ \hat {\mathbf S}} $), i.e., the collective spin fluctuations in the transverse directions are vanishingly small.
The phase-space representation is this state is given by the Bloch sphere portrait in Fig.\ref{fig_entDyn} (b,e).
More generally, quasiclassical states can be characterized as Gaussian 
phase-space distributions  centered around a point $\big(\mathbf q_{\text{cl}},\mathbf p_{\text{cl}} \big) \equiv
\big(
\braket{\hat{\mathbf q} },
\braket{\hat{\mathbf p} }
\big)
$, with a width of order $\heff$ per degree of freedom.  Ground states of collective models are typically in this class (see, e.g., Refs.\cite{SciollaBiroliMF,VidalLMGAntiferro}).

According to the standard semiclassical theory \cite{littlejohn,bhaduri,TWA},  quantum fluctuations around the classical trajectory $\big(\mathbf q_{\text{cl}}(t),\mathbf p_{\text{cl}}(t) \big)$ will remain approximately Gaussian for a diverging time scale as $\hbar_{\text{eff}}\to0$ (the so-called Ehrenfest time scale) during the  evolution. This will be further discussed in Sec.\ref{sec_tEhr}.

\subsection{The quantum kicked top}
\label{sec_introKT}

As a first illustrative model, we consider a driven model: the quantum kicked top.
The latter can be defined as an ensemble of quantum spins in a magnetic field periodically kicked via collective interactions.
The model is described by the Hamiltonian 
\begin{equation}
\label{eq:H}
\hat H(t) = \alpha \hat S_x + \frac {\beta}{2Ns} \, \hat S_{z}^2  \sum_{n=-\infty}^{\infty} \delta(t-n\tau)
 \ ,
 \end{equation}
where $\hat S_{x,y,z}$ are the collective spin operators in Eq.(\ref{Stot}) and $\tau$ the period of the periodic kicking. 
We fix $\tau=1$. Depending on the value of the kicking strength $\beta$, this
model is known to exhibit a transition between a regular regime and a chaotic one \cite{Haake1987, Haake2010}. 
Being a paradigmatic model for quantum chaotic behavior, its bipartite  \cite{zarum_quantum-classical_1998, miller_signatures_1999, chaudhury_quantum_2009, ghose_entanglement_2004, piga_quantum_2019, wang_entanglement_2004, trail_entanglement_2008, ghose_chaos_2008, kumari_untangling_2019, lombardi_entanglement_2011, stamatiou_quantum_2007}, multipartite entanglement \cite{ghose_chaos_2008, Madhok2014, Fiderer2018, pappaScrambling} and scrambling dynamics \cite{pappaScrambling, sieberer_digital_2019, PilatowskyCameo2020} have been intensively explored.

The stroboscopic time-evolution operator (namely, the time-evolution operator over one period) encodes the dynamical stability properties --- regularity or chaos --- of the system.
It can be written as
\beq
\label{eq:evKT}
\hat U = \hat U_\beta \hat U_\alpha \quad \text{with}\,\, \hat U_\alpha = e^{-i\alpha \hat S^x}, \,\, \hat U_\beta =
e^{-i\frac{\beta}{2Ns} (\hat S^z)^2} \ .
\eeq
%
Due to the collective nature of the interactions, for large $N$ the classical limit is approached. 
In this limit, the stroboscopic evolution from time $t=n$ to $t=n+1$ can be expressed as a discrete map on the Bloch sphere. 
This is obtained as the composition of the two following transformations
\begin{subequations}
\label{eq:classKT}
\begin{align}
\label{eq:KTevAngle1}
& \begin{dcases}
\phi' =   
\arctan \left[ \tan \phi  \cos \alpha
- \frac{\sin 
\alpha 
}{\tan \theta\cos\phi}   \right] + \pi\,  H(-\cos\phi) \\
\cos \theta' =
 \cos \theta \cos \alpha 
 + \sin\theta \sin \phi \sin 
 \alpha \ ,
\end{dcases} 
\\ 
& \begin{dcases}
\label{eq:KTevAngle2}
\phi'' = \phi' + \beta \cos\theta' \\
 \cos \theta'' = \cos \theta' 
\end{dcases}\ ,
\end{align} 
\end{subequations}
where $H(x)$ is the Heaviside step function. 
See Appendix \ref{app:eomKT1} for the derivation.

\subsection{The Dicke model}
\label{sec_dicke}

As a second illustrative example, we consider a conservative system: the Dicke model. 
The latter was originally defined \cite{dicke_coherence_1954} as an ensemble of two-level atoms collectively interacting with a single mode of the quantized electromagnetic field.
Representing the atoms as spins-$1/2$, one can write the Dicke Hamiltonian as
\footnote{The Dicke model is often used in cavity-QED setups, where photon pumping and leakage require a Lindblad description of the dynamics. Here, we will only be concerned with ideally isolated systems, as relevant, e.g., for trapped-ions experiments.}
\beq
\label{eq_Dicke}
\hat H =
\omega_0 \hat S^z+ \omega \hat b^\dagger \hat b + \frac{\gamma}{\sqrt{N}} \frac{\hat b^\dagger + \hat b}{\sqrt{2}} \hat S^x ,
\eeq
where $\hat S^{x,y,z}$ are  spin-$1/2$ collective operators as in Eq.(\ref{Stot}) and $\hat b^\dagger$, $\hat b$ are  creation and annihilation operators of a bosonic mode. For convenience, we define the real quadrature operators $\hat Q = (1/ \sqrt{2}) (\hat b + \hat b^\dagger )$, $\hat P = (1/ i\sqrt{2}) (\hat b - \hat b^\dagger )$.
%
The Dicke model has interesting equilibrium and nonequilibrium properties.
{At zero temperature, the system undergoes a phase-transition at $\gamma_c= \sqrt{\omega \omega_0}/2$,
 between a normal phase $(\gamma <\gamma_c)$ to a super-radiant one $(\gamma >\gamma_c)$ \cite{dicke_coherence_1954}. Furthermore, in the classical limit the accessible phase space may undergo a progressive regular-to-chaotic transition upon varying the energy $E$ and/or the coupling $\gamma$ \cite{emary_chaos_2003, deAguiar1992}. Its bipartite \cite{Furuya1998,lambert_entanglement_2004, lobez_entropy_2016, sinha_chaos_2019,lewis2019unifying}, multipartite entanglement  \cite{song_quantum_2012, wang_quantum_2014, zhang_large-n_2015, mirkhalaf_entanglement_2017, Gietka2019, bhattacherjee_spin_2016,lewis2019unifying} and scrambling dynamics \cite{buijsman_nonergodicity_2017, alavirad_scrambling_2019, dickeHirsch2010, lewis2019unifying} have been intensively explored. }

The dynamics of the Dicke model approach their classical limit for ${N\to\infty}$, described by the classical Hamiltonian ${\hat H / N \to \mathcal{H}_{\text{cl}}}$
\beq
\label{eq_HamDickeClass}
\mathcal{H}_{\text{cl}} = \omega_0 \mathcal{S}^z + \omega \frac{\mathcal{Q}^2 + \mathcal{P}^2}{2} + \gamma \mathcal{Q} \mathcal{S}^x \ ,
\eeq
with
\begin{gather}
\label{eq_scalingDicke}
\hat{\mathbf{S}}  \thicksim
  \frac N 2 \; \mathbf { Z}(t)   \ , \quad 
\hat Q \thicksim \sqrt{N} \mathcal{Q}(t)  \ ,\quad 
\hat P \thicksim \sqrt{N} \mathcal{P}(t)  \ ,
\end{gather}
where $\mathbf { Z}(t)$ represents the direction of the average collective spin  and it is parametrized by the time-dependent angles $\phi(t), \,\theta(t)$ [cf. Eq.(\ref{eq:rotating})].
The functions $\mathcal{Q}(t)$, $\mathcal{P}(t)$ describe the limiting classical dynamics of the bosonic mode. The $\sqrt{N}$ scaling may be understood as the occurrence that  all terms in the Hamiltonian are extensive and  balance each other in equilibrium.
The rescaling in Eq.\eqref{eq_scalingDicke} renders the emergence of the effective Planck's constant $\hbar_{\text{eff}}= 1/N$ manifest.

The classical limit of the Hamiltonian governs the coupled dynamics of the atoms and the radiation field via the Hamilton equations
\beq
\label{eq:DickeCl}
\begin{dcases}
\dot{\mathcal{Q}} = \omega \mathcal{P}  \\
\dot{\mathcal{P}} = - \omega \mathcal{Q} - \frac{\gamma}{2} \sin \theta \cos \phi   \\
 \dot \phi = - \omega_0 \, \tan\theta + \gamma \mathcal{Q} \,  \cos \phi  \\
\dot {\theta} =  - \gamma \mathcal Q \, \sin \phi 
\end{dcases} \ .
\eeq
See Appendix \ref{App:eomDI} for the derivation.

\section{Quantifiers of entanglement and chaos}
\label{sec_quantities}

In this section, we introduce the quantifiers of bipartite and multipartite entanglement and of dynamical chaos, which we will examine in the subsequent analysis.

\subsection{Entanglement entropies}
\label{sec_ee}
For a composite system with Hilbert space ${\mathcal H = \mathcal H_A \otimes \mathcal H_B}$ in a pure state $\ro = \ket{\psi}\bra{\psi}$, the bipartite entanglement between subsystems $A$ and $B$ is encoded in the reduced density matrix ${\ro_A = \Tr_{B}\,{\hat \rho}}$ 
\footnote{The nonvanishing eigenvalues of ${\ro_B = \Tr_{A}\,{\hat \rho}}$ are equal to those of $\ro_A$.}.  
The system is entangled with respect to the bipartition $(A,B)$ if $\ro_A$ (equivalently, $\ro_B$) is not pure.
The amount of bipartite entanglement can be quantified by the Renyi entropies
\beq
\label{eq:Renyi}
S_A^\alpha = - \frac{1}{1-\alpha} \ln \Tr \ro_A^\alpha \ ,
\eeq
parameterized by $\alpha>1$.
The von Neumann entropy is obtained as their limit for $\alpha\to1$, i.e.,  
   \beq
   \label{eq:SA}
   {S_{A} =   - \Tr\big( \ro_A \ln \ro_A \big) } \ .
   \eeq
   
In spatially extended systems with interactions depending on the distance between particles, it is natural to consider bipartitions where subsystem $A$ is constituted by degrees of freedom within a connected region of space, and $B$ its complement. However, in  fully-connected $N$-particle systems, the permutational symmetry makes spatial bipartitions meaningless. Hence, we consider bipartitions  specified by the number $N_A=f_A\, N$ of particles in subsystem $A$ (with $N_B=N-N_A=f_B\,N$).

In addition to spatial bipartitions, one can examine bipartitions between different types of degrees of freedom, irrespective of their spatial location. 
This notion is meaningful in collective models, as well.
For instance, for the Dicke model introduced in Sec.\ref{sec_dicke}, we will focus on the entanglement between the atoms and the cavity field.

For a pictorial representation of the bipartitions considered in this paper, see Fig.\ref{fig_entDyn} (a,d).

\subsection{Quantum Fisher information and spin squeezing}
\label{sec:qfi}

A different approach is to characterize the system via the multipartite entanglement properties of the time-evolving state, as given by the quantum Fisher information (QFI) $\mathcal F(\hat O, \hat \rho)$. This quantity 
was introduced in metrology to bound the precision of the estimation of a parameter $\phi$, conjugated to an observable $\hat O$ using a quantum state $\hat \rho$, via the so-called quantum Cramer-Rao bound $\Delta \phi^2 \geq 1/M \mathcal F(\hat O, \hat \rho)$, where $M$ is the number of independent measurements made in the protocol \cite{Pezz2018}. 
The QFI has key mathematical properties \cite{Braunstein1994,PETZ2011,Tth2014, Pezz2018}, such as convexity, additivity, monotonicity, and it
can be used to probe the multipartite entanglement structure of a quantum state~\cite{Hyllus2012,Tth2012}.  If, for a certain $\hat O$, the QFI density satisfies the inequality
\begin{equation}
 f_Q=\frac{\mathcal{F}(\hat O, \hat \rho)}{N} > m  \ ,
\end{equation}
then, at least $(m+1)$ parties in the system are entangled (with $1\leq m \leq N-1$ a divisor of $N$). Namely, $m$ represents the size of the biggest entangled block of the quantum state.
In particular, if $N-1 \leq f_Q(\hat O) \leq N$, then the state is called genuinely $N$-partite entangled. 
%
%
The QFI has an operational definition in terms of statistical speed of quantum states under external parametric transformations \cite{Braunstein1994, Wootters1981}.
For a general mixed state, described by the density matrix $\hat \rho=\sum_n p_n |n\rangle\langle n|$,  it reads 
\cite{Braunstein1994}
\begin{eqnarray}
\label{full_QFI}
{\mathcal F}(\hat O, \hat \rho)=2\sum_{n,m} \frac{(p_n-p_{m})^2}{p_n+p_{m}} |\langle n| \hat O|m\rangle|^2 {\leq 4 \, \langle \Delta \hat O^2 \rangle } ,
\end{eqnarray}
with $\langle  \Delta \hat O^2 \rangle =\text{Tr}(\hat \rho\, \hat O^2 ) -  \text{Tr}(\hat \rho\,\hat O )^2$. The equality holds for pure states $\hat \rho = \ket{\psi}\bra{\psi}$.
In general, different operators $\hat{O}$ lead to different bounds and there is no systematic method (without some knowledge on the physical system \cite{Hauke2016,Pezz2017Gabri}) to choose the optimal one. 

In this work, we study the dynamical QFI of pure states out of equilibrium $\ket {\psi( t)}$.
In the case of spin systems of Sec.\ref{sec:infiniteSpinSpyst}, we focus on collective spin projections ${\hat O=\hat{S}_{\mathbf n}= \sum_{i=1}^N \mathbf n \cdot \mathbf {\hat{s}}_i}$ in the direction of the $3d$ unit vector $ \mathbf n$, while for composite systems we consider $\hat O = \mathbb{1}_{\overline{S}} \otimes \hat{S}_{\mathbf n}$, where ${\overline{S}}$ is the complement of the spin subsystem. 
The optimal QFI is then given by the maximal fluctuation of the total spin as
\begin{equation}
    \label{QFI}
    f_Q(t) = 4 \max_{\mathbf n} \frac{\langle \Delta \hat S_{\mathbf n}^2(t)\rangle}N \ .
\end{equation}
In the case of composite systems, such as the Dicke model of Sec.\ref{sec_dicke},
Eq.\eqref{QFI} detects not only the correlations between the individual spins, but also the entanglement between the collective spin and the other degrees of freedom (see e.g., Ref.\cite{Gietka2019}).

A related experimentally relevant indicator of multipartite entanglement in spin systems is given by
\emph{spin squeezing}, a concept first introduced in Ref.\cite{HiroshimaSqueezing}.
This observable is associated with the reduction of collective spin quantum fluctuations along one direction at the expense of an enhancement of orthogonal fluctuations, due to the Heisenberg principle. 
Spin squeezing is usually quantified by the minimal transverse variance of collective spin fluctuations  \cite{WinelandSqueezing,SpinSqueezingReview} as
\beq
\xi^2  \equiv \Min_{\abs{\mathbf{u}}=1,\mathbf{u}\perp \mathbf{Z}} \frac{\Big\langle \big(\mathbf{u}\cdot \mathbf {\hat S}\big)^2 \Big\rangle}{ N/4} 
. 
\eeq
The squeezing parameter $\xi^2$ is
equal to $1$ for coherent states, and smaller for squeezed states (see, e.g., Refs.~\cite{HiroshimaSqueezing,SpinSqueezingReview}). 
It has long been known \cite{Sorensen1,Sorensen2, pezze2009entanglement} that collective spin squeezing is a witness of many-body quantum entanglement. 
It is possible to demonstrate that $\xi^2\geq 1/f_Q$ \cite{pezze2009entanglement}, namely there exists a class of states which are not spin-squeezed but can be maximally entangled.
In the following, we will show that a simple relation exists between the QFI and spin squeezing in the semiclassical regime.

\subsection{Scrambling and the square commutator}
\label{sec_sc}

Recently, the study of chaos in quantum systems has received a renewed attention with emphasis on the notion of scrambling. 
This revival has been triggered by Kitaev's proposal to characterize quantum chaotic properties in terms of
the growth in time of the squared non-equal time commutator of two initially commuting
observables~\cite{kitaevTalk}, i.e.,
\begin{equation}
\label{SC}
c(t) = - \langle [\hat B(t), \hat A]^2 \, \rangle \ ,
\end{equation}
where the expectation value is taken in a generic quantum state $\hat \rho$, i.e., $\langle \cdot \rangle = \Tr( \cdot \hat \rho)$. Note that $c(t)\ge 0$ if $\hat A$, $\hat B$ are hermitian. 
This object measures the non-commutativity induced by the
dynamics between two initially commuting operators $\hat A, \hat B$ and it contains out-of-time-order correlators $\langle \hat B(t) \hat A \hat B(t) \hat A\rangle$, characterized by the absence of time-ordering. 

The square commutator was originally introduced in 1969 by Larkin and Ovchinnikov in Ref.\cite{larkin1969quasiclassical}
to describe semi-classically the exponential sensitivity to
initial conditions and the associated Lyapunov exponent. In fact, in the classical limit, $c(t)$ encodes the square of the derivatives of the classical trajectory to respect to the initial
conditions \cite{cotler2018out}. Thus, whenever the classical limit is chaotic, $c(t)$ is expected to grow
exponentially in time, with a rate set by twice the classical Lyapunov exponent \cite{Jalabert2018, Rozenbaum2017, pappaScrambling, dickeHirsch2010, lewis2019unifying, craps2019Lyapunov, rozenbaum2019quantum, Rammensee2018, rautenberg2019classical, prakash2019scrambling, pappalardi2019quantum, craps2019Lyapunov, rautenberg2019classical, schmitt2019semiclassical, wang2019complexity}. In this context, several quantum generalizations of the classical Lyapunov spectrum (see below) have been proposed \cite{Gharibyan2019,Rozenbaum2019,yan2020quantum}.

In the present case, we will study the square commutator in Eq.(\ref{SC}) by taking the expectation value in pure quasiclassical initial states introduced in Sec.\ref{sec_is}. In the case of spin systems, we study the square commutator between two collective spin projections (\ref{Stot}), namely
\beq
\label{eq:cab}
c_{\alpha \beta}(t) = -  \bigg(\frac 1 {Ns} \bigg)^{2} \bra{\psi_0} \left [ \hat S^{\alpha}(t), \hat S^{\beta}(0)\right ]^2 \ket{\psi_0} \ ,
\eeq
where $\alpha, \beta=x,y,z$ and $\ket{\psi_0}$ is a fully polarized spin-coherent initial state.

\subsection{Lyapunov exponents}
\label{sec_LE}

Here we recall the definition of the characteristic Lyapunov exponents, while in the Appendix \ref{app:LE} we report a brief summary of their main properties.
We refer the reader to the abundant literature on this topic, e.g., Ref.~\cite{VulpianiChaosbook}.

The notion of deterministic chaos is associated with the strong sensitivity of the evolved state of a system on its initial condition.
Given a generic $d$-dimensional flow $\dot{\mathbf x} = \mathbf f (\mathbf x)$ in phase space, the measure of the instability of a trajectory $\mathbf x(t)$ is provided by the \emph{maximum Lyapunov exponent}.

Consider an initial condition $\mathbf x(0)$ and a neighboring point $\tilde{\mathbf x}(0)$ displaced by an infinitesimal amount $\tilde{\mathbf x}(0) = \mathbf x(0) + \boldsymbol{\delta}(0)$.
Chaos is defined by an exponential growth in time of the separation between the corresponding trajectories, $\delta(t)=\abs{\tilde{\mathbf x}(t)-\mathbf x(t)}
\sim \delta(0) \exp(\lambda t)$, with $\lambda>0$.
The rate $\lambda$ generally depends on the initial state and on the observation time $t$. 
A non-ambiguous definition thus requires ``time-averaging''
\beq
\label{eq_deflambda}
\lambda \coloneqq 
\lim_{t\to\infty}  \;
\lim_{\delta(0)\to0} \;
\frac 1 t  \ln \frac {\delta(t)}{  \delta(0)} \ .
\eeq
The inner limit $\delta(0)\to0$ translates the (nonlinear) evolution of small displacements away from the initial condition into the (linear) tangential map of the flow along the given trajectory.

The number $\lambda$ above does not exhaust all the possible information on the separation of nearby initial conditions. 
Consider an infinitesimal hypercube surrounding the initial condition $\mathbf x(0)$, identified by
$d$ independent infinitesimal displacements $\{\mathbf w^{(i)}\}_{i=1}^d$, which spans the {tangent space} at $\mathbf x(0)$.
The evolution transports this hypercube along the trajectory $\mathbf x(t)$, and simultaneously deforms it. 
The tangent vectors $\{\mathbf w^{(i)}\}$ evolve according to the so-called variational equation
\beq
\label{eq_clDispEv}
\dot{\mathbf w}(t) = A[\mathbf x(t)]\cdot \mathbf w(t) 
\ . 
\eeq
where $ A[\mathbf x(t)] = \frac{\partial \mathbf f}{\partial \mathbf x}\big|_{\mathbf x(t)}$ is usually called the linear stability matrix.
The formal solution to this linear equation is
\beq
\label{eq:evolutor}
\mathbf w(t) = U\big[\mathbf x(t)\big] \mathbf w(0)
\eeq
where $U\big[\mathbf x(t)\big] 
= \mathcal{T} \exp \int_0^t d\tau \, A[\mathbf x(\tau)]$ is the evolution operator, and $\mathcal{T} \exp$ denotes the time-ordered matrix exponential.
The deformation of the hypercube in time is captured by inspecting the (positive) eigenvalues $\nu_1(t) \ge \nu_2(t) \ge \dots \ge \nu_d(t) \ge 0$ of the symmetric matrix 
\beq
G(t) = U( t)^T \cdot U(t) \ ,
\eeq
{that we refer to as the {Oseledets} matrix} \footnote{in the absence of a standard terminology.}.
The asymptotic \emph{Lyapunov spectrum} is then defined as 
\beq
\label{eq_defLyapunov}
\lambda_k = \lim_{t\to\infty} \frac 1 t \ln \sqrt{\nu_k(t)} \ .
\eeq
The existence of this limit for almost all initial data is the content of the celebrated Oseledets multiplicative theorem \cite{Oseledets1968multiplicative}. In particular, one has $\lambda_1 \equiv \lambda$.
In nonergodic dynamics, the numbers $\{\lambda_k\}$ may still depend on the particular trajectory.

The Lyapunov spectrum allows one to access the Kolmogorov-Sinai entropy rate $\Lambda_{\text{KS}}$, a fundamental quantifier of irreversibility in dynamical systems. 
The latter is related to the asymptotic loss of information on the state of the system induced by an arbitrarily fine coarse-graining of the phase space \cite{KS1,KS2}.
By Pesin's theorem \cite{Pesin}, one has \beq
\Lambda_{\text{KS}} = \sum_{k \; : \; \lambda_k>0} \lambda_k \ .
\eeq

It is important to stress that the characteristic Lyapunov exponents are defined by a long-time limit [see Eq.\eqref{eq_defLyapunov}]. Accessing the latter may be challenging in numerical simulations.
The by now standard algorithm for a robust computation of the Lyapunov spectrum has been proposed by Benettin, Galgani and Strelcyn in a series of papers around 1980 \cite{Benettin1976, Benettin1980P1, Benettin1980P2}. 
The convergence of the computations is typically quite slow in Hamiltonian systems. This is especially relevant in those undergoing an order/chaos transition, on which we will be concerned in the following.
For finite observation-time windows, one naturally defines the \emph{local} or \emph{finite-time Lyapunov exponents} $\{\lambda_k(t)\}$ as in Eq.\eqref{eq_defLyapunov} without taking the long-time limit.
This notion is particularly important in semiclassical dynamics due to the relatively short time window before saturation, as we will extensively discuss in Sec.\ref{sec_EChaos}.

In Appendix \ref{app:LE}, we recall further properties of the Lyapunov spectrum with emphasis on Hamiltonian systems, and in Appendix \ref{app:benettin} we briefly review the algorithm of Benettin \emph{et al.} in view of its importance later on in this work.

 \section{Relationship between entanglement growth and chaos}
 \label{sec_theory}
 
In this section, we begin by reviewing how quantum fluctuations evolve around the limiting classical trajectory. We then show how they yield the evolution of $S_A(t)$, $f_Q(t)$ and $c(t)$ in the semiclassical regime. Hence, we discuss how the dynamics of entanglement and chaos 
is determined by the structure of the underlying classical phase-space and its chaoticity.
Finally, we outline the range of validity of the semiclassical description and discuss the saturation due to finite-size effects.

 \subsection{Dynamics of quantum fluctuations}
 \label{sec_scEom}

 As reviewed in Sec.\ref{sec_models}, collective interactions allow for a reformulation of the nonequilibrium dynamics as an effective few-body  system in the semiclassical regime, where the impact of quantum fluctuations is controlled by the system size $N$ via the relation $\hbar_{\text{eff}}=1/N$.
 The generality of this approach has been discussed in Sec.\ref{sec_generality}.
 
 A system in this class is thus described by $n$ degrees of freedom, compactly denoted ${\hat{\boldsymbol \xi}=(\hat q_1,\dots,\hat q_n,\hat p_1,\dots,\hat p_n)}$, satisfying the canonical commutation relations ${[\hat q_i , \hat p_j] = i \hbar_{\text{eff}} \delta_{ij}}$, or $[\hat{\boldsymbol \xi},\hat{\boldsymbol \xi}]=i \hbar_{\text{eff}} \mathbb{J} $. 
 Here we have introduced the symplectic unit $\mathbb{J}$, given by the $2n \times 2n$ antisymmetric matrix {\scriptsize $\mathbb{J} = {\begin{pmatrix} \mathbb{0}_n & \mathbb{1}_n \\ -\mathbb{1}_n & \mathbb{0}_n  \end{pmatrix}}$}, which satisfies $\mathbb{J}^2=-\mathbb{1}_{2n}$.
 The evolution is governed by the Hamiltonian $\hat H = \heff^{-1} \; \mathcal{H}_\cl(\hat{\boldsymbol \xi})$
 and the Heisenberg equations read
 $
 \dot{\hat{\boldsymbol \xi}} =
 \mathbb{J} \; \partial \mathcal{H}_\cl(\hat{\boldsymbol \xi})
 $
 \footnote{subtleties related to the ordering of the operators are not relevant in the following discussion}.
 As discussed in Sec.\ref{sec_is}, the relevant initial states $\ket{\Psi_0}$ in our study of entanglement dynamics are quasiclassical states, i.e., states which satisfy 
 \beq
 \label{eq_QFinstate}
 \braket{\Psi_0|
 \Big(\hat{\boldsymbol \xi} - \boldsymbol \xi(0)\Big)
 \Big(\hat{\boldsymbol \xi} - \boldsymbol \xi(0)\Big)
 |\Psi_0}
 = \mathcal{O}(\heff)
 \ ,
 \eeq
 with
 $
 \boldsymbol \xi(0) \equiv
 \braket{\Psi_0|
 \hat{\boldsymbol \xi}
 |\Psi_0} = \mathcal{O}(1)
 $. 
 The meaning of this condition is that initial quantum fluctuations around the average are of the order of the minimal uncertainty allowed by the Heisenberg principle.

 We now aim at describing the evolution of quantum fluctuations around the average.
 We observe that, by virtue of Eq.\eqref{eq_QFinstate}, the average $\braket{\hat{\boldsymbol \xi}(t)}$ moves along the classical trajectory to the leading order in $\heff$,
 \beq
 \frac d {dt} 
 \braket{\Psi(t)|\hat{\boldsymbol \xi}|\Psi(t)} = 
 \mathbb{J} \; \partial \mathcal{H}_\cl\Big(\braket{\Psi(t)|\hat{\boldsymbol \xi}|\Psi(t)}\Big) + \mathcal{O}(\heff)\ ,
 \eeq
 i.e., $\braket{\Psi(t)|\hat{\boldsymbol \xi}|\Psi(t)} = \boldsymbol \xi_\cl(t) +\mathcal{O}(\heff) $.
 Quantum fluctuations around the average are encoded in the dimensionless variables
 \beq
 \label{eq_defQF}
 \delta \hat{\boldsymbol \xi} \equiv \heff^{-1/2} \big(
 \hat{\boldsymbol \xi}-\boldsymbol \xi_\cl(t)
 \big)\ ,
 \eeq
 which satisfy the commutation relations $[\delta \hat{\boldsymbol \xi},\delta \hat{\boldsymbol \xi}]=i\mathbb{J}$, and, by construction, $\braket{\delta \hat{\boldsymbol \xi}(t)}=\mathcal{O}(\heff^{1/2})$ \cite{Benatti2017}. 
 
In systems of collectively interacting spins, 
the quantum fluctuations $\delta \hat{\boldsymbol{\xi}} = (\delta \hat q, \delta \hat p)$ as in Eq.\eqref{eq_defQF} describe the collective spin fluctuations transverse to the instantaneous spin polarization direction, cf. Sec.\ref{sec:infiniteSpinSpyst}.
These spin fluctuations can be introduced in the formalism by performing a time-dependent Holstein-Primakoff trasformation around the instantaneous average spin \cite{ vidalHP, LeroseShort} (see also \cite{LeroseLong, LeroseKapitza, SacredLog}).
This standard transformation \cite{wannier} maps the transverse fluctuations of a quantum spin to a canonical bosonic mode. 
When these fluctuations are small compared to the size of the collective spin, one can approximate the transformation to the quadratic order, obtaining 
 \begin{align}
 \label{eq:spinRot}
 \begin{split}
\frac{\hat S^\alpha}{Ns}& 
 =    {X}_{\alpha}(t)\,
 \, \bigg(\frac \heff s \bigg)^{1/2}\, \delta \hat q
 +  {Y}_{\alpha}(t) \,\,
 \bigg(\frac \heff s \bigg)^{1/2}\, \delta \hat p \\
 & \quad  
 + {Z}_{ \alpha}(t) \;
 \left(   1 -\bigg(\frac \heff s \bigg) \, \frac{ \delta \hat q^2 +  \delta \hat p^2 - 1}{2}   \right)   \\
 & \quad 
 + \mathcal O \left( (\heff/s)^{3/2}\right) \ ,
 \end{split}
 \end{align}
 with 
 $\alpha=x,y,z$. 
 %
 Here, the time-dependent unit vector  $\mathbf { Z}(t)$ represents the classical dynamics of the collective spin polarization direction. It can be parameterized through the spherical angles $\phi(t)$ and $\theta(t)$ as in Eq.\eqref{eq:rotating}. 
 The transverse directions identified by the unit vectors $\mathbf X(t),\,\mathbf Y(t)$ can be parameterized as
 \begin{equation}
\label{eq:rotaFrame}
\mathbf{X}(t) \equiv 
\left( \begin{matrix}
\cos\theta \cos\phi \\
\cos\theta\sin\phi \\
-\sin\theta
\end{matrix} \right) ,  \quad
\mathbf{Y}(t) \equiv 
\left( \begin{matrix}
-\sin\phi \\
\cos\phi \\
0 
\end{matrix} \right) \ ,
\end{equation}
and span the orthogonal space to $\mathbf { Z}(t)$.
The short-hand notation in Eq.(\ref{eq:spinRot}) $X_{\alpha}(t)$, $Y_{\alpha}(t)$, $Z_{\alpha}(t)$  denotes the $\alpha$-th components of the basis vectors $\mathbf X(t)$, $\mathbf Y(t)$, $\mathbf Z(t)$ in Eqs.\eqref{eq:rotating} and \eqref{eq:rotaFrame} (i.e., $X_z=-\sin\theta$, $Y_z=0$, $Z_z= \cos\theta$, \dots).
One can check that the bosonic operators $\delta \hat q, \delta \hat p$ introduced via the Holstein-Primakoff transformation, correspond to the rescaled fluctuations $\delta \hat{\boldsymbol \xi}$ introduced above for the collective spin when the Bloch sphere is parametrized through the canonical phase-space variables $q=\phi$ and $p=\cos\theta$.
 {When the system comprises $M>1$ collective spins, of magnitude $N_j s \gg 1$, $j=1,\dots,M$, one can perform the analogous transformation \eqref{eq:spinRot} on their components $\hat S^{\alpha}_j$ to obtain the joint semiclassical description.} 

The general transformation \eqref{eq_defQF} is time-dependent. The exact evolution equations for the quantum fluctuations $\delta \hat{\boldsymbol \xi}$ are thus generated by the modified Hamiltonian
\beq
\hat{\widetilde{H}}(t)=
\heff^{-1} \; \mathcal{H}_\cl
\Big(\boldsymbol{\xi}_\cl(t)+ \heff^{1/2}\delta \hat{\boldsymbol \xi} \Big)
- 
\heff^{-1/2} \; 
\dot{\boldsymbol \xi}_\cl (t) \, \mathbb{J} \,
\delta \hat{\boldsymbol \xi} \ .
\eeq
{We can now expand the Hamiltonian with respect to the small parameter $\heff$, obtaining} {the time-dependent Hamiltonian}
\beq
\label{eq:HamiTrunca}
\hat{\widetilde{H}}(t) \, = \, 
\heff^{-1} \; \hat{{H}}_0(t)
\,+ \,
\heff^{-1/2} \; \hat{{H}}_1(t)
\,+\, \hat{{H}}_2(t) 
\,+\, \mathcal{O}(\heff^{1/2})\ .
\eeq
 Here, $\hat{{H}}_0(t) = \mathcal{H}_\cl
\big(\boldsymbol \xi_\cl(t)\big)$ is just a classical quantity (the classical energy along the classical trajectory), and the linear term $\hat{{H}}_1(t)=
\Big[
\partial \mathcal{H}_\cl \big(\boldsymbol \xi_\cl(t)\big)  -  \dot{\boldsymbol \xi}_\cl(t) \mathbb{J}
\Big]
\delta \hat{\boldsymbol \xi}
$
 vanishes identically by construction, consistently with the vanishing of $\braket{\delta \hat{\boldsymbol \xi}(t)}$ to the leading order in $\heff$.
 The operator expansion thus starts from the (finite) quadratic order.
 Within the semiclassical regime, and for a time scale that diverges as $\heff\to0$ (the so-called Ehrenfest time scale, see below), we can neglect the remainder $\mathcal{O}(\heff^{1/2})$ in the expansion. 
 The evolution of the quantum fluctuations in this regime is determined by a linear homogeneous differential equation,
 \beq
 \label{eq:xiDot}
 \frac d {dt} \delta \hat{\boldsymbol \xi} =
 A(t)
 \; \delta \hat{\boldsymbol \xi}
 \equiv
 \mathbb{J}     \, \partial^2 \mathcal{H}_\cl \big(\boldsymbol \xi_\cl(t)\big) \; \delta \hat{\boldsymbol \xi}\ ,
 \eeq
 identical with the classical variational equation for the evolution of infinitesimal displacements away from the classical trajectory [cf. Sec.\ref{sec_LE} Eq.\eqref{eq_clDispEv}]. In fact, the classical and quantum evolutions generated by a quadratic Hamiltonian coincide, as is well known.

 The solution to this equation is formally written as 
 \beq
 \begin{split}
 \label{eq:deltaXi}
 \delta \hat{\boldsymbol \xi} (t) &=
  U (t)
 \; 
  \delta \hat{\boldsymbol \xi} (0) \ ,
   \end{split}
  \eeq
  where $U (t)
   \equiv \mathcal{T} \exp \int_0^t d\tau \, 
  A(\tau)$ is the tangential map, which encodes the evolution of infinitesimal classical displacements. 
Due to the asymptotic Gaussian description for small $\hbar_{\text{eff}}$, all the information on the quantum state is encoded in the \emph{correlation matrix}
 \beq
 \label{eq:Gdef}
 [G(t)]_{ij}=
 \frac 1 2
 \braket{\Psi(t)|
 \delta \hat{ \xi}_i \,
 \delta \hat{ \xi}_j+
 \delta \hat{ \xi}_j \,
 \delta \hat{ \xi}_i
 |\Psi(t)} \ ,
 \eeq
 with $i,j=1,\dots,2n$. This matrix is symmetric and positive definite; the square root of its eigenvalues quantify the width of the quantum fluctuations around the classical average, and are constrained from below by the Heisenberg principle (see, e.g., Ref. \cite{littlejohn}). Notice that the rescaling by $\hbar_{\text{eff}}^{1/2}$ in Eq.\eqref{eq_QFinstate} is equivalent to the statement that $G(t) = \mathcal{O}(1)$. 
The evolution of the correlation matrix $G(t)$ 
can be directly expressed via Eq.\eqref{eq:deltaXi} as
 \beq
  \label{eq:Gevolve}
 G(t) = U(t)^T \; G(0) \; U(t) \ .
 \eeq

 \subsection{Semiclassical expressions of  entanglement and chaos quantifiers}
 \label{sec_scEnt}
 
In this section we will analytically derive the relation between the entanglement quantifiers of Sec.\ref{sec_quantities} and the chaos indicators in the semiclassical regime.

\subsubsection{Semiclassical entanglement entropies}
\label{sec_ee_sc}

 We consider a quantum collective model and introduce a bipartition $(A,B)$ of its degrees of freedom as discussed in Sec.\ref{sec_ee}.
 Within the semiclassical description, the bipartite system can be represented by two sets of semi-classical variables $\hat{\boldsymbol \xi}=(\hat{\boldsymbol \xi}_A,\hat{\boldsymbol \xi}_B)$, with $n_A$ and $n_B$ collective degrees of freedom, respectively ($n_A+n_B=n$) \footnote{When bipartitions of a permutationally invariant system are considered, one has $n_A=n_B=n$, where $n=q-1$ is the number of collective degrees of freedom [cf. Secs.\ref{sec_models}, \ref{sec_generality}]}.
 In this regime, the entanglement between the two subsystems is encoded in the entanglement between their bosonic quantum fluctuations $\delta \hat{\boldsymbol \xi}_A$, $\delta \hat{\boldsymbol \xi}_B$. The extent of these quantum fluctuations is collected in their correlation matrix $G(t)$ in Eqs. \eqref{eq:Gdef}-\eqref{eq:Gevolve}.
  It is convenient to define the subsystem \emph{reduced correlation matrix} $G_A(t)$ as the $2n_A \times 2 n_A$ sub-matrix of $G(t)$ built out of the coordinates of subsystem $A$ alone, i.e.,
  \beq
 \label{eq:GAdef}
 [G_{A}(t)]_{ij}=
 \frac 1 2
 \braket{\Psi(t)|
  \delta \hat{ \xi}_{i} \,
 \delta \hat{ \xi}_{j}+
 \delta \hat{ \xi}_{j} \,
 \delta \hat{ \xi}_{i}
 |\Psi(t)}_{\substack{ 1\leq i\leq 2n_A \\  1\leq j\leq 2n_A}} \ .
 \eeq 
  Due to the asymptotic Gaussian description for small $\hbar_{\text{eff}}$, the reduced density matrix $\hat \rho_A(t)$ is also Gaussian and fully determined by the matrix $G_A(t)$.
  The entanglement entropies can thus be computed via standard techniques \cite{vidal2007enta}. 
  
  The dynamics of the entanglement entropies in bosonic systems governed by quadratic Hamiltonians
  has been derived and discussed in full generality in Refs. \cite{Bianchi2018, BianchiModakRigolEntanglementBosons}. 
  It is shown therein, that the second Renyi entropy \eqref{eq:Renyi} can be expressed as the logarithm of the phase-space volume spanned by the time-evolving phase-space distribution associated with the reduced state of the subsystem, i.e.,
  \beq
  \label{eq:S2Gaussian}
  S^{(2)}_A(t)= 
  \frac 1 2 
  \ln 
  \; \det \big(2 G_A(t)\big)
  \ .
  \eeq
  While the global evolution preserves the total volume, i.e., $\det \big(2 G(t)\big) \equiv 1$, the information loss generated by projecting the collective quantum fluctuations onto a subsystem with $n_A < n$ yields an increase of entropy, whose origin is rooted in the development of quantum entanglement. By Eq.\eqref{eq:S2Gaussian}, this increase may be visualized as an enhancement of the projected volume spanned by the reduced quantum fluctuations within the subsystem's phase space, due to the progressive stretching of the global phase-space volume spanned by the quantum fluctuations. 
  Similarly, the von Neumann entanglement entropy \eqref{eq:SA} can be computed as
\begin{align}
\begin{split}
    \label{eq:SaGAussian}
        S_A(t) = \sum_{i=1}^{n_A} S(\nu_i(t))  &\quad \text{with} \\
        S(\nu) = \frac{\nu+1}2 &\ln \frac{\nu+1}2 - \frac{\nu-1}2 \ln\frac {\nu-1}2 \ ,    
\end{split}
\end{align}
  where $\pm \nu_i(t)$ ($\nu_i(t) \ge 1$) are the so-called symplectic eigenvalues of $2G_A(t)$ \footnote{From the correlation matrix $G$, one defines $J = - 2G \mathbb{J}$, where $\mathbb{J}$ is the $2n\times 2n$ symplectic unit. The matrix $[iJ]_A$ restricted to $A$ can be shown to have pairs of opposite real eigenvalues $\pm\nu_i$ (with $\nu_i>1$ as follows from the Heisenberg
relations). The $\{\nu_i\}_{i=1,\dots n_A}$ are referred as the symplectic eigenvalues of $G_A$ and determine the entanglement entropy via Eq.\eqref{eq:SaGAussian} }.
The entanglement entropy $S_A(t)$ is bounded above and below by the second Renyi entropy up to a constant, and hence their  growths are superimposed after a finite transient, $S_A(t) \underset{t\gg1}{\thicksim} S^{(2)}_A(t)$, see Ref.\cite{Bianchi2018}. Their common asymptotic behavior generically depends on the subsystem only via its number $n_A$ of degrees of freedom, and their evolution is completely determined by that of $G(t)$. 
 
As discussed in Sec.\ref{sec_ee}, in many interesting semi-classical models, 
the relevant subsystem $A$ is made of only one collective degree of freedom, i.e., $n_A=1$.
These include both the paradigmatic models discussed below, namely the quantum kicked top and the Dicke model.
In this case Eq.\eqref{eq:SaGAussian} simplifies and $S_A(t)$ can be expressed as a function of the determinant of $G_A(t)$, i.e.,
\begin{align}
    \begin{split}
     \label{eq:SAnA1}
    S_A(t) & = 2 \sqrt{\det G_A} \, \arccoth\left(2 \sqrt{\det G_A}  \right) 
    \\ & \quad 
    + \frac 12 \ln \left ( \det G_A - \frac 14 \right ) \ . 
    \end{split}
\end{align}
From this equation, the asymptotic result of Ref.\cite{Bianchi2018} immediately follows, i.e., $S_A(t) \underset{t\gg1}{\thicksim} \frac 12\, \ln 
\; \det G_A(t) 
$, since $\det G_A(t){\gg} 1$, 
in agreement with the second Renyi entropy in Eq.\eqref{eq:S2Gaussian}.

In the case of collective spin systems of the form of Eq.\eqref{eq:Hgeneralcollective}, one considers bipartitions between two sets of $N_A=f_A\, N$ and $N_B=f_B\, N$ spins ($f_A+f_B=1$), and a further simplification occurs. By performing a change of variables to the dynamical collective fluctuations and the frozen relative fluctuations of the two spins 
\footnote{
Explicitly,
\begin{equation*}
\begin{dcases}
\delta \hat q = +\sqrt{f_A} \; \hat q_A + \sqrt{f_B} \; \hat q_B \\
\delta \hat q_{\text{rel}} = -\sqrt{f_B} \; \hat q_A + \sqrt{f_A} \; \hat q_B 
\end{dcases} 
\begin{dcases}
\delta  \hat p = +\sqrt{f_A} \; \hat p_A + \sqrt{f_B} \; \hat p_B \\
\delta  \hat p_{\text{rel}} = -\sqrt{f_B} \; \hat p_A + \sqrt{f_A} \; \hat p_B 
\end{dcases}  \ .    
\end{equation*}}, it is easy to compute that \cite{SacredLog}
\beq
\label{eq:detASpin}
\det G_A = \frac 14 +  f_A\, f_B \langle \hat n_{\text{exc}} \rangle \ ,
\eeq
where $\hat n_{\text{exc}} = (\delta \hat q^2 + \delta \hat p^2  -1 )/ 2 $ represents  the number of bosonic excitations of the collective spin $\mathbf{\hat S}$. 
This allows to compute $S_A(t)$ in a closed form, without the need to compute the reduced correlation matrix $G_A(t)$ \cite{SacredLog, Ingo}.
It is then clear that $S_A$ vanishes for $\langle \hat n_{\text{exc}} \rangle \to 0$ and grows as $\frac{1}{2} \ln \langle\hat n_{\text{exc}} \rangle$ for $\langle\hat n_{\text{exc}}\rangle \gg 1$.
Hence, Eqs.(\ref{eq:SAnA1}-\ref{eq:detASpin})
clarify that the state of subsystem $A$ (or $B$) is pure only if $ \langle  \hat n_{\text{exc}} \rangle = 0$, i.e., if the spin system is fully polarized (coherent), as occurs in the absence of interactions. 
{Conversely, the state is entangled in the presence of collective quantum excitations.} 

As we will see in the next sections, the entanglement entropy of a collective spin system can been quantitatively related to the quantum Fisher information and to the spin squeezing.

\subsubsection{Semiclassical quantum Fisher information and spin squeezing}
\label{sec_qfi_sc}

The quantum Fisher information for collective spin systems is given by the maximal variance of the collective spin operators [cf. Eq.(\ref{QFI})]. This information is encoded in the correlation matrix $G(t)$ \eqref{eq:Gdef}, which describes the dynamics of the fluctuations in the transverse direction. Therefore, the semiclassical QFI is given by the maximum eigenvalue of the correlation matrix $G(t)$
\begin{equation}
    \label{eq:QFIsm}
    f_Q(t) = 4 \Max \left [\text{Eigvals} \, G(t)\right ] \ .
\end{equation}

In the case of a fully-connected spin system (Sec.\ref{sec:infiniteSpinSpyst}), one can determine the QFI explicitly, by computing the eigenvalue of the $2\times 2$ spin correlation matrix. This yields the equation
\begin{equation}
    \label{eq:QFIsm1Dof}
    f_Q(t) = 1 + 2 \langle \hat n_{\text{exc}}(t) \rangle  + 2 \sqrt{\langle \hat n_{\text{exc}}(t) \rangle (\langle \hat n_{\text{exc}}(t) \rangle+1)} \ ,
\end{equation}
where $\langle \hat n_{\text{exc}}(t) \rangle$ is the number of bosonic excitations of the collective spin $\mathbf{\hat S}$ [cf. Eqs.\eqref{eq:spinRot} and \eqref{eq:detASpin}].

As discussed in Sec.\ref{sec:qfi}, spin squeezing represents a convenient indicator of multipartite entanglement in spin systems.
At the semiclassical level relevant here, quantum fluctuations are Gaussian, and one derives \cite{gaussianQI}
\beq
\label{eq:squeezing}
\xi^2(t) = 1+2 \langle  \hat n_{\text{exc}}(t) \rangle - 2 \sqrt{ \langle  \hat n_{\text{exc}}(t)\rangle( \langle  \hat n_{\text{exc}}(t)\rangle +1)} \; .
\eeq
 {Equations (\ref{eq:SAnA1}-\ref{eq:detASpin}), \eqref{eq:QFIsm1Dof} and \eqref{eq:squeezing} express the quantitative link} --- pictorially illustrated in Fig.\ref{fig_entDyn} ---  between the entanglement entropy $S_A$, the quantum Fisher information $f_Q$, and the spin squeezing parameter $\xi$, in collective spin models in the semiclassical regime in and out of equilibrium.
In particular, in this regime the inequality discussed in Sec.\ref{sec:qfi} is saturated, i.e.,  $f_Q = 1/ \xi^2$.


 \subsubsection{Semiclassical square commutator}
\label{sec_sc_sc}

 Along similar lines, we can compute the semiclassical evolution of the out-of-time-order square commutator defined in Sec.\ref{sec_sc} for a system initialized in a quasiclassical state.
 Starting from the definition in Eq.\eqref{SC} and expanding the operators up to the quadratic order in the quantum fluctuations, one readily finds
 \beq
 \begin{split}
 c_{ij}(t) &\equiv
 - {\heff^{-2}}\braket{ \Psi_0 |
 \Big[\hat \xi_i(t), \hat \xi_j(0)\Big]^2
 |\Psi_0 } \\
 & =  \Big(U_{i \bar{j}}(t)\Big)^2 + 
 \mathcal{O}(\heff),
 \end{split}
 \eeq
 where $\bar{j} \equiv (j+n)\mod 2n$. The semiclassical square commutator thus directly probes the sensitivity of the classical trajectories to infinitesimal perturbations.

In the case of fully-connected spin systems, the square commutator between two collective spin operators \eqref{eq:cab} reads  
\begin{align}
\begin{split}
\label{eq:cabGauss}
c_{\alpha\beta}(t) & = 
\Big[  {X}_{\alpha}(t)  \big (\, U_{qq}(t)\,\, 
 {Y}_{\beta} (0)
- U_{qp}(t)\,\, {X}_{\beta}(0)
\, \big) \\
& \quad \quad 
+ {Y}_{\alpha}(t) \, \big (\, U_{pq}(t)\, Y_{\beta}(0)  - U_{pp}(t)\, X_{\beta}(0)
\,\big)\Big]^2 
\\
& \quad \quad
+ \mathcal O(\heff) \ ,
\end{split}
\end{align}
with the same notation as in Eq.\eqref{eq:spinRot}. 
In order to get this result, one first
plugs the expansion of the spin operators \eqref{eq:spinRot} into the definition \eqref{eq:cab}. Then, after substituting the formal solution for the spin fluctuations at time $t$, i.e., $ {\delta\hat q(t) = U_{qq}(t)\,\delta\hat q(0) + U_{qp}(t)\,\delta\hat p(0)}$ and ${\delta\hat p(t) = U_{pq}(t)\,\delta\hat q(0) + U_{pp}(t)\,\delta\hat p(0)}$ [cf. Eq.(\ref{eq:deltaXi})], the equal-time commutators between the conjugate variables yield the above Eq.\eqref{eq:cabGauss}.

\subsection{Entanglement growth and chaos}
\label{sec_EChaos}

In the previous section, we have established how the semiclassical dynamics of quantum fluctuations 
determine the evolution of the entanglement quantifiers of interest, via the time-dependent correlation matrix $G(t)$.
This connection highlights that the entanglement growth is determined by the chaoticity properties of the semiclassical dynamics, in turn dictated by the stability of the underlying phase-space trajectories.

The correlation matrix $G(t)$ in Eq.\eqref{eq:Gdef} is equivalent to the Oseledets matrix that defines the Lyapunov spectrum in Eq.\eqref{eq_defLyapunov} 
\footnote{The presence of $G(0)$ instead of the identity matrix is immaterial for the definition of the Lyapunov exponents: these are intrinsic quantities associated with the flow that do not depend on the arbitrary choice of the phase-space metric.}, as  the quantum fluctuations 
evolve in the same way as the linearized displacements.
 Hence, the spectrum of the growth rates of the quantum fluctuations encoded in $G(t)$ coincides with the \emph{finite-time} Lyapunov spectrum $\{ \lambda_k(t) \}$ of the underlying semiclassical trajectory within the Ehrenfest time scale $T_{\text{Eh}}(N)$, and converges to the proper asymptotic Lyapunov spectrum $\{ \lambda_k \}$ as $N\to\infty$.
On the other hand, for short times, one can consider the limit $t\to 0$ of the above expressions and retrieve the correct early-time expansions, see e.g. Ref.\cite{Sorelli2019}.

When the classical dynamics is integrable,
the collective motion of the system is orderly and takes place along regular trajectories in phase space, meaning that  nearby initial conditions separate linearly in time (generically).
This implies that all Lyapunov exponents vanish.
This scenario largely persists under weak integrability-breaking Hamiltonian perturbations, as established by KAM theory \cite{PoschelKAM}, whereby regular trajectories gradually leave room to chaotic portions of the phase space arising from dynamical resonances.
Thus, in integrable or near-integrable semiclassical systems,
the temporal growth of the quantum correlations is at most polynomial, $G_{ij}(t) \sim t^2$, as can be shown explicitly by switching to action-angle variables. 
Conversely, in systems with far-from-integrable semiclassical dynamics featuring fully developed chaos in phase space, the Lyapunov spectrum is nonvanishing. 
This implies an asymptotic exponential growth of quantum fluctuations, generically given by $G_{ij}(t) \sim e^{2 \lambda t}$, where $\lambda$ is the maximal Lyapunov exponent.
%
An immediate consequence of the above observations concerns the asymptotic growth rate of the square commutator. 
In fact, the latter results to be twice the maximum Lyapunov exponent of the underlying semiclassical dynamics, 
\beq
c(t) \sim e^{ 2 \lambda t}.
\eeq


Crucially, the chaoticity properties of the semiclassical dynamics determine the speed of the entanglement growth, as determined by Eqs.\eqref{eq:SaGAussian} and \eqref{eq:QFIsm}.
In fact, by Eq.\eqref{eq:QFIsm}, we immediately derive that the QFI grows as 
\beq
f_Q(t) \sim e^{2\lambda t}.
\eeq
The determination of the bipartite entanglement entropies growth requires a more elaborate analysis. In Refs. \cite{Bianchi2018, BianchiModakRigolEntanglementBosons}
Bianchi, Heykl, \emph{et al.} discuss the bipartite entanglement dynamics generated by quadratic bosonic Hamiltonians.
As thoroughly shown therein,
the growth of $S_A(t)$ is generically linear in time with a rate set by the sum of the largest $2n_A$ Lyapunov exponents, \beq
S_A(t) \underset{t\gg1}{\thicksim} S^{(2)}_A(t) \thicksim \Lambda_{A} t =\bigg(\sum_{k=1}^{2n_A} \lambda_k\bigg) t.
\eeq
For $n_A=n/2$, the rate coincides with the classical Kolmogorov-Sinai entropy rate $\Lambda_{KS}=\sum_{\lambda_k \, : \, \lambda_k>0} \lambda_k$ \cite{KS1,KS2,VulpianiChaosbook}.
Analogous equations to the three above apply to the phase-space \emph{separatrices} when the classical dynamics is integrable; in this case, the Lyapunov spectrum is given by the linearized dynamics around the unstable fixed point on which the trajectory terminates \cite{SacredLog,Ingo}.

By contrast, for generic trajectories, integrable systems have $\Lambda_{KS}=0$. In this case, one has
\begin{subequations}
\begin{align}
   S_A(t) & \underset{t\gg1}{\thicksim} S^{(2)}_A(t) 
   \thicksim  c \ln t \ ,  \\
   f_Q(t) & \sim t^2 \ , \\
   c(t) & \sim t^2 \ ,
\end{align}
\end{subequations}
with $c$ an integer. 

The classification is concluded by the case of stable equilibrium configurations, the linearized dynamics of which is equivalent to that of coupled harmonic oscillators. 
Accordingly, all the quantities of interest perform bounded (periodic or quasiperiodic) oscillations. (Note that the same applies to effective \emph{linear} semiclassical dynamics with suppressed anharmonic contribution, as in the recently discovered mechanism in Refs. \cite{ChoiPerfectScars,Ho2019}.)
A summary of the above discussion is presented in Table \ref{tab:esempio}.

 Since quadratic Hamiltonians describe the dynamics of quantum fluctuations around the limiting classical trajectory in the limit $\hbar_{\text{eff}}\to0$ to the leading order [cf. the discussion in Sec.\ref{sec_scEom}], Bianchi, Heykl, \emph{et al.}  conjecture that their analysis applies to generic quantum systems in the appropriate semiclassical regime. In particular, $S_A(t) \sim \Lambda_A t$, where $\Lambda_A$ is the Kolmogorov-Sinai entropy rate determined by the Lyapunov spectrum as above. 
 It is one of the main purposes of the present work to thoroughly assess this conjecture and firmly establish its range of validity in quantum many-body systems possessing a relevant and controlled semiclassical limit. 
The asymptotic results of Refs. \cite{Bianchi2018,BianchiModakRigolEntanglementBosons} ideally describe the \emph{average} asymptotic growth at long times. However, typical semiclassical systems generally present 
strong additional finite-time fluctuations
 in the entanglement quantifiers.
For example, when the limiting classical trajectory is periodic with period $T_{\text{cl}}$, for integrable (chaotic) dynamics one has $T_{\text{cl}}$-periodic oscillations superimposed to the logarithmic (linear) growth of $S_A(t)$ and to the polynomial (exponential) growth of $f_Q(t)$ and $c(t)$.
For general aperiodic classical trajectories, the time-dependence can be much more complicated. 
These effects can obscure the asymptotic growth until the saturation due to the finite $\heff$.
Accordingly, deviations from the asymptotic result of Refs. \cite{Bianchi2018,BianchiModakRigolEntanglementBosons} can be observed.

  In Secs.\ref{sec_KT} and \ref{sec_Dicke} below, we will concentrate on systems exhibiting a progressive transition to chaos as a parameter is varied. In such systems, finite-time fluctuations play a major role, due to the complexity of the phase space, featuring a fractal structure of regular trajectories (KAM tori) and chaotic regions. 
 For this reason, the correct semiclassical identification holds between the growth rate of quantum entanglement and the \emph{finite-time} Lyapunov spectrum $\{ \lambda_k(t) \}$, rather than the proper asymptotic one.
 The discrepancy may be particularly severe, due to the relatively short Ehrenfest time scale 
 in chaotic systems. 
The long-time convergence of the rate of growth 
of the relevant entanglement and chaos quantifiers to the asymptotic 
ones compatible with the Lyapunov spectrum competes with their saturation 
in a finite system at the Ehrenfest time scale $ T_{\text{Eh}}(N) \sim \ln N$.
Hence, the theoretical long-time rates of growth 
will hardly be accessible in practice.
 This point is  often overlooked in the recent literature on OTOC and its relation to chaos.
 

We conclude the discussion by commenting that
not only the entanglement entropy $S_A(t)$ has a finite limit as $\heff\to0$, but this limiting quantity has a natural classical interpretation in terms of the loss of information under phase-space coarse-graining during the time-evolution \cite{zurek1995quantum,Casati_2012} --- which is the meaning of the classical Kolmogorov-Sinai entropy in dynamical systems.
 It is also interesting to remark that the growth of entanglement entropy in the semiclassical regime is sensitive to the full Lyapunov spectrum, unlike the growth of the OTOCs, which is sensitive to the maximum Lyapunov exponent only.

\subsection{Ehrenfest time and finite-size effects} 
\label{sec_tEhr}
 At this stage, it is natural to comment on the time scale of validity of the 
 semiclassical description outlined above.
 The latter is the well known \textit{Ehrenfest time scale}, and is estimated as the time at which the size of quantum fluctuations becomes comparable with the typical length in phase space, i.e. $\mathcal O(G(t))=\mathcal O(\heff^{-1})$.
 For orderly, integrable-like motion, quantum fluctuations grow polynomially in time as $G(t) \sim t^2$, which yields $T_{\text{Eh}} \sim \heff^{-1/2} = \sqrt{N}$. In the presence of unstable, chaotic evolution, instead, one has $G(t) \sim e^{2\lambda t}$, where $\lambda$ is the maximum Lyapunov exponent defined in Sec.\ref{sec_LE}. In this case, thus, $T_{\text{Eh}} \sim (1/\lambda)\ln \heff^{-1/2} = (1/2\lambda) \ln N$.

{At this time scale, the semi-classical analysis described before breaks down and a full quantum regime takes place, dominated by interference. 
From the numerical simulations for finite systems, we find that the entanglement descriptors saturate to values compatible with their statistical-mechanical predictions: 
in particular, we find
\begin{align}
    S_A^{\infty} \propto \ln \heff^{-1} \ , \quad
    f_Q^{\infty} \propto \heff^{-1} \ ,
\end{align}
which is also compatible with the results of Sec.\ref{sec_ee_sc}-\ref{sec_qfi_sc} evaluated at $T_{\text{Eh}}$.
In other words, the asymptotic state is genuinely multipartite entangled $f^{\infty}_Q\propto N$, while the bipartite entanglement entropy saturates to $S^{\infty}_A\propto \ln N_A$. This is actually related to the usual volume-law scaling of entanglement out of equilibrium \footnote{In fact, the stationary states after a quantum quench explore all the allowed Hilbert space, and their entanglement is upper-bounded by $S_A\leq \text{dim}(\mathcal H_A)$. In collective models under consideration here, however, the conservation of the collective spin magnitude $|\mathbf S|^2$ reduces the dimension of the allowed Hilbert space to $\text{dim}(\mathcal H_A)$.}. 
For the chaotic driven dynamics, the value of the QFI is compatible with the values of the infinite
temperature state: $f_Q^{\infty} = 1 + N/3 + \mathcal O(1/N)$. Likewise, the entanglement entropy saturates to the value expected for a random state, 
derived by Page in Ref.\cite{Page1993}
$S_{\text{Page}}=\ln m -m/2n + \mathcal O(1/mn)$, with $m,n$ the dimensions of the Hilbert space of the two subsystems.
On the other hand, in this regime the square commutator \eqref{SC} is characterized by a fully quantum
nonperturbative growth which leads to saturation only in the case of a fully chaotic dynamics, while it grows polynomially in the case of integrable systems. For a discussion of this effect see, e.g., Ref.\cite{pappaScrambling}.}

\section{The quantum kicked top} 
\label{sec_KT}
In this section, we will apply the theoretical analysis developed in Sec.\ref{sec_theory} to study the quantum kicked top, previously introduced in Sec.\ref{sec:models}.  
We will start by deriving the semiclassical evolution of quantum fluctuations  in Sec.\ref{sec:KT_SFev}. Subsequently, we numerically compare  the semiclassical results with the exact dynamics in finite-size systems in Sec.\ref{sec_KTnum} and discuss the results in Sec.\ref{sec_KTdis}.

\subsection{Evolution of the spin fluctuations}
\label{sec:KT_SFev}
We derive the semiclassical evolution of the Gaussian spin fluctuations ${\delta \hat{\boldsymbol{\xi}}=(\delta \hat q, \delta \hat p)}$ around the classical solution as a discrete map.
We first perform the bosonization of spin fluctuations around the time-dependent polarization direction
${ \vec{\mathcal{S}}(t) \equiv \big\langle {\mathbf S}(t) \big\rangle \propto {\mathbf Z}}$ via the Holstein-Primakoff transformation in Eqs.\eqref{eq:spinRot}. 
The stroboscopic evolution from time $t=n$ to $t=n+1$ of the 
$2\times2$ correlation matrix $G(n) = \langle \delta \boldsymbol{\xi}(n) \delta \boldsymbol{\xi}(n) \rangle $
is given by the composition of the following two maps 
\begin{subequations}
\label{eq_KTfluc}
\begin{align}
    & \begin{dcases}
    G'_{qq} = \cos^2(\psi-\psi')\, G_{qq} + \sin [2(\psi-\psi')] G_{qp} \\
    \qquad\qquad\qquad\qquad\qquad\qquad\qquad + \sin^2(\psi-\psi') \, G_{pp}\\
    G'_{pp} = \sin^2(\psi-\psi')\, G_{qq} - \sin [2(\psi-\psi')] G_{qp} \\
    \qquad\qquad\qquad\qquad\qquad\qquad\qquad
    + \cos^2(\psi-\psi') \, G_{pp}\\
    G'_{qp} = -\cos[2(\psi-\psi')]\, G_{qq} + \cos [2(\psi-\psi')] G_{qp} \\
    \qquad\qquad\qquad\qquad\qquad\qquad\qquad
    + \sin[2(\psi-\psi')] \, G_{pp}\\
    \end{dcases}
    \\\nonumber\\
   &  \begin{dcases}
    G''_{qq} = G'_{qq}\\
    G''_{pp} = G'_{pp} 
        - 2\beta\sin^2\theta'\,  G'_{qp}
        +(\beta \sin^2\theta')^2\, G'_{qq}
    \\
    G''_{qp} =G'_{qp} - \beta \sin^2\theta' \, G'_{qq}
    \end{dcases}
\end{align}
\end{subequations}
where we have defined the angles ${\psi = - \arctan \left(
 {\tan \phi}/{\cos\theta}
\right)}$,
${\psi' = - \arctan \left(
 {\tan \phi'}/{\cos\theta'}
\right)}$, with $\theta'$, $\phi'$ given by the intermediate classical point before the kick, cf. Eq.\eqref{eq:classKT}.
The details of the calculation are reported in Appendix \ref{app:eomKT1}.
Together with Eqs.\eqref{eq:classKT} and the appropriate initial conditions, they give a complete description of the semiclassical dynamics of the quantum kicked top at stroboscopic times. This analysis is valid before the Ehrenfest time scale $T_{\text{Eh}}$.

\onecolumngrid

\begin{figure}[H]
\centering
\begin{tabular}{lr}
\includegraphics[width=0.42\textwidth]{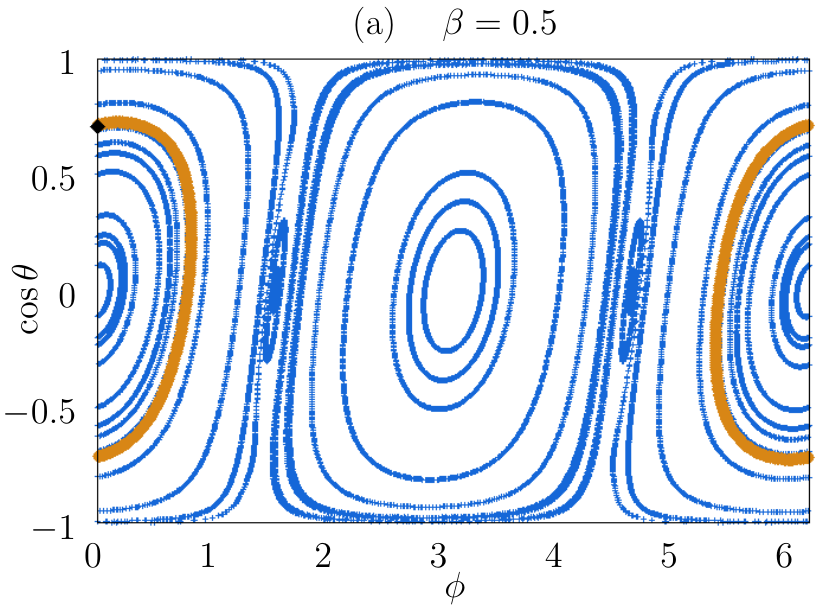} &
\includegraphics[width=0.42\textwidth]{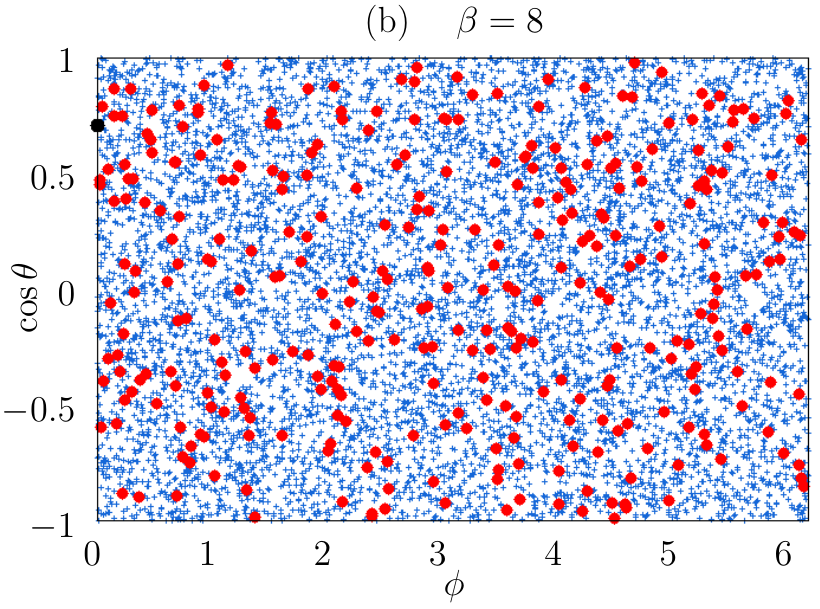} \\
\includegraphics[width=0.44\textwidth]{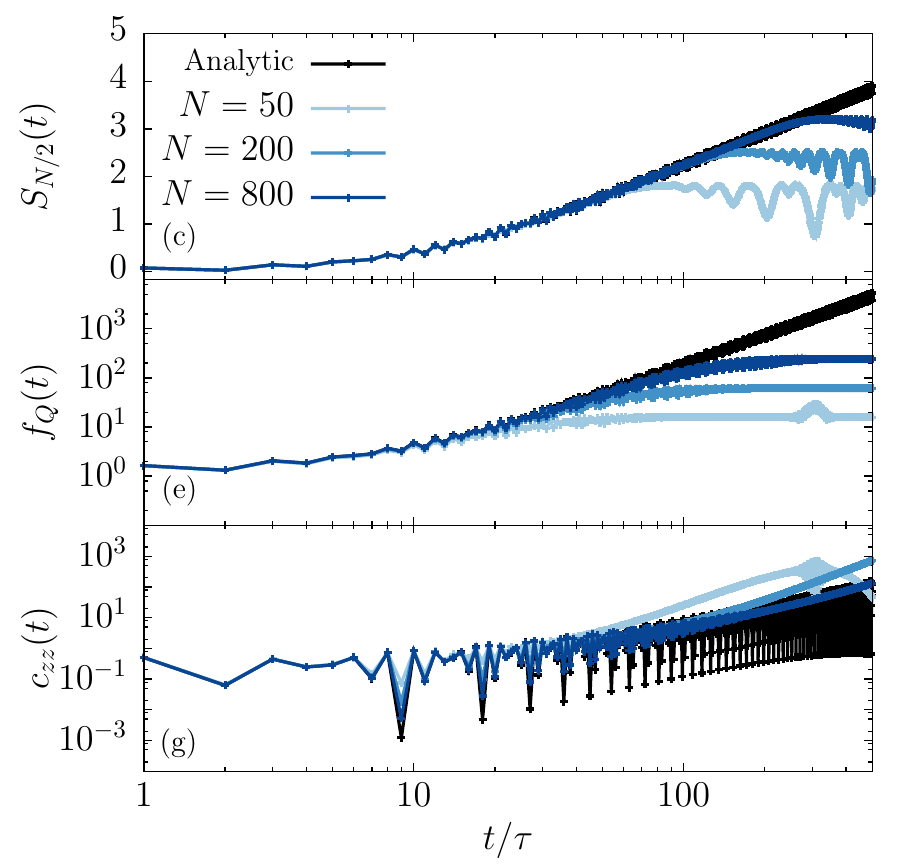} &
\includegraphics[width=0.43\textwidth]{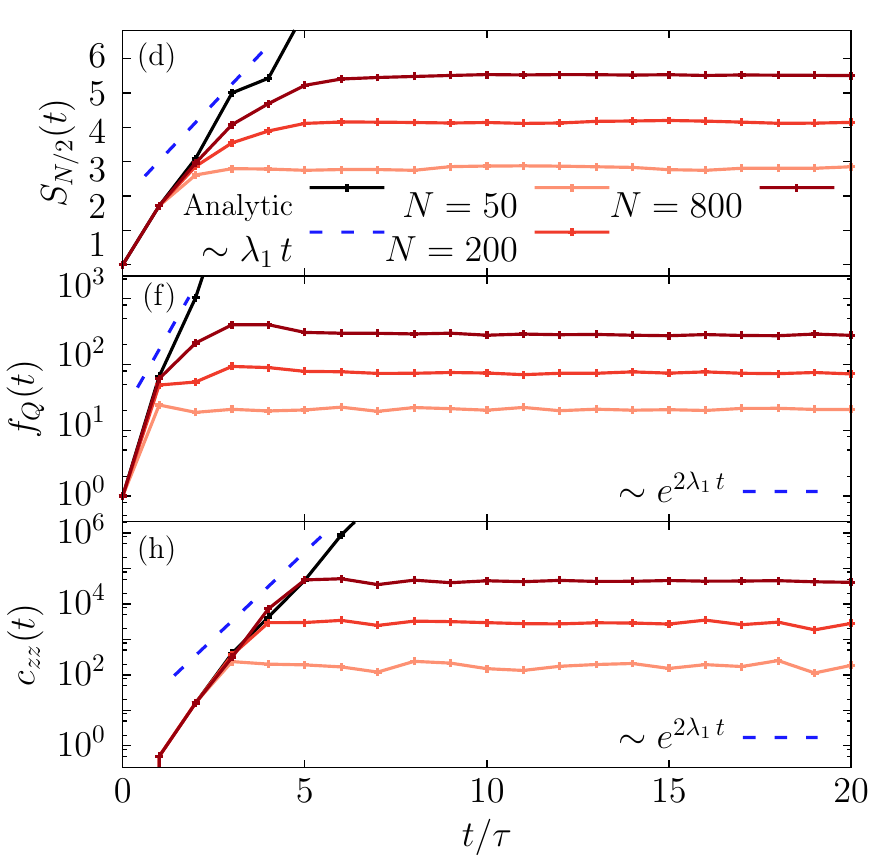} 
\end{tabular}
\caption{
Dynamics of the quantum kicked top with $\alpha=\pi/2$, in the predominantly ordered and chaotic regimes with
$\beta=0.5$ and $8$, respectively.
Top panels: 
Poincar\`e map (stroboscopic phase-space trajectory) for the regular and chaotic dynamics, on the left (a) and right (b) panels, respectively. The black diamond and dot represent the initial condition of the orange (full line) and red (dotted) trajectory, respectively. This initial condition ($\phi_0=0$ and $\theta_0=\pi/4$) corresponds to the initial state for the quantum simulations in the bottom panels via Eq.\eqref{eq:is}. 
Bottom panels: We compare the corresponding analytical prediction [the black (upper) line] for the entanglement entropy (c,d), quantum Fisher information (e,f) and square commutator (g,h), valid in the thermodynamic limit, with exact computations at finite system size $N=50, \,200, \,800$. Here $\lambda_1=1.12$ is the maximal Lyapunov exponent computed in the Appendix \ref{app_lyapuExamples}}.
\label{fig:entasKT}
\end{figure}

\twocolumngrid

 \subsection{Numerical simulations}
 \label{sec_KTnum}
We compare  the predictions of the semiclassical dynamics with the entanglement and chaos indicators obtained via exact numerical computations, specifically via exact diagonalization (ED). 

Our general scheme is the following. We start from an initially polarized state on the Bloch sphere, which corresponds to a spin-coherent state parametrized by the two spherical angles angles $(\theta_0, \phi_0)$ as
\begin{equation}
    \label{eq:is}
    \ket{\psi_0} = \ket{\theta_0, \phi_0} =  e^{i\phi_0 \hat S_z} \, e^{i \theta_0 \hat S_y}\, \ket{S, S} \ ,
\end{equation}
where $\mathbf{\hat{S}}$ are the collective spin operators in Eq.(\ref{Stot}) and  $\ket{S, S}=\ket{S=N/2, S_z=N/2}$ is the fully polarized state in the $z$ direction.

Then, we let it evolve with the Floquet operator \eqref{eq:evKT} generated by the Hamiltonian (\ref{eq:H}), and we compute the stroboscopic time-evolution of the entanglement entropy \eqref{eq:SA}, the QFI \eqref{QFI} and the square commutator (\ref{SC}), at times $t_n=n\tau=0,1,2,\dots$ (recall that we have fixed $\tau=1$). In all our simulations, we fix $\alpha=\pi/2$, while $\beta$ ranges in a sufficiently large interval to appreciate the  order/chaos transition in the classical limit.\\
Let us provide a few details on the ED simulations.
We construct the initial state in Eq.(\ref{eq:is}) following Ref.\cite{LeeLoh2015} and compute the entanglement entropy using the decomposition
in Ref.\cite{Latorre2005}. The numerical QFI is given by the maximal eigenvalue of the covariance matrix ${\text{Cov}(\hat A, \hat B) = 4\langle \hat A \hat B\rangle - 4\langle \hat A\rangle \langle \hat B\rangle}$ with $\hat A, \hat B=\hat S^{x,y,z}$ \cite{Gabbrielli2019}. For the square commutator  (\ref{SC}), we choose ${\hat A=\hat B= \hat S^z/S}$,  for definiteness.

\onecolumngrid

\begin{center}
\begin{figure}[H]
\centering
\includegraphics[width=0.42\textwidth]{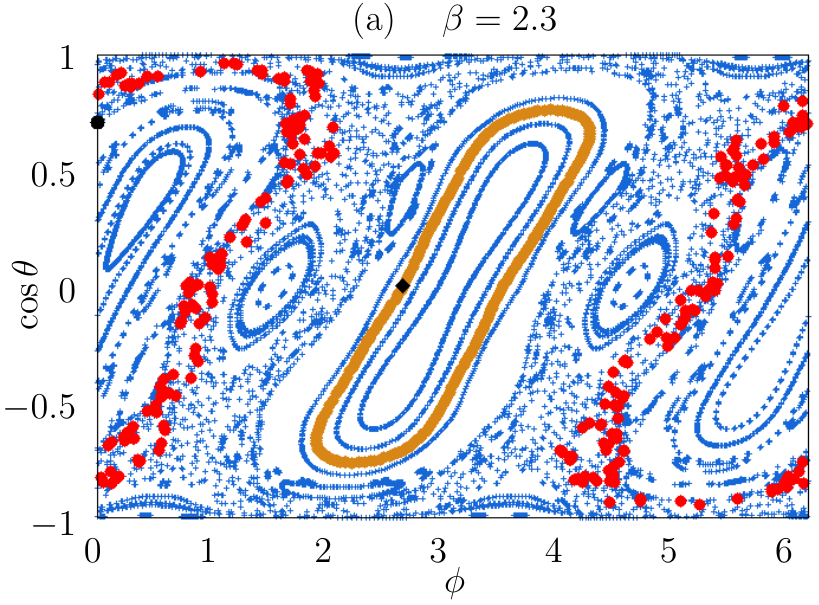} \\
\includegraphics[width=0.43\textwidth]{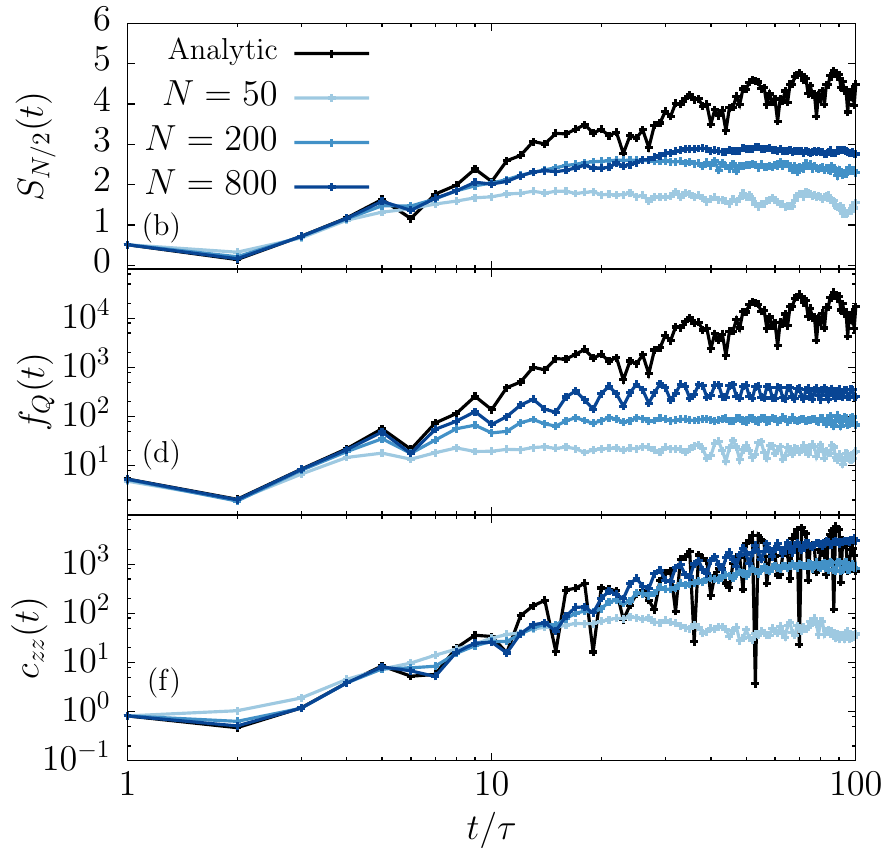} 
\includegraphics[width=0.43\textwidth]{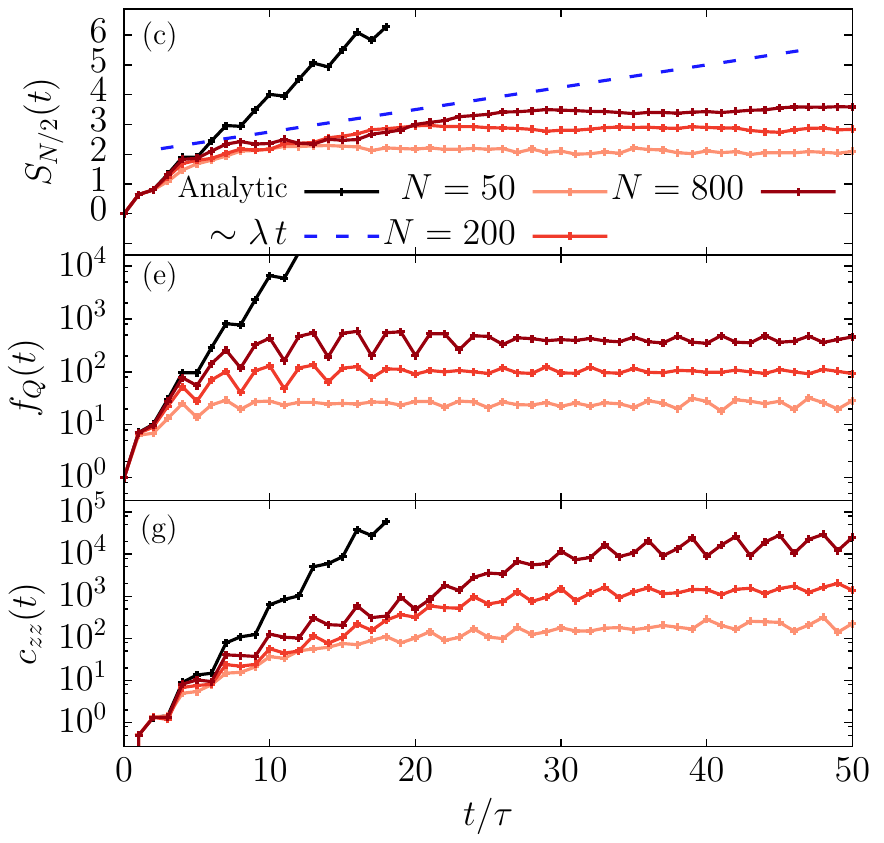} 
\caption{
Dynamics of the quantum kicked top with $\alpha=\pi/2$, in the intermediate regime across the order/chaos transition, with $\beta=2.3$.
Top panel (a): Poincar\`e map
(stroboscopic phase-space trajectories). 
The black dot and diamond, giving rise to the red (dotted) and orange (full line) trajectories, indicate the initial condition of the simulations in the bottom panels. 
Bottom panels: Comparison between the analytical prediction [the black (upper) line] for the entanglement entropy (b,c), quantum Fisher information (d,e) and square commutator (f,g), valid in the thermodynamic limit, and exact computations at finite $N=50, \,200, \,800$.
Left panels: 
Initial condition $\theta_0=\pi/4$, $\phi_0=0$ corresponding to a regular trajectory.  Right panels: Initial condition $\theta_0=\pi/4$, $\phi_0=2.7$ corresponding to a chaotic trajectory. Here $\lambda_1=0.08$ is the maximal Lyapunov exponent computed in the Appendix \ref{app_lyapuExamples}.}
\label{fig:entasKT_mixed}
\end{figure}    
\end{center}

\twocolumngrid

For the semiclassical analysis, we apply the discrete-time map in Eqs.(\ref{eq:KTevAngle1}-\ref{eq:KTevAngle2}) for the classical phase space --- the Bloch sphere, parameterized by the canonically conjugated variables $\cos \theta$ and $\phi$ as in Eqs.\eqref{eq:classKT} --- and in Eqs.\eqref{eq_KTfluc} for the quantum fluctuations. 
The initial conditions are $(\, \cos\theta(0), \phi(0), G_{qq}(0), G_{pp}(0), G_{qp}(0)\, ) = (\cos\theta_0, \phi_0, 1/2, 1/2, 0)$, which represent the state in Eq.\eqref{eq:is}. 
From the time evolution, we directly compute the entanglement entropy, QFI and $c_{zz}(t)$ from Eqs.(\ref{eq:SAnA1}-\ref{eq:detASpin}), Eq.\eqref{eq:QFIsm1Dof} and Eq.\eqref{eq:cabGauss}, respectively. 

A remark is in order concerning the semiclassical numerical methods. 
For these kinds of simulations, it is crucial that the numerical integration is symplectic.
For single degrees of freedom, simplecticity reduces to the conservation of the volume in phase space, i.e., $\det G(t) \equiv 1/4$. 
Although the map in Eqs.\eqref{eq_KTfluc} is exact, we find violations of this conservation law after a few kicks in the chaotic regime, due to machine-precision errors. To the aim of presenting accurate results for the time windows shown in the figures below, we have resorted to a multi-precision arithmetic library \cite{mpmath} and fixed the precision to at least $400$ digits.

\subsection{Discussion}
\label{sec_KTdis}
We study, as a function of the kicking strength $\beta$, how the qualitative change in the semi-classical phase space across the order/chaos transition determines a change in the dynamics of the entanglement. 

In Fig.\ref{fig:entasKT} we present the numerical results deep in the two orderly and chaotic phases. 
For small $\beta$ (left panels), the phase-space trajectories are mostly regular KAM tori. 
In this case, the classical Lyapunov exponent is vanishing.
Accordingly, the asymptotic growth of quantum fluctuations in the semiclassical regime is polynomial in time.
The theory in Sec.\ref{sec_theory} predicts a logarithmic growth of the bipartite entanglement entropy and an exponential growth of the quantum Fisher information and of the square commutator.
As shown in panels (c,e,g), the ED numerical data follow the semiclassical curves for a time window $T_{\text{Eh}}(N) \sim \sqrt{N}$ that increases with the system size.\\
Conversely, for large $\beta$ (right panels), chaos is fully developed in the classical phase space, and the motion is practically ergodic. 
The Lyapunov exponent $\lambda$ is thus positive and almost uniform.
The theory in Sec.~\ref{sec_theory} predicts a linear growth of the bipartite entanglement entropy, with an asymptotic average slope $\lambda$, and an exponential growth of the quantum Fisher information and of the square commutator, with an asymptotic average rate $2\lambda$.
As shown in panels (d,f,h), the ED numerical data follow the semiclassical curves for a time window $T_{\text{Eh}}(N) \sim \ln {N}$ that increases slowly with the system size. \\
Hence, we turn to the intriguing intermediate regime across the order/chaos transition, characterized by a complex structure of phase-space trajectories featuring persisting KAM tori forming stability islands in a growing chaotic sea (we adopt here the standard figurative terminology in the literature).
It is widely known that the point-to-point and finite-time fluctuations of the Lyapunov spectrum are typically strong in Hamiltonian systems with a mixed phase space.
The comparison in Fig.~\ref{fig:entasKT_mixed} allow us to test the theory of Sec.~\ref{sec_theory}.
Even in this case, the finite-size numerical data of the quantum evolution approach the result of the semiclassical computation as $N\to\infty$ for an increasing time window.
The behavior of the entanglement and chaos indicators for both the sample regular and chaotic initial states are partially masked by enhanced oscillations as compared to the corresponding evolution in Fig.~\ref{fig:entasKT_regChao}. Despite this effect, the distinction between the two qualitative behaviors is apparent.\\
In all cases, we observe some extent of discrepancy between the slope or rate of the transient growth of our indicators, and those compatible with the asymptotic Lyapunov exponent. This discrepancy is typically more pronounced when the phase space is complex and mixed [cf. Fig.~\ref{fig:entasKT_mixed}] than in a fully chaotic phase space
[cf. Fig.~\ref{fig:entasKT}].
In Appendix \ref{app_lyapuExamples} we show that this is reflected in the rate of convergence of the numerical computations of $\lambda$.

\section{The Dicke model}
\label{sec_Dicke}
In this section, we will apply the theoretical analysis of Sec.\ref{sec_theory} to the Dicke model introduced in Sec.\ref{sec_dicke}. We will first derive the semiclassical evolution of the quantum fluctuations in Sec.\ref{sec_scEom}. 
Then, in Sec.\ref{sec_numdicke} we compare our analytical predictions with exact numerics in finite-size systems only for the entanglement entropy dynamics. Note that the QFI and the square commutator have been explored in the same context  recently \cite{song_quantum_2012, wang_quantum_2014, zhang_large-n_2015, mirkhalaf_entanglement_2017, Gietka2019, bhattacherjee_spin_2016, buijsman_nonergodicity_2017, alavirad_scrambling_2019, dickeHirsch2010, lewis2019unifying}. \\

 \subsection{Evolution of the quantum fluctuations}
 The evolution of the quantum fluctuations around the classical coupled evolution of the collective spin and of the cavity mode \eqref{eq:DickeCl} can be obtained by adapting the method of Secs.\ref{sec_scEom} and \ref{sec_scEnt}.
 The collective spin fluctuations may be described via a Holstein-Primakoff expansion around the time-dependent direction of the average spin orientation ${ \vec{\mathcal{S}}(t) \equiv \big\langle \hat {\mathbf S}(t) \big\rangle \propto {\mathbf Z}}$, i.e. Eq.\eqref{eq:spinRot}.
The cavity-mode fluctuations are represented by deviations  away from its macroscopic expectation value \eqref{eq_scalingDicke}
\begin{gather}
\left\{
\begin{split}
\label{eq_classVarDicke}
\hat Q & = \sqrt{N} \mathcal{Q}(t) + \delta \hat Q \ , \\
\hat P & = \sqrt{N} \mathcal{P}(t) + \delta \hat P    \ .
\end{split} 
\right .
\end{gather}
The quantum fluctuations are thus compactly denoted by ${\delta \hat{\boldsymbol{\xi}} = (\delta \hat Q, \delta \hat P, \delta \hat q, \delta \hat p)}$. As explained in Sec.\ref{sec_dicke}, the $\sqrt{N}$ scaling of classical variables may be understood as the occurrence that  all terms in the Hamiltonian are   extensive and  balance each other in equilibrium.
Conversely, typical quantum fluctuations in equilibrium, quantified by the expectation values of quadratic bosonic operators, are of order $\mathcal{O}(1)$, i.e., subextensive.
This corresponds to having an effective Planck's constant $\hbar_{\text{eff}}= 1/N$.

The semiclassical equations of motion are found by applying the method of Sec.\ref{sec_scEom}.  Substituting the expansions in Eqs. \eqref{eq:spinRot} and \eqref{eq_classVarDicke} into the Dicke Hamiltonian \eqref{eq_Dicke} and truncating it at the quadratic order, one finds the same structure as in Eq. \eqref{eq:HamiTrunca}. The classical trajectory $\mathcal Q(t), \mathcal P(t)$ and $\mathbf Z(t)$ is determined by the vanishing of the linear term in the quantum fluctuations, i.e., $\hat H_1(t) \equiv 0$. 
Their dynamics are thus regulated by the quadratic Hamiltonian $\hat H_2(t)$, from which we find Eq.\eqref{eq:xiDot}, i.e.
\beq
\label{eq_evolvixiDicke}
\frac{d\,}{dt}\delta \hat {\boldsymbol{\xi}} = A(t)\, \delta \hat {\boldsymbol{\xi}} \ ,
\eeq
with 
\beq
\label{corre_mat}
A(t) = 
\begin{pmatrix}
0 & \omega & 0 & 0 \\
- \omega & 0 & -\frac {\gamma}{\sqrt{2}} \cos\theta \cos \phi & \frac {\gamma}{\sqrt 2}\, \sin \phi \\
- \frac {\gamma}{\sqrt{2}} \sin \phi & 0 & 0 & - \gamma \mathcal Q \frac{\cos\phi }{\sin \theta} \\
- \frac {\gamma}{\sqrt 2}\, \cos\theta \cos \phi & 0 & + \gamma \mathcal Q \frac{\cos\phi }{\sin \theta}  & 0 
\end{pmatrix} \ .
\eeq

\onecolumngrid

\begin{figure}[H]
\centering
\includegraphics[width=0.42\textwidth]{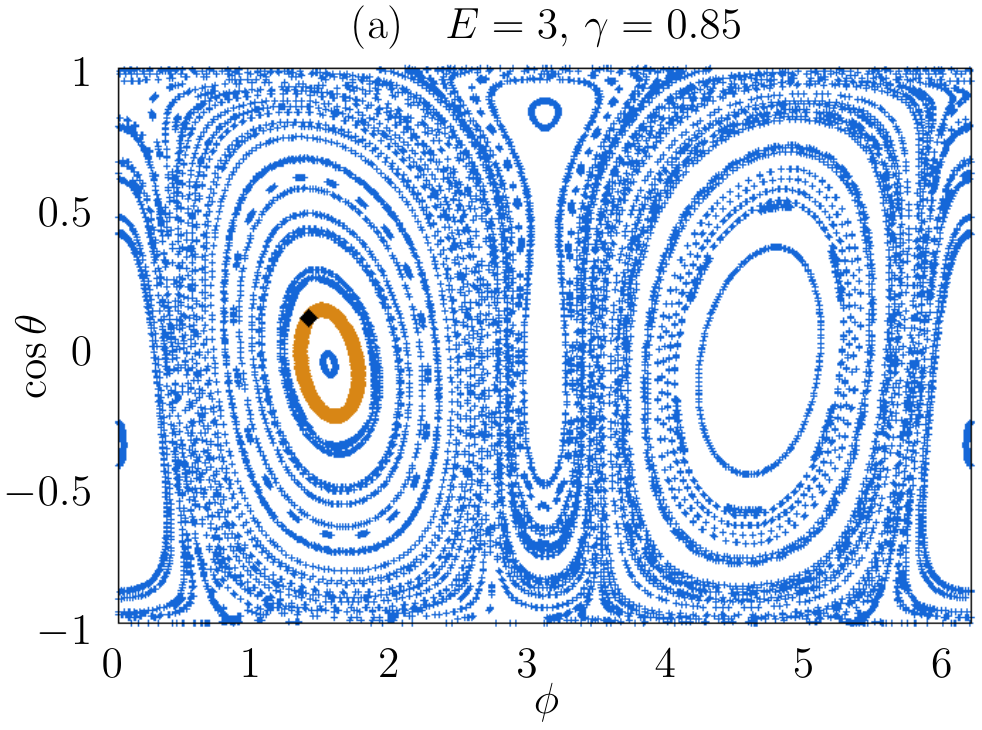} 
\includegraphics[width=0.42\textwidth]{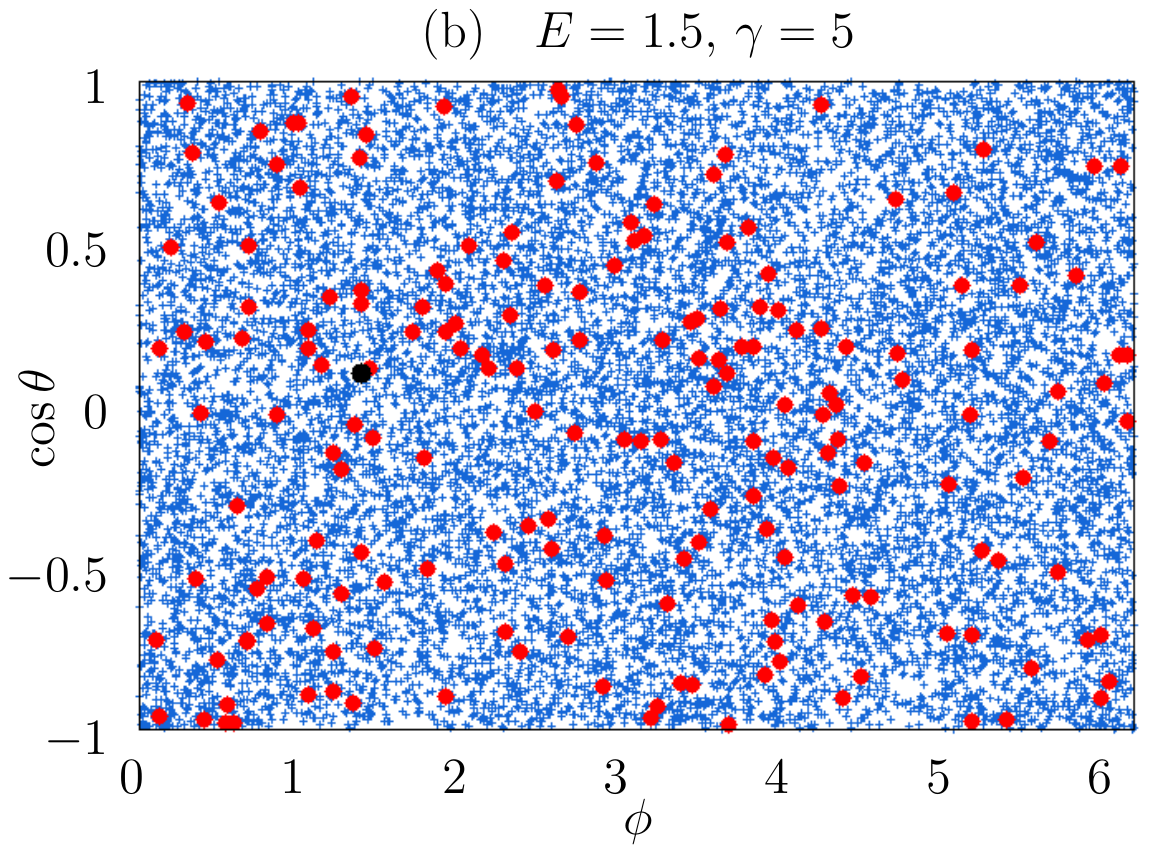} \\
\includegraphics[width=0.43\textwidth]{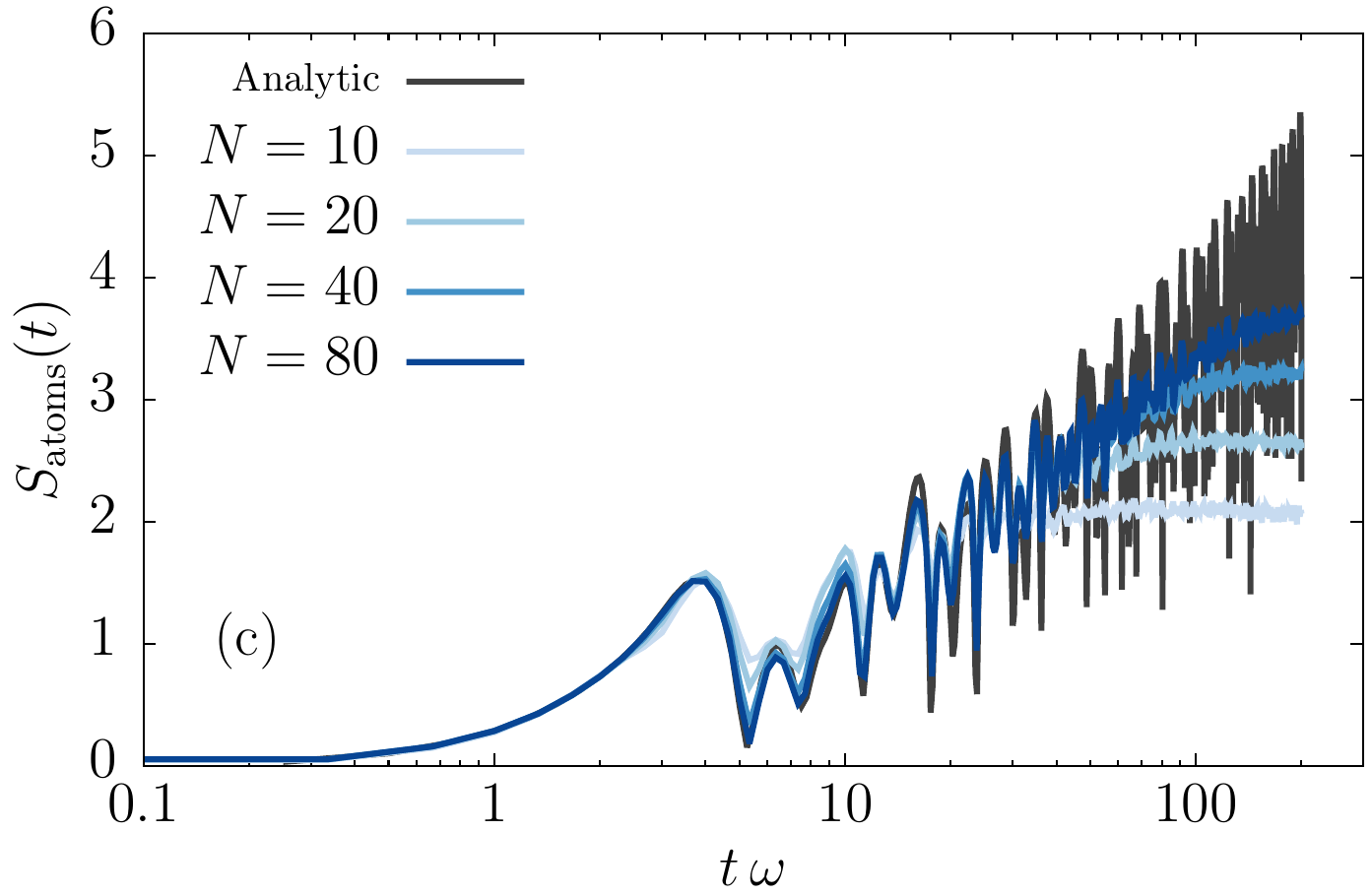} 
\includegraphics[width=0.44\textwidth]{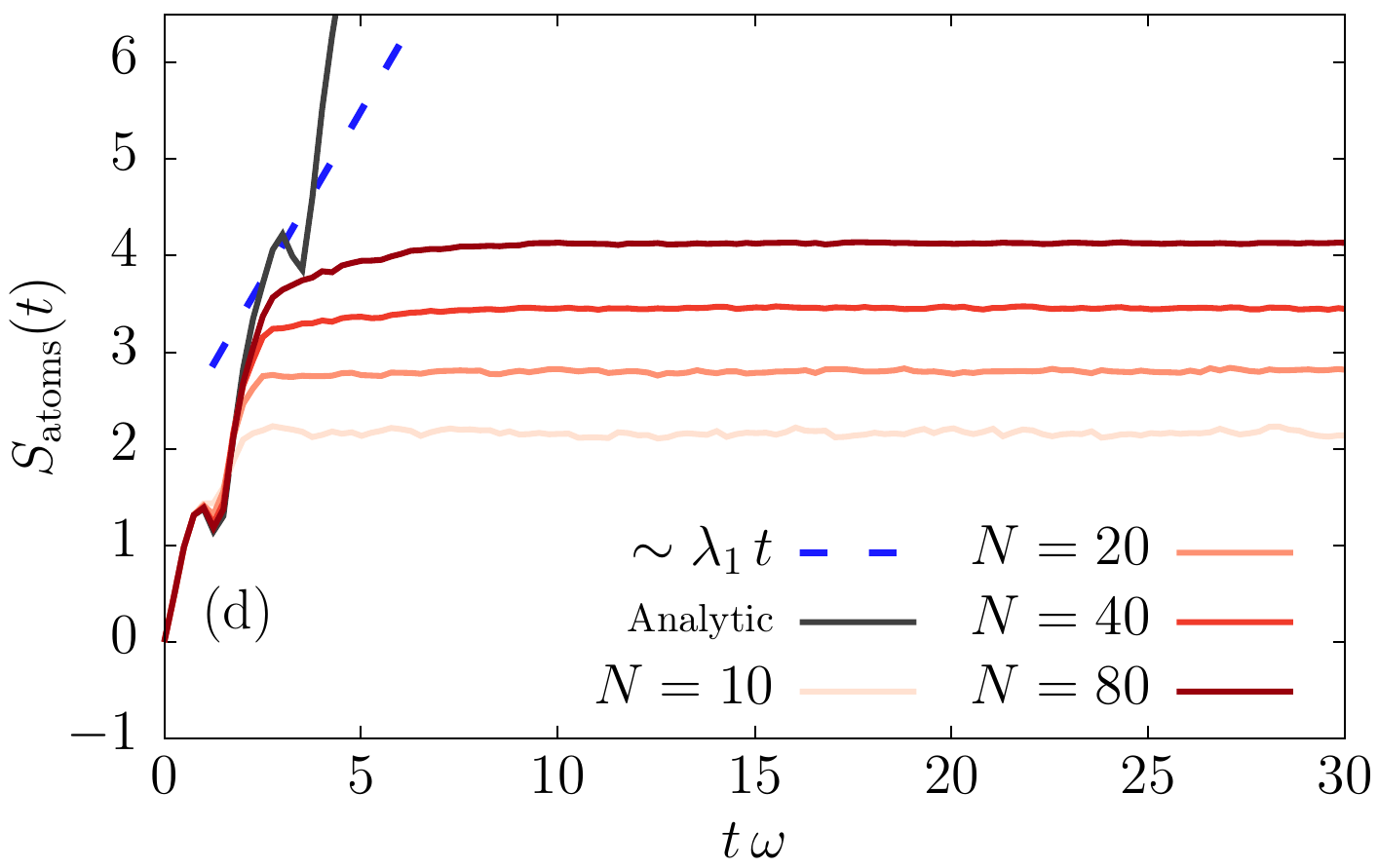} 
\caption{Entanglement dynamics for the Dicke model in the regular (a) and chaotic (b) regime with $E=3$, $\gamma=0.85$ and $E=1.5$, $\gamma=5$. (a-b) Poincarè maps with $\mathcal P=0$ and $\mathcal Q>0$ at fixed energies. The black diamond and dot correspond to the initial condition $(Q_0, P_0, \theta_0, \phi_0)$ $=(Q(E, \theta_0, \phi_0),\, 0.,\, \theta_0, \phi_0)$ with $\theta_0=\text{arcos}(0.1)$ and $\phi_0=1.4$ chosen for the simulation of the entanglement entropy below. (c-d) Comparison between the semi-classical result [the black (upper) line] with exact ED computations at finite $N=10, \,20, \,40,\,80$. (c) Dynamics in regular phase-space $E=3$, $\gamma=0.85$. (d) Dynamics in chaotic phase-space $E=1.5$, $\gamma=5$. Here $\lambda_1=0.7$ is the maximal Lyapunov exponent computed in the Appendix \ref{app_lyapuExamples}.}
\label{fig:entasDicke_RegCha}
\end{figure}

\twocolumngrid
Hence, the evolution of the correlation matrix $G(t)$ is determined via Eq.\eqref{eq:Gevolve} from $A(t)$ by integrating Eq.\eqref{eq_evolvixiDicke}.
The details of the calculation to obtain Eq.\eqref{corre_mat} are reported in Appendix \ref{App:eomDI}. Together with Eqs.\eqref{eq:DickeCl} and with the appropriate initial conditions, Eqs.(\ref{eq_evolvixiDicke},\ref{corre_mat}) give a complete description of the semiclassical dynamics of the Dicke model, before the Ehrenfest time scale $T_{\text{Eh}}$.

\subsection{Numerical simulations}
\label{sec_numdicke}
We now compare the semiclassical predictions for entanglement and chaos indicators with the numerical results obtained via exact diagonalization (ED) of the Hamiltonian. 

We start from an initial state, which is a tensor product of a spin coherent state of the atomic ensemble and a bosonic coherent state for the cavity, namely
\begin{equation}
    \label{eq:isDicke}
    \ket{\Phi_0} = \ket{\theta_0, \phi_0} 
    \otimes
    \ket {\alpha}
    \quad \text{with}
    \quad \alpha = \frac{\mathcal Q_0 + i \mathcal P_0}{\sqrt 2} \ ,
\end{equation}
where $\ket{\theta_0, \phi_0}$ is the spin coherent state defined in Eq.(\ref{eq:is}), while the bosonic coherent state $\ket{\alpha} = e^{i (\alpha \hat b^\dagger + \alpha^* \hat b)} \ket{0}$ is obtained by displacing the standard bosonic coherent vacuum $\ket{0}$ (defined by $\hat b \ket{0} = 0$, $\braket{0|0}=1$) by the complex vector $\alpha$. 
This quantum initial state corresponds to a minimal-uncertainty Gaussian distribution in the classical phase space, centered around the point $(\mathcal Q_0, \mathcal P_0, \cos\theta_0, \phi_0)$ (see e.g. Refs.\cite{deAguiar1992, Furuya1998}). 
Then, we let evolve the system with the Dicke Hamiltonian \eqref{eq_Dicke} and we study the temporal development of quantum correlations.
    
We perform exact diagonalization using QuTip, an open-source software for quantum optics dynamics \cite{Johansson2012, Johansson2013}. 
The spin Hilbert space is treated exactly, while we set a large cutoff $N_{\text{cut}}$ on the photon Hilbert space, checking that the results are converged upon increasing $N_{\text{cut}}$. 
In all simulations, we take a maximum ${N_{\text{cut}} = \Delta \times N}$, where $N$ is the number of spins and $\Delta \sim 4 \div 8$ varies depending on the trajectory. 
A convenient way to a priori estimate the needed magnitude of $\Delta$ is to evaluate the maximum of $(\mathcal Q^2(t) + \mathcal P(t)^2)/2 $ along the reference classical trajectory in the target time window.

In the semiclassical simulations, we start from the classical initial conditions corresponding to the quantum state (\ref{eq:isDicke}). 
We fix $\mathcal P_0=0$ and the value of the energy $E$. 
The classical initial condition is then $(\mathcal Q_0(E, \phi_0, \theta_0), 0, \cos\theta_0, \phi_0)$.


\onecolumngrid

\begin{figure}[H]
\centering
\includegraphics[width=0.42\textwidth]{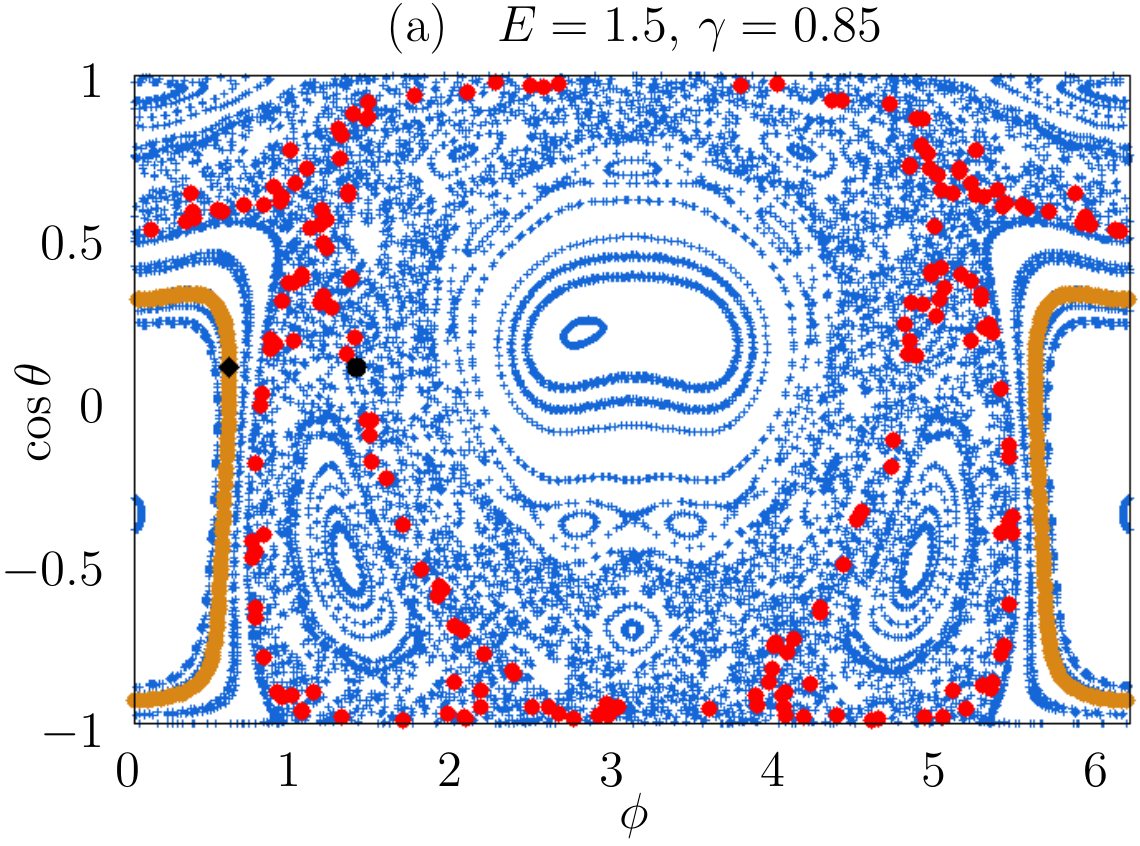}\\ 
\includegraphics[width=0.43\textwidth]{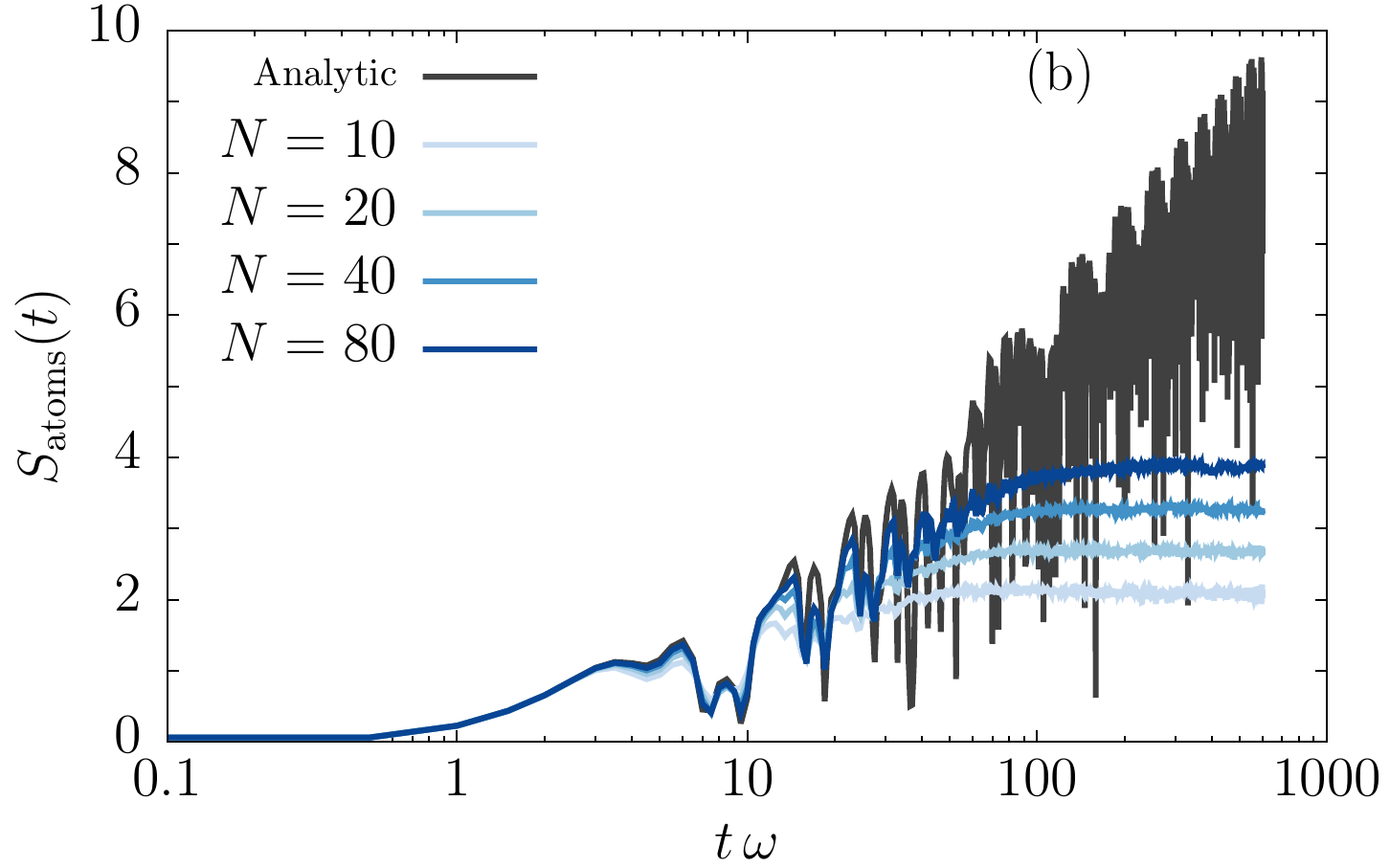} 
\includegraphics[width=0.43\textwidth]{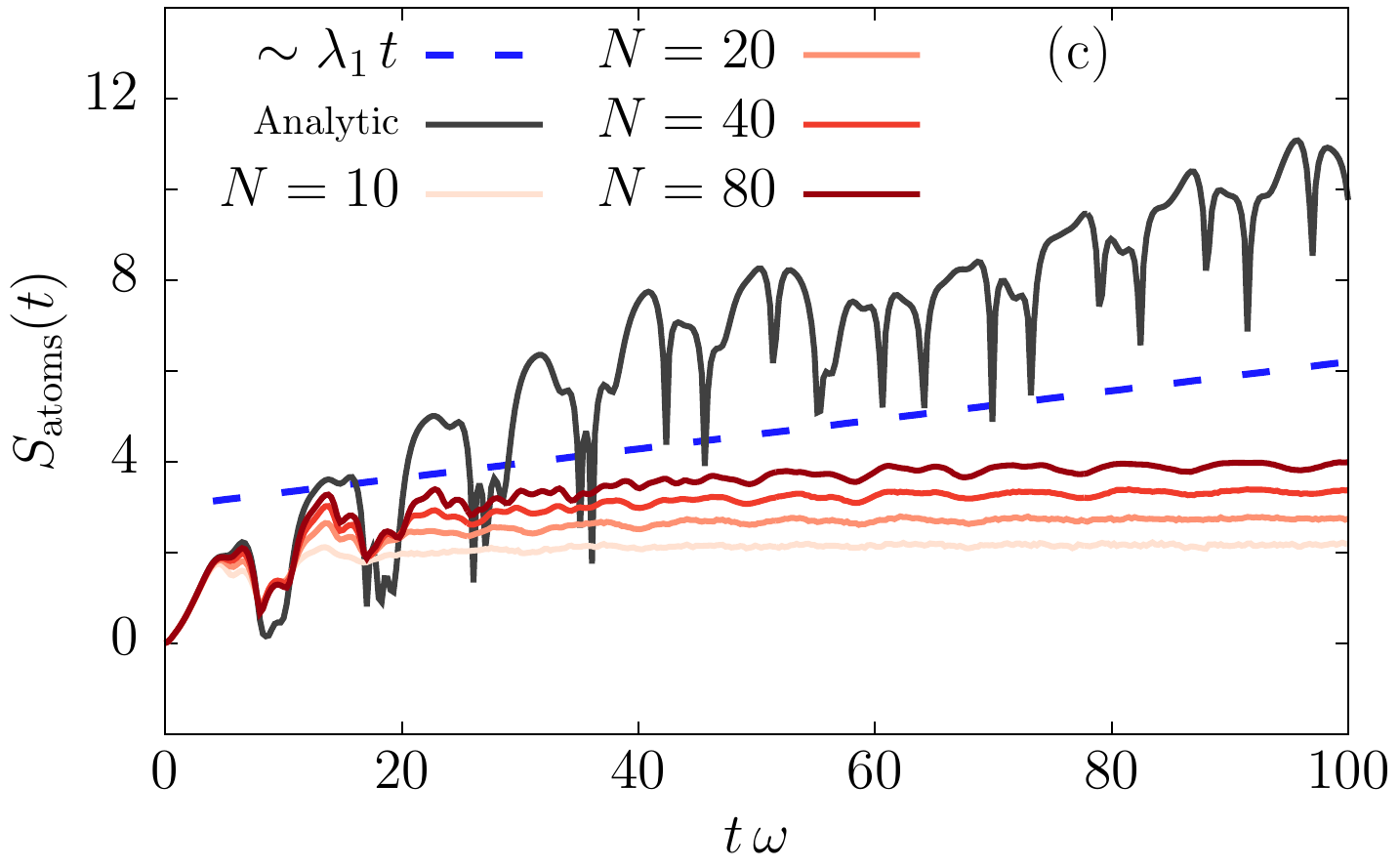} 
\caption{Entanglement dynamics for the Dicke model in the mixed regime with $E=1.5$ and $\gamma=0.85$. (a) Poincarè map with $\mathcal P=0$ and $\mathcal Q>0$ at fixed energy $E=1.5$ with $\gamma=0.85$. The black diamond (dot) correspond to regular (chaotic) initial conditions $(Q_0, P_0, \theta_0, \phi_0)$ $=(Q(E, \theta_0, \phi_0),\, 0.,\, \theta_0, \phi_0)$ with $\theta_0=\text{arcos}(0.1)$ and $\phi_0=0.6$ ($\phi_0=1.4$). (b-c) Comparison between the semi-classical entanglement entropy [the black (upper) line] with exact ED computations at finite $N=10, \,20, \,40,\,80$. (b) Dynamics starting from the regular initial condition $\phi_0=0.6$ (diamond in (a)). (c) Dynamics starting from the chaotic initial condition $\phi_0=1.4$ (dot in (a)). Here $\lambda_1=0.03$ is the maximal Lyapunov exponent computed in the Appendix \ref{app_lyapuExamples}.
\label{fig_dickemixed}
}
\end{figure}

\twocolumngrid

We then numerically integrate  Eqs.\eqref{eq:DickeCl},\eqref{eq_evolvixiDicke}. Since the Dicke Hamiltonian \eqref{eq_Dicke} is non-separable --- i.e., it \emph{cannot} be decomposed as $\mathcal H(\mathbf q, \mathbf p) = K(\mathbf p) + V(\mathbf q)$ --- efficient symplectic integrators are not available. 
For this reason, we employ an auto-adaptive fourth-order Runge-Kutta algorithm, fixing the relative and absolute accuracy to $10^{-14}$.
The lack of symplecticity of the numerical integration is witnessed, e.g., by violations of the phase-space volume conservation. This limitations restricts   the validity of the classical simulations to relatively short times in the chaotic regimes.

Time-evolution is visualized via Poincar\'e sections at fixed energy $E=H(\mathcal{Q}_0,0,\cos\theta_0,\phi_0)$ and $\mathcal P =0$ in the four-dimensional phase space: the diagrams trace out the sequence of points in the $\cos\theta-\phi$ plane where the trajectory pierces the Poincar\'e section with $\mathcal Q > 0$. 
The natural entanglement bipartition in the Dicke model consists in subdividing the degrees of freedom of the atoms and the cavity mode.
For any initial state (\ref{eq:isDicke}), the semiclassical entanglement entropy is thus computed from Eq.(\ref{eq:SAnA1}). 

\subsection{Discussion}
\label{sec_discdicke}
Similarly to the analysis of the quantum kicked top, we investigate all the qualitative dynamical regimes of the Dicke model and validate the correspondence between the entanglement dynamics and chaoticity properties in the semiclassical regime.
Unlike the quantum kicked top, the Dicke model represents an isolated (undriven) system, so the energy is conserved.
As its value of $E$ and/or of the coupling $\gamma$ are varied,
the accessible phase space may undergo a progressive order/chaos transition \cite{emary_chaos_2003}. This allows us to test the theoretical conclusions of Sec.\ref{sec_theory} for autonomous dynamics.

In Fig.\ref{fig:entasDicke_RegCha}, we show the Poincar\'e sections in two limiting cases of predominantly regular and chaotic behavior, in the top left and right panels, respectively.
The initial state in Eq.\eqref{eq:isDicke} associated with the classical phase-space point denoted by a black marker is selected and the corresponding time-evolution of the von Neumann entanglement entropy between atoms and cavity mode is shown in the bottom panels.
As it is apparent, the relation between orderly collective motion and slow logarithmic growth of entanglement on one side, and between collective chaos and fast linear growth of entanglement on the other side, is strongly corroborated by the outcome of the simulations.

In Fig.\ref{fig_dickemixed} we turn to the intermediate regime of mixed classical phase-space across the order/chaos transition.
The system is prepared in the two initial states corresponding to the phase-space points marked in black in the Poincar\'e section (top panel), representative of regular and chaotic trajectories, and the relative nonequilibrium dynamics of the entanglement entropy between atoms and cavity is displayed in the bottom panels. 
Similarly to the case of the quantum kicked top, the asymptotic growth of  the entanglement entropy is partly obscured by pronounced oscillations and strong finite-time fluctuations. 
However, convergence to the semiclassical prediction upon increasing the number $N$ of atoms is observed over an increasing time window.

As in the case of quantum kicked top, we observe deviations between the slope or rate of the transient growth of the entanglement entropy, and that compatible with the asymptotic Lyapunov spectrum. Even in this case, this effect tends to be more pronounced when the phase space is complex and mixed [cf. Fig.\ref{fig_dickemixed}] than in a fully chaotic phase space [cf. Fig.\ref{fig:entasDicke_RegCha}].
Appendix \ref{app_lyapuExamples} discusses the rate of convergence of the numerical computations of the Lyapunov spectrum, highlighting the connection with the discrepancies presented in Figs.\ref{fig:entasDicke_RegCha} and \ref{fig_dickemixed}.

\section{Conclusions and perspectives}
\label{sec_conclusions}

In this work, we presented a unifying framework underlying the growth of entanglement in systems characterized by a well-defined classical limit, in agreement with previous suggestions in the literature \cite{Zurek1994,Bianchi2018,lewis2019unifying}.
Overall, the established picture that the transient entanglement growth happens via decoherence was confirmed, and the exact relationship between the notions of bipartite entanglement, multipartite entanglement and scrambling was clarified in the semiclassical regime.
Quantum entanglement indicators approach a finite limit as the effective Planck's constant vanishes, $\heff\to0$ \cite{zurek1995quantum,Casati_2012}, and this limit possesses a clean interpretation in terms of the subsystem quantum fluctuations around the classical trajectory.
Their temporal growth is associated with the chaoticity properties of the underlying classical phase space.
This allows to make clear quantitative predictions on the asymptotic entanglement growth, based on the knowledge of the classical limit: 
Before the Ehrenfest time, for regular dynamics the entanglement entropy $S_A(t)$ grows only logarithmically in time, while the QFI and the square commutator polynomially; for chaotic dynamics, $S_A(t)$ undergoes a linear growth with a coefficient given by the classical Kolmogorov-Sinai entropy rate, while the QFI and the square commutator grow exponentially with a rate set by twice the largest classical Lyapunov exponent. This discussion is summarized in Table \ref{tab:esempio}.
For the entanglement entropy dynamics, this classification builds on the results of Ref. \cite{Bianchi2018,BianchiModakRigolEntanglementBosons} for quadratic bosonic Hamiltonians.
To the best of our knowledge, these results constitute the first general prediction for the dynamics of the quantum Fisher information and collective spin-squeezing in the semiclassical limit.
{We further discussed the finite-time fluctuations of entanglement quantifiers, crucial in finite quantum systems with a relatively short saturation time, and relate them to the underlying classical trajectories.}
We finally corroborated our analysis via detailed numerical computations in paradigmatic many-body collective quantum systems of current experimental relevance which undergo an order/chaos transition, namely the quantum kicked top and the Dicke model, finding excellent agreement with the analytical predictions in all dynamical regimes.

The semiclassical analysis presented here  underlies  the slow growth of entanglement in spin systems with algebraically-decaying interactions \cite{SacredLog}.
The same approach could be applied to the entanglement growth in open systems, where it has been already shown that quantum fluctuations around the mean-field observables are the responsible for the growth of the entanglement negativity \cite{Benatti2017}.

A very challenging problem is to understand how quantum interference effects enter the game after the Ehrenfest time, causing a saturation of the entanglement quantifiers \cite{Rammensee2018}. 
In fact, the intriguing occurrence that the long-time average of entanglement quantifiers bears signatures of the underlying classical phase space even for intermediate $\heff$ is still incompletely understood and is a matter of ongoing debate \cite{bhosale2018, ruebeck2017}.

%
It is worth stressing that our results contribute to establish a clear predictive framework for the study of entanglement dynamics in more general semiclassical approaches, such as those based on time-dependent variational principles \cite{Hallam2019,Michailidis2020,Ho2019}. {Also, they can likely be extended to match complementary approaches to entanglement dynamics such as that in Ref. \cite{kumari_untangling_2019}.}

We finally reiterate that the connection between entanglement dynamics and chaos studied here, has direct experimental relevance for the detection of entanglement and its dynamics via measurements of collective quantities \cite{ghose_chaos_2008,Bohnet2016,Grttner2017}, the experimental accessibility of which is well-established with standard techniques and tools of quantum atomic experiments --- see Fig.\ref{fig_entDyn} and the relative discussion. 

\acknowledgments

 We acknowledge useful discussions with L. Pezzè and A. Smerzi. 




\appendix

\section{\\ Dynamics in fully connected models: \\ Mapping to an
effective classical dynamics}
\label{App:MF_cl}

In this appendix, we review the general mapping, due to Sciolla and Biroli \cite{SciollaBiroliMF}, of the
quantum dynamics of permutationally-symmetric systems onto the effective semiclassical dynamics of their collective variables in the thermodynamic limit.

With reference to the setting and notations of Sec.\ref{sec_models}, one observes that possible off-diagonal transitions 
governed by the permutationally-symmetric Hamiltonian $\hat H$, are uniquely identified by a set of integers $m_1,\dots,m_q$
\beq
\Ket{N_1,\dots,N_q} \to \Ket{N_1+m_1,\dots,N_q+m_q}.
\eeq
For convenience, we turn the occupation numbers $N_\alpha$ into fractions $x_\alpha\equiv N_\alpha/N$, with $0\le x_\alpha \le 1$ and $\sum_{\alpha=1}^q x_\alpha = 1$,
and denote basis states by $\Ket{\mathbf x}$, where $\mathbf x=(x_1,\dots,x_q)$.
Hence, we write the matrix elements of $\hat H$ as \footnote{For simplicity, we assume time-reversal invariance, which results in real matrix elements $T_{\mathbf m}(\mathbf x) \in \mathbb{R}$.}
\beq
\label{eq_matrixelementsemiclassical}
H_{\mathbf x, \mathbf x'} \equiv \braket{\mathbf x| \hat H | \mathbf x'} = V(\mathbf x) \, \delta_{\mathbf x, \mathbf x'} - \sum_{ \mathbf m \ne \mathbf 0 } T_{\mathbf m}(\mathbf x) \delta_{\mathbf x, \mathbf x' + \mathbf m/N} \ ,
\eeq 
with $\mathbf m = (m_1,\dots,m_q) \in \mathbb{Z}^q$.
 Terms in the Hamiltonian $\hat H$ involving up to $k$ bodies yield ``local'' transitions in the TSS basis, characterized by $\abs{\mathbf m} \equiv \sum_\alpha \abs{m_\alpha} \le 2k$. 
By the extensivity of the Hamiltonian $\hat H$, both  $V(\mathbf x)$ and $T_{\mathbf m}(\mathbf x) $ are extensive,
 \beq
 V(\mathbf x) \sim N \, v(\mathbf x), \qquad T_{\mathbf m}(\mathbf x) \sim N \, t_{\mathbf m}(\mathbf x)\ .
 \eeq
Crucially, the densities $v$ and $t$ are smooth functions of $\mathbf x$, as they generally result from combinatoric factors of the occupation numbers which are insensitive to small changes $N_\alpha \mapsto N_\alpha \pm 1,2,\dots$ to the leading order in the thermodynamic limit $N\to\infty$ \cite{SciollaBiroliMF}.
These properties allow one to rewrite the Schr{\oe}dinger equation for the TSS wavefunction $\psi(\mathbf x,t)$ as
\beq
\label{eq_sciolla}
\frac{i}{N} \frac{\partial }{\partial t} \psi(\mathbf x,t) = \bigg\{ v(\mathbf x) - \sum_{ {\mathbf m}} t_{\mathbf m}(\mathbf x) \cosh \bigg( \frac{\mathbf m}{N} \cdot \frac{\partial}{\partial \mathbf x} \bigg) \bigg\} \psi(\mathbf x,t)\ .
\eeq
%
%
Defining the operators
\beq
x_\alpha \mapsto \hat q_\alpha, \qquad -i \hbar_{\text{eff}}  \frac{\partial}{\partial  x_\alpha} \mapsto \hat p_\alpha\ , 
\eeq
one recognizes that the evolution  in Eq. \eqref{eq_sciolla} is governed by the effective Hamiltonian
\beq
\label{eq_semiclassical}
\mathcal{H}_{\cl}(\hat {\mathbf q}, \hat {\mathbf p}) \equiv v(\hat {\mathbf q}) - \sum_{\mathbf m} t_{\mathbf m}(\hat{\mathbf q}) \cos ( \mathbf m \cdot \hat {\mathbf p} )\ ,
\eeq
with an effective Planck's constant 
\beq
\hbar_{\text{eff}} \equiv \frac{1}{N} \qquad \text{($\hbar=1$ in our units)}
\eeq
that approaches zero in the thermodynamic limit.
Thus, the dynamics of the original system of all-to-all interacting $q$-level units is equivalent to the semiclassical dynamics of $q-1$ collective degrees of freedom via Eq. \eqref{eq_semiclassical} (due to the constraint $\sum_\alpha^q x_\alpha \equiv 1$).

\section{\\ Derivation of the semiclassical evolution equations}
\label{App:eom}

In this appendix we derive the equations of motion of the classical collective variables and of the quantum fluctuations, for the quantum kicked top and the Dicke model.

\subsection{Kicked top}
\label{app:eomKT1}

We start by deriving the stroboscopic map for the classical limit of the kicked top, cf. Eq. \eqref{eq:classKT}.
With reference to the setting and notations of Sec.\ref{sec_introKT},
we adopt a convenient parametrization of the spin via spherical coordinates along the $z$ axis via \eqref{eq:rotating},
so that the nonlinear part of the evolution --- the kick $\hat U_\beta$ --- looks simple. 
The discrete classical map  that describes the stroboscopic evolution of the collective spin  on the Bloch sphere is the composition of two maps, respectively generated by $\hat U_\alpha$ and $\hat U_\beta$. 

The classical map generated by $\hat U_\beta$ reads
\beq
\begin{dcases}
\theta''=\theta'  \\ 
\phi'' = \phi' + \beta \cos\theta' 
\end{dcases}\ .
\eeq
Due to the our choice of coordinates, obtaining the free precession described by $\hat U_\alpha$ is less straightforward.
One strategy is to work it out in spherical coordinates with polar axis along $x$, and to transform into the original coordinates before and after the application of $\hat U_\alpha$.
To this aim, we reparameterize the time-dependent  collective spin direction as 
\beq
\label{eq_sphericalx}
{\mathbf Z} = 
\begin{pmatrix}
\cos \eta \\
\sin \eta \cos \xi \\
\sin \eta \sin \xi 
\end{pmatrix}\ ,
\eeq
where $\eta$ and $\xi$ are respectively the polar and azimuthal angle in spherical coordinates with respect to the $x$ axis. 
With this choice, the classical precession is described as
\beq
\label{eq_classicalevolutionsphericalx}
\begin{dcases}
\eta' =\eta \\
\xi' = \xi + \alpha  
\end{dcases}\ .
\eeq
The expression in the original coordinates is obtained by mapping $(\eta,\xi)$ one-to-one to $(\theta,\phi)$ by equating the two expressions of $\mathbf Z$ in Eqs. \eqref{eq:rotating} and \eqref{eq_sphericalx}. This transformation yields Eqs.\eqref{eq:classKT}.

Let us now determine the evolution of the quantum fluctuations.
The transformation generated by $\hat U_\beta$ can be obtained straightforwardly following the procedure described in Sec.\ref{sec_scEom}.
One gets $\widetilde{H}_2 = \frac 1 2 \beta \sin^2\theta \; \delta \hat q^2 $ in Eq. \eqref{eq:HamiTrunca}, and hence
\begin{align}
\label{eq_KTmap1}
\begin{dcases}
\delta \hat q'' = \delta \hat q'\\
\delta \hat p'' = \delta \hat p' - \beta \, \sin^2 \theta'\,  \delta \hat q'
\end{dcases} \ .
\end{align}
To obtain the discrete transformation generated by $\hat U_\alpha$,
we can again resort to the adapted coordinates $(\eta,\xi)$.
We define the rotated frame $(\bar{{\mathbf X}},\bar{{\mathbf Y}},\mathbf Z)$ with the new spherical angles $\theta \to \eta$, $\phi \to \xi$, i.e.,
\beq
\bar{{\mathbf X}} \equiv \partial_\eta {\mathbf Z} / {\abs{\partial_\eta {\mathbf Z}}}\ , 
\qquad \bar{{\mathbf Y}} \equiv \partial_\xi {\mathbf Z} / {\abs{\partial_\xi {\mathbf Z}}}\ ,
\eeq
 such that $(\bar{{\mathbf X}},\bar{{\mathbf Y}},{{\mathbf Z}})$ is an orthonormal frame adapted to the $(\eta,\xi)$-parametrization of the sphere.
 Along these lines, we define the corresponding transverse spin components and the associated bosonic variables via the truncated Holstein-Primakoff transformation,
 \beq
 \hat S^{\bar{X}} \equiv \bar{{\mathbf X}} \cdot \hat {\mathbf S} \simeq \sqrt{Ns} \, \delta \bar{q}\ , \qquad
  \hat S^{\bar{Y}} \equiv \bar{{\mathbf Y}} \cdot \hat {\mathbf S} \simeq \sqrt{Ns} \, \delta\bar{p}\ .
 \eeq
 In this description, the free precession around $x$ generated by $\hat U_\alpha$ is exactly canceled by the inertial term, and one obtains  
\beq
\label{eq_Gevolutionsphericalx}
\begin{dcases}
\delta\bar{q}'=\delta\bar{q} \\
\delta\bar{p}'=\delta\bar{p} 
\end{dcases} \ .
\eeq
Now, we only need to find the relation between $(\delta\bar{q},\delta\bar{p})$ and $(\delta\hat{q},\delta\hat{q})$.
This can be obtained by noting that both $(\bar{{\mathbf X}},\bar{{\mathbf Y}})$   and $({{\mathbf X}},{{\mathbf Y}})$ are orthonormal bases of the tangent plane to the unit sphere at the point ${{\mathbf Z}}$. 
Therefore, they must be related via a rotation, i.e.,
\beq
\label{eq_relationbetweentangents}
\begin{dcases}
\bar{{\mathbf X}}  &= + \cos \psi \,  {{\mathbf X}}  +  \sin \psi \,  {{\mathbf Y}} , \\
\bar{{\mathbf Y}}  &= -\sin \psi \,  {{\mathbf X}}  +  \cos \psi  \, {\mathbf Y}
\end{dcases} \ .
\eeq
for some angle $\psi \in [0,2\pi)$. 
This angle can be determined by noting that, by construction, $\bar{{\mathbf X}}$ belongs to the plane generated by ${\mathbf x}$ and ${\mathbf Z}$, and hence the equation 
\beq
\bar{{\mathbf X}} \cdot ({\mathbf x} \times {\mathbf Z}) =0
\eeq 
holds. 
Substituting the first of Eqs. \eqref{eq_relationbetweentangents} as well as the third of Eqs. \eqref{eq:rotating}, we find
\beq
\label{eq_anglebetweentangents}
\psi = - \arctan \left(
 \frac{\tan \phi}{\cos\theta}
\right)\ ,
\eeq
which determines $\psi$ up to the ambiguity $\psi \leftrightarrow \psi+\pi$.
Equation \eqref{eq_relationbetweentangents} immediately yields
\beq
\begin{dcases}
\delta \bar{{q}}  &= + \cos \psi \, \delta \hat {{q}}  +  \sin \psi \,  \delta \hat {{p}} , \\
\delta \bar{{p}}  &= -\sin \psi \, \delta \hat {{q}}  +  \cos \psi  \, \delta \hat {{p}} , \\
\end{dcases}
\ .
\eeq
hence one finds
\begin{align}
\label{eq_KTmap2}
\begin{dcases}
\delta \hat q' = +  \cos(\psi-\psi') \, \delta \hat q + \sin(\psi-\psi') \, \delta \hat p \\
\delta \hat p' = -  \sin(\psi-\psi') \, \delta \hat q+  \cos(\psi-\psi') \, \delta \hat p 
\end{dcases} \ .
\end{align}
Substituting the two maps in Eqs. \eqref{eq_KTmap1} and \eqref{eq_KTmap2} into the definition \eqref{eq:Gdef} of the correlation matrix $G(t)$, one directly obtains the desired, ambiguity-free, discrete-time evolution equations \eqref{eq_KTfluc} for the quantum fluctuations.

\subsection{Dicke model}
\label{App:eomDI}
Here, we derive the equations for the classical trajectory \eqref{eq:DickeCl} and for the evolution of the quantum fluctuations around it \eqref{corre_mat} generated by the Dicke Hamiltonian \eqref{eq_Dicke} for large $N$.

 Collective spin fluctuations can be described via a Holstein-Primakoff expansion around the time-dependent direction of the average orientation, as discussed in Sec.\ref{sec_ee_sc}.
Cavity mode fluctuations are represented by deviations  away from its macroscopic expectation value.
With reference to the setting and notations of Sec.\ref{sec_dicke}, one has:
\begin{gather}
\hat S^\alpha \simeq
 { X} _{ \alpha}(t) \; \sqrt{\frac N 2} \; \delta  \hat q \; + { Y}_{ \alpha}(t) \; \sqrt{\frac N 2} \; \delta  \hat p  \\
 \qquad\qquad\qquad\qquad + { Z} _{ \alpha}(t) \; \left(   \frac N 2- \frac{\delta \hat q^2 + \delta  \hat p^2 - 1}{2}   \right) \nonumber \\
\hat Q = \sqrt{N} \mathcal{Q}(t) + \delta \hat Q \\
\hat P = \sqrt{N} \mathcal{P}(t) + \delta \hat P
\end{gather}
with $\alpha = x,y,z$.
The classical functions $\mathcal{Q}(t)$, $\mathcal{P}(t)$ and $\mathbf { Z}(t)$ are chosen in such a way that they account for  the classical dynamics of the system. 
As a consequence, the quantum bosonic  operators $(\delta  \hat q, \delta  \hat p)$ and $(\delta \hat Q, \delta \hat P)$ have vanishing expectation values and describe quantum fluctuations around the classical dynamics.
The $\sqrt{N}$ scaling of classical variables may be understood as the occurrence that  all terms in the Hamiltonian are   extensive (and  balance each other in equilibrium).
Conversely, typical quantum fluctuations in equilibrium, quantified by the expectation values of quadratic bosonic operators, are of order $\mathcal{O}(1)$, i.e., subextensive.
This corresponds to having an effective Planck's constant $\hbar_{\text{eff}}= \hbar/N$.

The semiclassical time-evolution of the system can be obtained by substituting the time-dependent expansion above into the Hamiltonian and truncating to quadratic order, cf. Eq.\eqref{eq:HamiTrunca}.
We obtain
\beq
\hat H =  \;   N \,  \mathcal{H}_{\text{cl}}
\;  + \sqrt{N} \, \hat H_{1} \; + \, \hat H_{2} \; + \, \mathcal{O}\bigg(   \frac{1}{\sqrt{N}} \bigg)\ ,
\eeq
with $\mathcal{H}_{\text{cl}}$ given by Eq.\eqref{eq_HamDickeClass},
\beq
\begin{split}
\hat H_{1} =   &
 \sqrt{s} \big(    \omega_0 \,  { X} _{ z}  + \gamma \mathcal{Q} \, { X}_{ x} \big) 
\; \delta \hat q  \\
&  
 + \sqrt{s} \big(    \omega_0 \,   { Y} _ { z}  + \gamma \mathcal{Q} \, { Y} _{ x} \big)
 \;  \delta  \hat p  \\
 &  
 + \big( \omega \mathcal{Q} + s \gamma \, { Z}_{ x} \big)
 \; \delta\hat{Q}  
  \; \; +(\omega \, \mathcal{P})
 \; \delta \hat P
\end{split}
\eeq
and
\beq
\begin{split}
\hat H_{2} =   &
- \big( \omega_0 \,  { Z} _{ x} +\gamma \mathcal{Q} \, { Z} _{ x} \big)
 \frac{\delta \hat q^2 + \delta \hat p^2 -1}{2}
 \\ & \quad 
+  \omega \frac{\delta \hat Q^2 + \delta \hat P^2 -1}{2} \\
& \quad 
+ \sqrt{s} \gamma \Big( 
{ X} _{ x} \; \delta  \hat q \, \delta \hat Q +  { Y} _{ x} \; \delta  \hat p \, \delta \hat Q 
\Big) \ .
\end{split}
\eeq
The dynamics of quantum fluctuations are generated by the modified Hamiltonian $\widetilde{H} = \hat H - i \dot{\hat V}(t) \hat V^\dagger(t)$, which includes the inertial terms, due to the time-dependence of the transformation:
\beq
\begin{split}
& \widetilde{H}_{1} =  \; \hat H_{1} - \Big(
\sqrt{s} \; \dot{ \mathbf { Y}} \cdot  \mathbf { Z}  \; \delta \hat q
+ \sqrt{s} \; \dot{ \mathbf { Z}} \cdot  \mathbf { X}  \; \delta \hat p
- \dot { \mathcal{P} } \; \delta \hat Q 
+ \dot { \mathcal{Q} } \; \delta \hat P 
\Big)
  \\
& \widetilde{H}_{2} =   \; \hat H_{2} + 
\dot{ \mathbf { X}} \cdot  \mathbf { Y} \,  \frac{\delta \hat q^2 + \delta \hat p^2 -1}{2} \ .
\end{split}
\eeq
In order for the quadratic approximation to be self-consistent, one must appropriately choose the classical functions $\mathcal{Q}(t)$, $\mathcal{P}(t)$ and $\mathbf { Z}(t)$ in such a way that linear terms in the bosonic variables vanish, i.e., $\widetilde{H}_{1} \equiv 0$.
This results in the classical dynamics of the collective spin and the radiation field:
\beq
\label{eq_cl}
\begin{dcases}
\dot{\mathcal{Q}} = \omega \mathcal{P}  \\
\dot{\mathcal{P}} = - \omega \mathcal{Q} - \frac{\gamma}{2}  { Z} _ { x}  \\
\dot{ \mathbf { Y}} \cdot  \mathbf { Z} = \omega_0 \,  { X} _ { z} + \gamma \mathcal{Q} \,  { X} _{ x}\\
\dot{ \mathbf { Z}} \cdot  \mathbf { X} =  \omega_0 \,  { Y} _ { z} + \gamma \mathcal{Q} \, { Y} _{ x} 
\end{dcases} \ .
\eeq
The dynamics of quantum fluctuations is regulated by the equations of motion generated by the quadratic Hamiltonian $\widetilde{H}_{2} $ :
\beq
\label{eq_q}
\begin{dcases}
\dot{\delta \hat Q} =+ \omega  \delta \hat P \\
\dot{\delta \hat P} = -  \omega  \delta \hat Q - \sqrt{s} \gamma ({ X} _ { x}  \, \delta \hat q +  { Y}_ { x} \,  \delta \hat p )  \\
\delta \dot{ \hat q}  = - (\omega_0 \,  { Z} _ { z} + \gamma \mathcal{Q} \, { Z} _ { x} - \dot{ \mathbf { X}} \cdot  \mathbf { Y} ) \delta \hat p + \sqrt{s} \gamma  \,  { Y} _ { x} \, \delta \hat Q  \\
\delta \dot{ \hat p}  = + (\omega_0 \, { Z} _ { z} + \gamma \mathcal{Q} \,  { Z} _ { x} - \dot{ \mathbf { X}} \cdot  \mathbf { Y} ) \delta \hat q  - \sqrt{s} \gamma \, { X} _ { x} \, \delta \hat Q   \\
\end{dcases} \ .
\eeq
With the usual choice of parametrization of the rotating frame \eqref{eq:rotating}-\eqref{eq:rotaFrame} one has 
\beq
\dot{ \mathbf { Y}} \cdot  \mathbf { Z} = - \sin\theta \dot{\phi}\ ,
\qquad
\dot{ \mathbf { Z}} \cdot  \mathbf { X} = \dot{\theta}\ ,
\qquad
\dot{ \mathbf { X}} \cdot  \mathbf { Y} =  \cos\theta \dot{\phi}\ .
\eeq
From these equations, 
%
by substituting explicitly the coordinates \eqref{eq:rotating}-\eqref{eq:rotaFrame},
one gets the classical equations of motion in Eq.\eqref{eq:DickeCl} in the main text, and, for the fluctuations ${\delta \hat{\boldsymbol{\xi}}= (\delta \hat Q, \delta \hat P, \delta \hat q, \delta \hat p)}$, 
Eq. \eqref{eq_q}
%
can be written as 
\beq
\frac{d\,}{dt}\delta \hat{\boldsymbol{\xi}} = A(t)\, \delta \hat{\boldsymbol{\xi}} \ ,
\eeq
with the $4 \times 4$ matrix $A(t)$ expressed by Eq.\eqref{corre_mat} in the main text.

\section{\\ Lyapunov exponents: \\ Theory and numerical applications}

In this appendix, we will first recall the main definitions and properties of the Lyapunov spectrum in Sec.\ref{app:LE}. Then, in Sec.\ref{app:benettin}, we will review the standard algorithm of Benettin \emph{et al.} 1980 \cite{Benettin1980P2} for computing it numerically. We conclude in Sec.\ref{app_lyapuExamples}, by showing the application of the algorithm to the quantum kicked top and the Dicke model.

\subsection{The Lyapunov spectrum and the maximum Lyapunov exponent}
\label{app:LE}

We recall here some elementary but important properties of the Lyapunov spectrum concerning the $K$-dimensional ordiented volumes delimited by $K$ tangent vectors, i.e. $\text{Vol}_K(t) = \text{Vol}[\mathbf w^{(1)}(t), \mathbf w^{(2)}(t), \dots \mathbf w^{(K)}(t)]$. An important consequence of the Oseledets theorem states that the expansion/contraction rate of $\text{Vol}_K(t)$ is given by the sum of the first $K$ exponents as 
\beq
\label{eqA_eesubsystem}
\Lambda_K=\sum_{k=1}^K \lambda_k 
= \lim_{t\to \infty} \frac 1t \, \ln \left[ \frac{\text{Vol}_K(t)}{\text{Vol}_K(0)} \right ]\ .
\eeq
This corresponds to the total expansion rate of a (generic) $K$-dimensional sub-manifold corresponding a subsystem of $K\leq d$ degrees of freedom.
In particular, $\Lambda_d$ is the total expansion rate of the flow, i.e., the average of $\text{div} \, \mathbf f(\mathbf x(t))$ along the trajectory, which vanishes in conservative systems.
For time-independent (autonomous) systems, one has $\lambda_k=0$ for some $k$, because the direction of the trajectory is neither stretched nor shrunk. 
For Hamiltonian systems, which are the focus of this work, Lyapunov exponents come in conjugate pairs $\lambda_k=-\lambda_{2n-k}$ due to the symplectic nature of the phase space flow. In this case, Liouville-integrability of the dynamics is signalled by $\lambda_k\equiv 0$ for all $k$ and in the whole phase space. By contrast, generic Hamiltonian systems do not possess analytic integrals of motion beyond their energy, and their constant-energy surfaces may present a complex structure with invariant submanifolds (KAM tori) intertwined by chaotic regions. In these cases, the Lyapunov spectrum presents strong phase-space and temporal fluctuations.

\subsection{\emph{Benettin \emph{et al.}} algorithm for computing the Lyapunov spectrum}
\label{app:benettin}
The by now standard numerical algorithm for a robust computation of the Lyapunov spectrum has been proposed by Benettin, Galgani and Strelcyn in a series of papers around 1980, see Refs. \cite{Benettin1976, Benettin1980P1, Benettin1980P2}.
Its central idea is based the evolution of $K$ tangent vectors $(\mathbf w^{(1)}, \dots \mathbf w^{(K)} )$ and the use of Eq.\eqref{eqA_eesubsystem} to compute the volumes $\text{Vol}_K(t)$ and the resulting Lyapunov exponents $\{\lambda_k\}_{k=1}^K$.  
In chaotic systems, numerical  errors grow exponentially fast in time and infinitesimal displacements $\mathbf w^{(k)}(t)$ might result in computer overflows at large $t$.
To solve these issues, the method relies on the periodic orthonormalization of the evolved tangent-space basis, after a suitable time interval $s$. 
%
{(This allows one to disregard the numerical instability due to the use of non-symplectic integrators.)}
In Ref.\cite{Benettin1980P2}, the authors show that by choosing the Gram-Schmidt orthonormalization procedure, one can evaluate all the volumes $\{\text{Vol}_k\}_{k=1}^K$ at once.

Fixing initial condition $\mathbf{x}(0)=\mathbf x_0$, the procedure goes as follows. Choose $K$ independent tangent vectors at random at $t=0$, i.e. $\{\mathbf w^{(k)}_{0}\}_{k=1}^K$. Then, for $1\leq i\leq n$:
\begin{enumerate}
    \item evolve the vectors $\{\mathbf w^{(k)}_{(i-1)s}\}$ for a time interval $s$ via Eq.\eqref{eq_clDispEv} and initial conditions $\mathbf x_{(i-1)s}$; this yields $\{\mathbf w^{(k)}_{is}\}$;
    \item apply Gram-Schmidt procedure  
    \begin{equation}
    \alpha^{(1)}_i = | \mathbf w^{(1)}_{is} | \ , \quad
    \mathbf w'^{(1)}_{is} = \mathbf w^{(1)}_{is} / \alpha^{(1)}_i  
    \end{equation}
    \noindent where $|\cdot |$ is the {euclidean norm} \footnote{Note that the specific choice of the phase space metric is actually immaterial.}. For $2\leq k\leq K$
    \begin{subequations}
    \begin{align}
        \alpha_i^{(k)} & = \left | \mathbf w^{(k)}_{is} - \sum_{l=1}^{k-1} (\mathbf w'^{(l)}_{is} \cdot \mathbf w^{(k)}_{is}) \, \mathbf w'^{(l)}_{is}
        \right | \ ,
        \\
        \mathbf w'^{(k)}_{is} & = \frac 1{\alpha_i^{(k)}}\,\left ( \mathbf w^{(k)}_{is} - \sum_{l=1}^{k-1} (\mathbf w'^{(l)}_{is} \cdot \mathbf w^{(k)}_{is}) \, \mathbf w'^{(l)}_{is}\right ) \ ;
    \end{align}    
    \end{subequations}
    \item re-initialize the vectors $\mathbf w^{(k)}_{is}= \mathbf w'^{(k)}_{is}$ for $1\leq k\leq K$.
\end{enumerate}
From this, the finite-time Lyapunov spectrum $\{\lambda_k(\mathbf x_0)\}$ is computed as 
\beq
\label{eqA_numeLambdak}
\lambda^{(n,s)}_k(\mathbf x_0) = \frac 1{ns}\, \sum_{i=1}^n\, \ln \alpha_i^{(k)} \ ,
\eeq
for $k\leq1\leq K$. 
Convergence as $n\to \infty$ yields the proper, asymptotic Lyapunov spectrum.
Notice that  $\lambda^{(n,s)}_k(\mathbf x_0)$ should not depend on the time-interval $s$ and on the number of iterations $n$ independently, but rather via the product $r=s\,n$, i.e. $\lambda^{(r)}_k(\mathbf x_0)$. As $r$ increases, $\lambda^{(r)}_k$ approaches a well defined limit, the $k$-th Lyapunov exponent $\lambda_k=\lim_{r\to\infty}\lambda_k^{(r)}$.

\begin{figure}[b]
\centering
\includegraphics[width=0.45\textwidth]{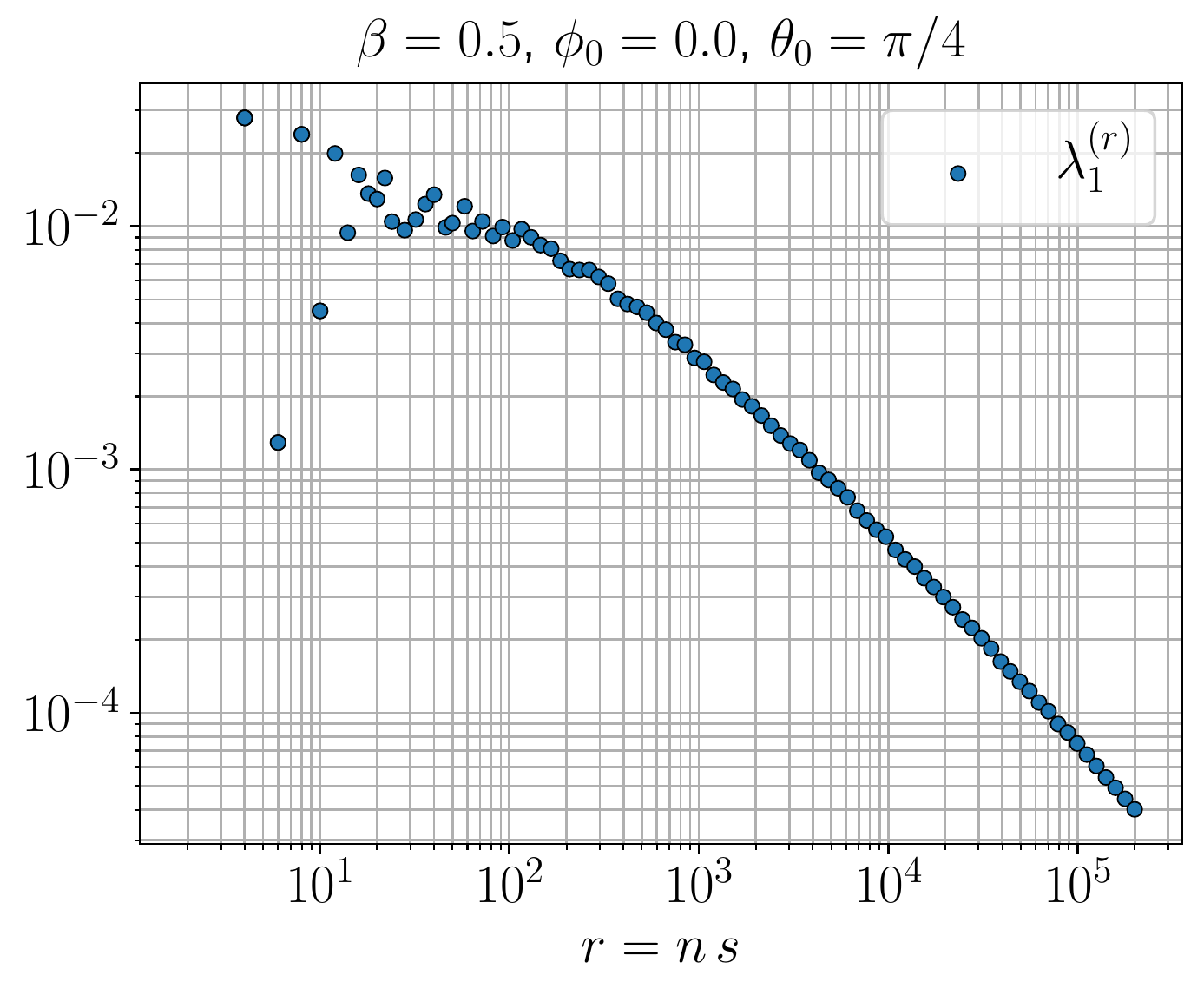} 
\includegraphics[width=0.45\textwidth]{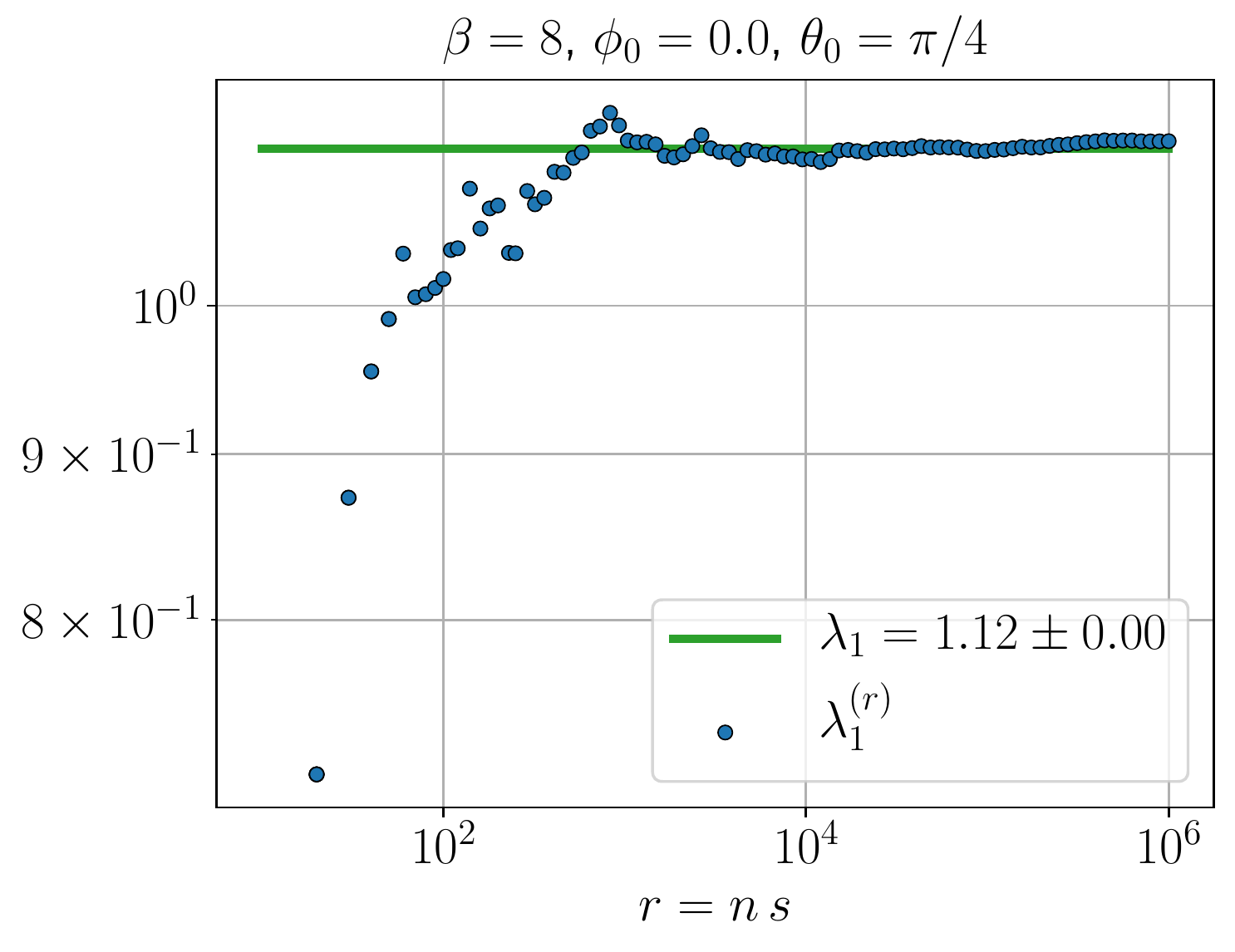} 
\label{fig:entasKT_regChao}
\caption{Convergence of the maximum Lyapunov exponent \eqref{eqA_numeLambdak} for the kicked top dynamics in the predominantly regular and chaotic regimes.
The trajectories shown here correspond to those in Fig.\ref{fig:entasKT}, with initial condition $\phi_0=0$ and $\theta_0=\pi/4$:
a regular one for $\beta=0.5$, with $s=2$ (top panel) and a chaotic one for $\beta=8$, with $s=10$ (bottom panel). 
}
\end{figure}

\subsection{Lyapunov exponents for the kicked top and the Dicke model}
\label{app_lyapuExamples}

We report the computation of the Lyapunov exponents of the kicked top (see Sec.\ref{sec:KT_SFev}) and of the Dicke model (see Sec.\ref{sec_Dicke}) obtained via  the algorithm described in App.\ref{app:benettin}.
%
\begin{figure}[t]
\centering
\includegraphics[width=0.45\textwidth]{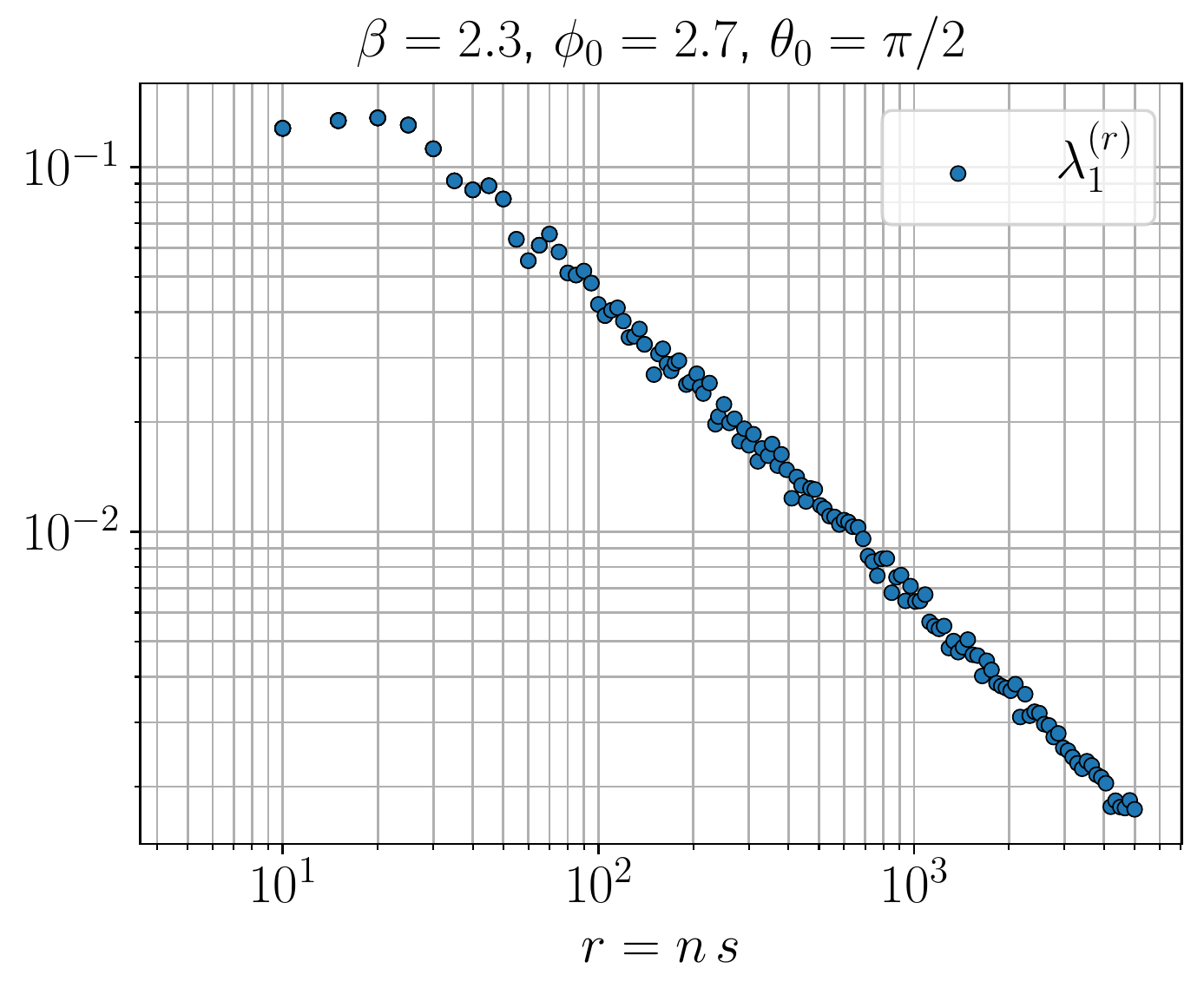} 
\includegraphics[width=0.45\textwidth]{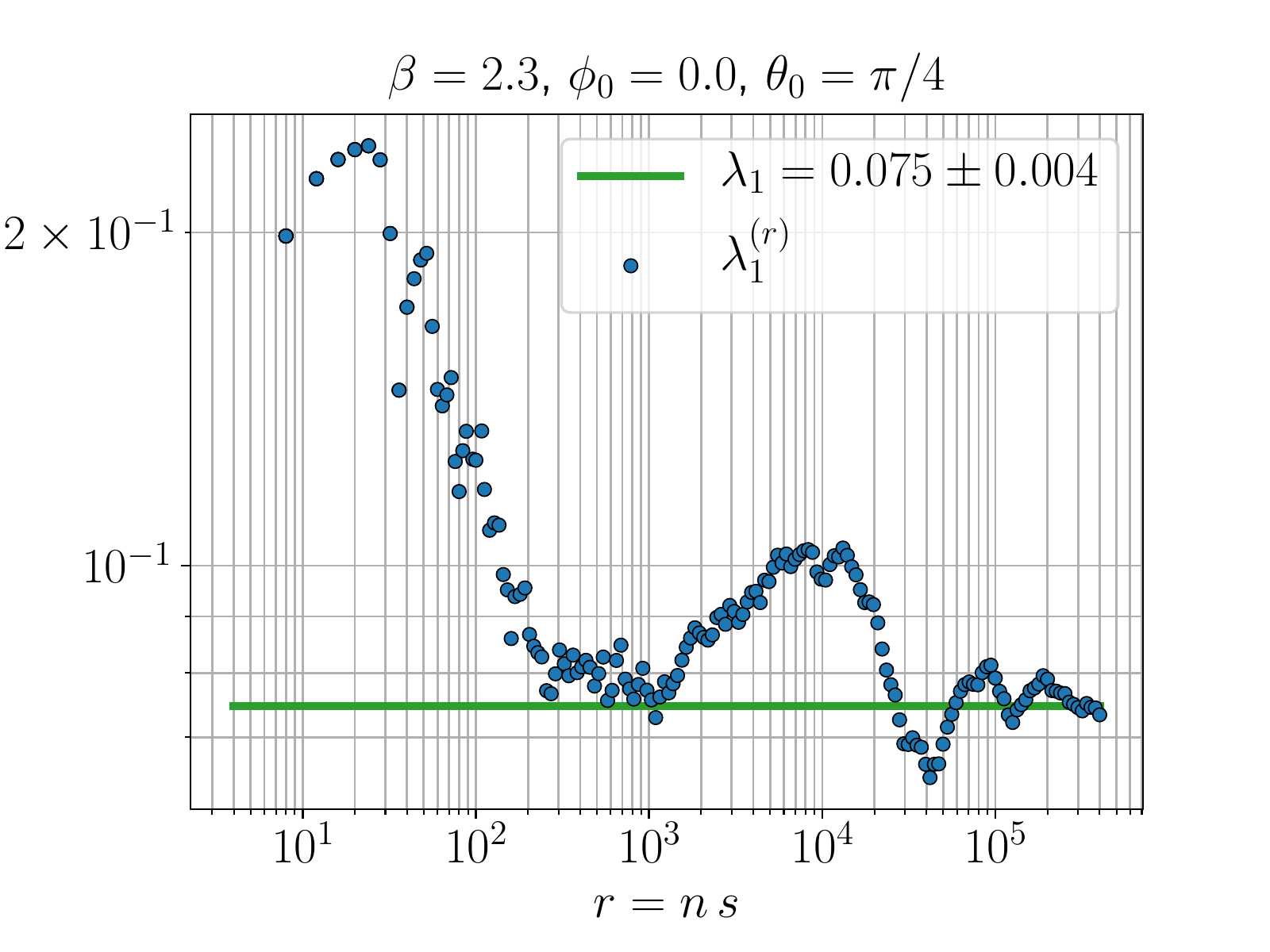} 
\label{fig:entasKT_Mixed}
\caption{Convergence of the maximum Lyapunov exponent \eqref{eqA_numeLambdak} for the kicked top dynamics in the intermediate regime with a mixed phase space.
The trajectories correspond to those in Fig.\ref{fig:entasKT_mixed}, with $\beta=2.3$:
a regular one with $\theta_0=\pi/2$ and $\phi_0=2.7$ (top panel),
and a chaotic one with $\theta_0=\pi/4$ and $\phi_0=0$ (bottom panel). 
Here we have fixed $s=5$.
}
\end{figure}
%
We apply that procedure to the kicked top evolution at stroboscopic times \eqref{eq:classKT}, by evolving the linear displacements via the map in  Eqs.(\ref{eq_KTmap1}-\ref{eq_KTmap2}). We fix a number $s$ of kicks  and we study the black trajectories in Fig.\ref{fig:entasKT}-\ref{fig:entasKT_mixed}. The results are shown in Fig.\ref{fig:entasKT_regChao} and Fig.\ref{fig:entasKT_Mixed} respectively. 
We plot the finite-time maximum Lyapunov exponent $\lambda_1^{(r)}$ in Eq.\eqref{eqA_numeLambdak} as a function of $r$. The maximum Lyapunov exponent $\lambda_1$ (green in the plots) is extracted numerically by averaging over the last two decades of the time window. For regular initial conditions it approaches zero in the long-time limit $r\to\infty$, while for chaotic trajectories it clearly converges to a finite value, at very large times $r\gg10^4$. As expected, the Lyapunov exponent for the chaotic trajectory in the intermediate regime with a mixed phase space is much smaller than the one for the fully chaotic phase, and convergence to the asymptotic value is much slower.

\begin{figure}[t]
\centering
\includegraphics[width=0.45\textwidth]{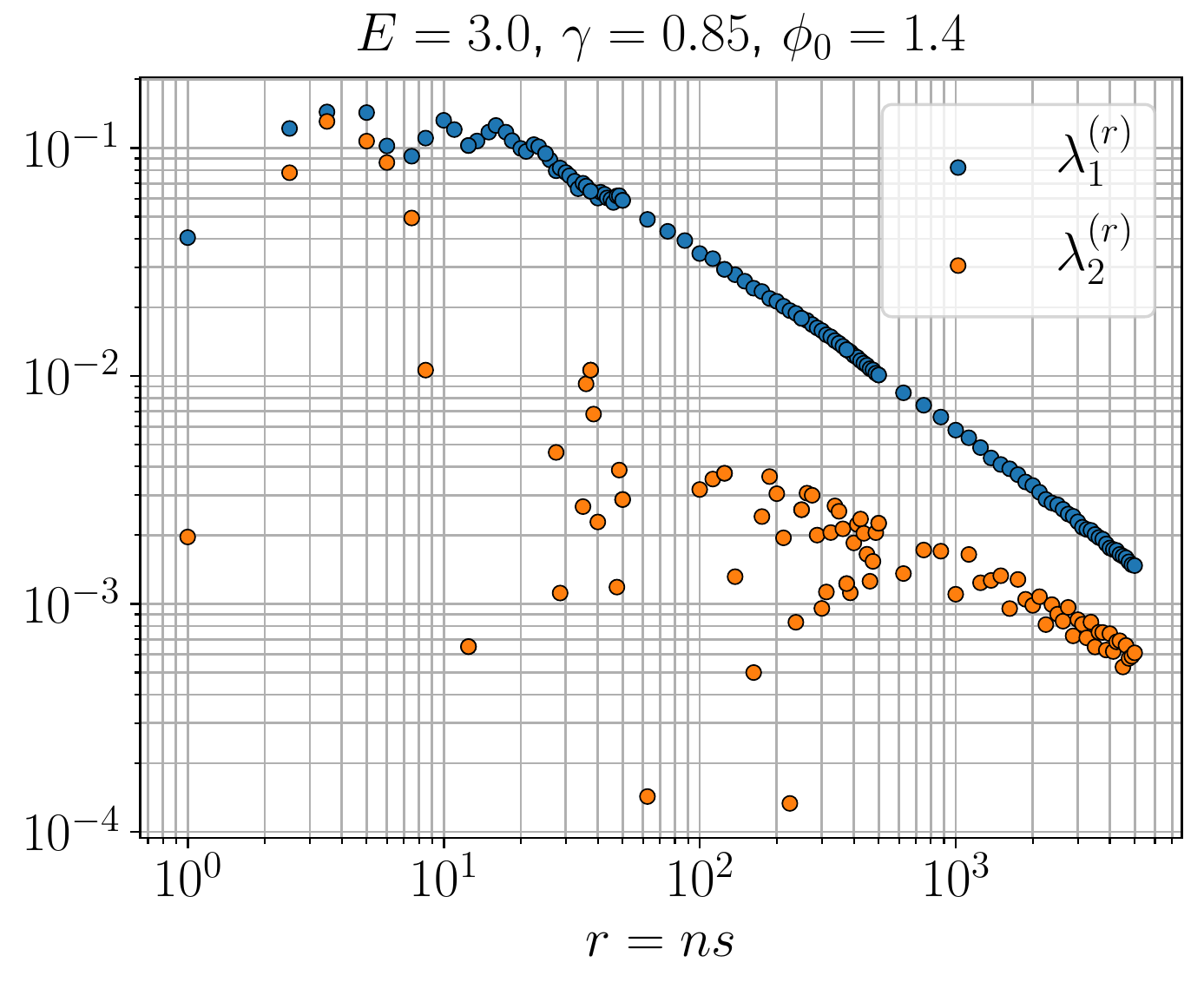} 
\includegraphics[width=0.45\textwidth]{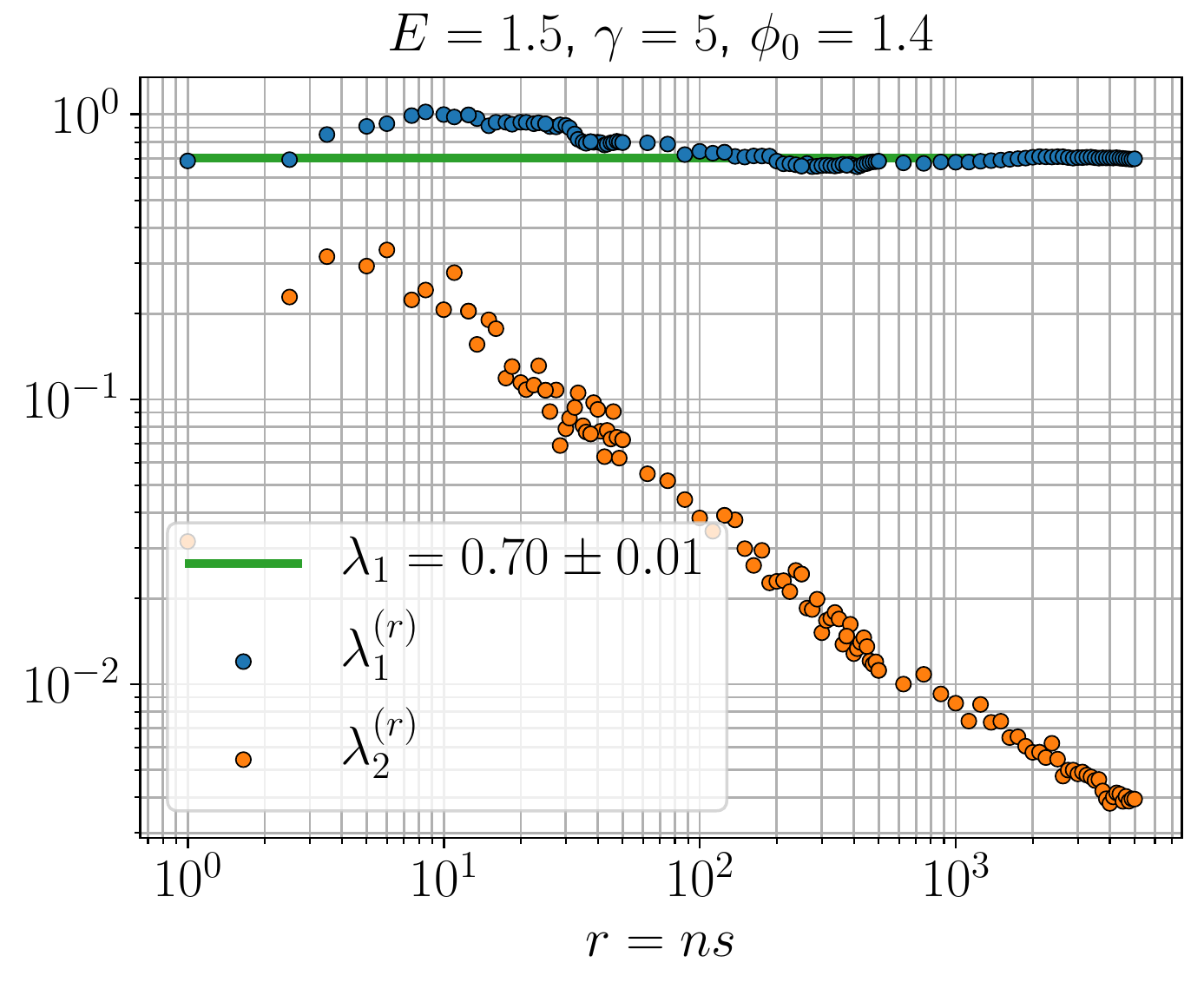} 
\caption{Convergence of the maximum Lyapunov exponent \eqref{eqA_numeLambdak} for the Dicke model dynamics in the predominantly regular and chaotic regimes. 
Top panel: regular trajectory with $E=3$, $\gamma=0.85$ Bottom panel: chaotic trajectory with $\beta=1.5$, $\gamma=5$. 
The common initial condition $\phi_0=1.4$ and $\cos \theta_0=0.1$ corresponds to the two trajectories in Fig.\ref{fig:entasDicke_RegCha}.
Here we have set $s=0.5$.}
\label{fig:entasDicke_regChao}
\end{figure}

\begin{figure}[t]
\centering
\includegraphics[width=0.45\textwidth]{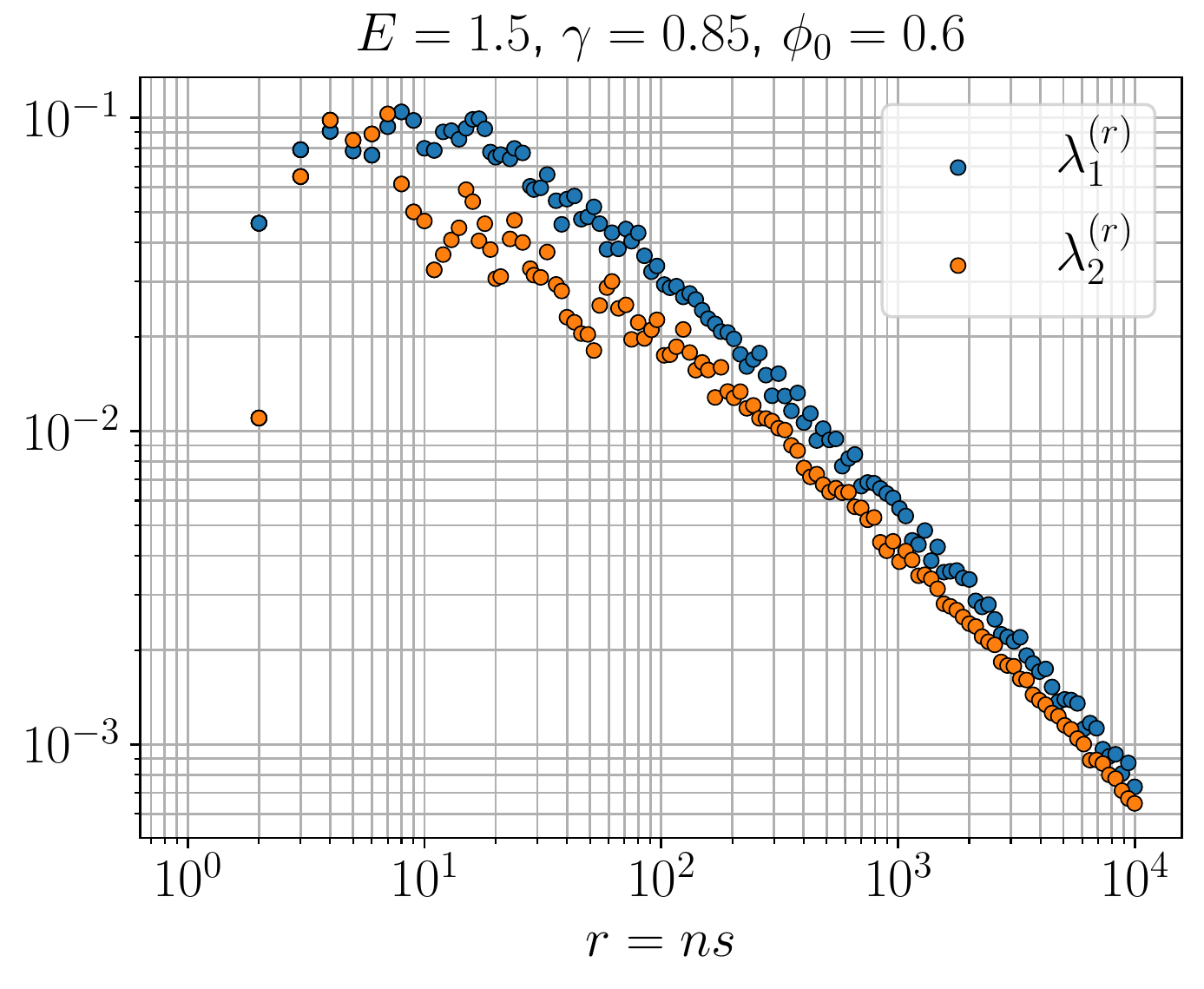} 
\includegraphics[width=0.45\textwidth]{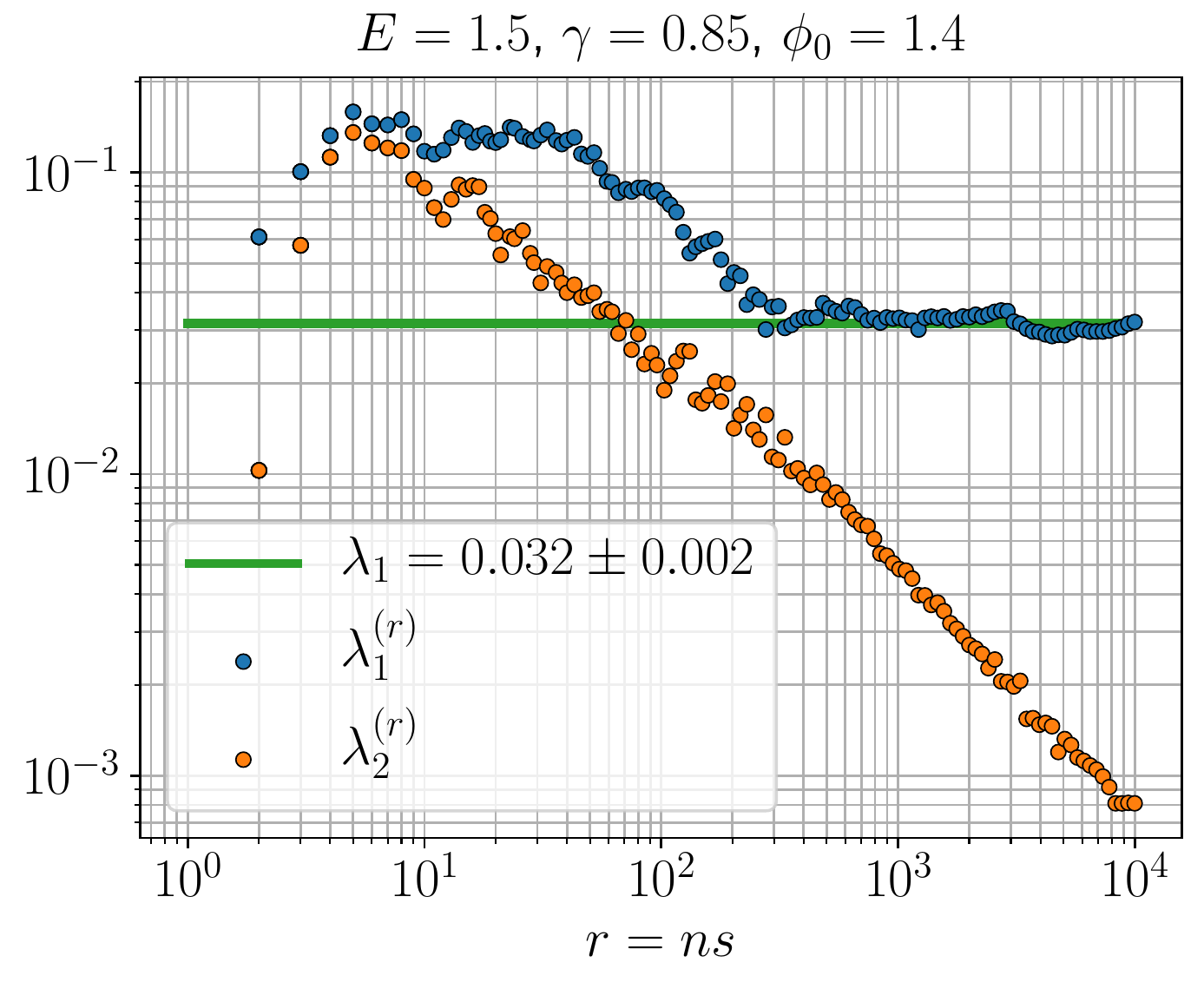} 
\caption{Convergence of the maximum Lyapunov exponent \eqref{eqA_numeLambdak} for the Dicke model dynamics in the intermediate regime with a mixed phase space. 
Here, $E=1.5$, $\gamma=0.5$.  
Top panel: Regular trajectory with initial condition $\cos\theta_0=0.1$ and $\phi_0=0.6$. 
Bottom panel: chaotic trajectory with initial condition  $\cos\theta_0=0.1$ and $\phi_0=1.4$. 
The parameters and initial conditions chosen here correspond to the two highlighted trajectories in Fig.\ref{fig:entasDicke_mixed}.
Here we have set $s=1$.
}
\label{fig:entasDicke_mixed}
\end{figure}

The same procedure is applied to the classical dynamics of the Dicke model \eqref{eq:DickeCl}, fixing $s=1$. 
The results for the regular and chaotic regimes are plotted in Fig.\ref{fig:entasDicke_regChao} and for the intermediate regime with a mixed phase space in Fig.\ref{fig:entasDicke_mixed}. 
Because of the conservation of energy, the second Lyapunov exponent $\lambda_2$ always vanishes. As for $\lambda_1$, similar remarks to the case of the kicked top apply.

\clearpage 
\bibliography{biblio2}  

\end{document}